\documentclass[%
reprint,
superscriptaddress,
frontmatterverbose,
showpacs,
preprintnumbers,
nofootinbib,
nobibnotes,
amsmath,amssymb,
aps,
prc,
]{revtex4-2}
\usepackage[pdftex]{graphicx}
\usepackage{dcolumn}
\usepackage{bm}
\usepackage[pdftex]{color}
\usepackage{CJK}
\usepackage[T1]{fontenc}
\usepackage{mathrsfs}
\def\brakket#1#2#3{\left\langle{#1}\middle|{#2}\middle|{#3}\right\rangle}
\def\ve#1{{\bm{#1}}}
\def\nuc#1#2#3{{}^{#2}_{#3}\mathrm{#1}}
\def\urm#1{\scriptstyle{\text{\textrm{\textmd{\textup{#1}}}}}}

\def\ca#1{{\mathcal{#1}}}

\let\temp\epsilon
\let\epsilon\varepsilon
\let\varepsilon\temp
\let\temp\relax
\let\temp\phi
\let\phi\varphi
\let\varphi\temp
\let\temp\relax
\DeclareMathOperator{\laplace}{\Delta}

\begin{document}
%
\begin{CJK*}{UTF8}{}
  \preprint{RIKEN-iTHEMS-Report-23}
  \title{
    Effects of Coulomb and isospin symmetry breaking interactions
    on neutron-skin thickness}
  \author{Tomoya Naito (\CJKfamily{min}{内藤智也})}
  \email{
    tnaito@ribf.riken.jp}
  \affiliation{
    RIKEN Interdisciplinary Theoretical and Mathematical Sciences Program (iTHEMS),
    Wako 351-0198, Japan}
  \affiliation{
    Department of Physics, Graduate School of Science, The University of Tokyo,
    Tokyo 113-0033, Japan}
  \author{Gianluca Col\`{o}}
  \email{
    colo@mi.infn.it}
  \affiliation{
    Yukawa Institute for Theoretical Physics, Kyoto University,
    Kyoto 606-8502, Japan}
  \affiliation{
    Dipartimento di Fisica, Universit\`{a} degli Studi di Milano,
    Via Celoria 16, 20133 Milano, Italy}
  \affiliation{
    INFN, Sezione di Milano,
    Via Celoria 16, 20133 Milano, Italy}
  \author{Haozhao Liang (\CJKfamily{gbsn}{梁豪兆})}
  \email{
    haozhao.liang@phys.s.u-tokyo.ac.jp}
  \affiliation{
    Department of Physics, Graduate School of Science, The University of Tokyo,
    Tokyo 113-0033, Japan}
  \affiliation{
    RIKEN Interdisciplinary Theoretical and Mathematical Sciences Program (iTHEMS),
    Wako 351-0198, Japan}
  \author{Xavier Roca-Maza}
  \email{
    xavier.roca.maza@mi.infn.it}
  \affiliation{
    Dipartimento di Fisica, Universit\`{a} degli Studi di Milano,
    Via Celoria 16, 20133 Milano, Italy}
  \affiliation{
    INFN, Sezione di Milano,
    Via Celoria 16, 20133 Milano, Italy}
  \author{Hiroyuki Sagawa (\CJKfamily{min}{佐川弘幸})}
  \email{
    sagawa@ribf.riken.jp}
  \affiliation{
    Center for Mathematics and Physics, University of Aizu,
    Aizu-Wakamatsu 965-8560, Japan}
  \affiliation{
    RIKEN Nishina Center, Wako 351-0198, Japan}
  \date{\today}
  \begin{abstract}
    Both the Coulomb interaction and isospin symmetry breaking (ISB) parts of the nuclear interaction break the isospin symmetry in atomic nuclei.
    Effects of these two kinds of interaction on properties of atomic nuclei,
    especially,
    the mass difference of mirror nuclei and 
    the neutron-skin thickness of $ N = Z $ and $ N \ne Z $ nuclei,
    are discussed.
    It is found that corrections to the Hartree-Fock-Slater approximation for the Coulomb interaction negligibly affect the neutron-skin thickness,
    while the charge-symmetry breaking term originating from the strong interaction might affect it non-negligibly.
    According to our calculations, the ISB terms other than the Coulomb interaction affect the estimation of the density dependence of the symmetry energy, $ L $, by about $ 0 $--$ 12 \, \mathrm{MeV} $
    using the correlation with the neutron-skin thickness.
  \end{abstract}
  \maketitle
\end{CJK*}
%
%
\section{Introduction}
\par
The isospin invariance of strong interaction was firstly proposed by Heisenberg in 1932~\cite{
  Heisenberg1932Z.Phys.77_1}.
If the isospin symmetry of the strong interaction is fully valid, 
the charge symmetry and the charge independence of nuclear interaction hold.
Here, the former denotes the case that the proton-proton nuclear interaction $ v_{pp} $ is the same as the neutron-neutron interaction $ v_{nn} $,
and the latter denotes the case that the $ T = 1 $ channel of the proton-neutron nuclear interaction $ v_{pn} $ is identical to the average of $ v_{pp} $ and $ v_{nn} $
for each $ L $, $ S $, \ldots channel.
However, the isospin symmetry of atomic nuclei is partially broken
due to the isospin symmetry breaking (ISB) terms of nuclear interaction together with the Coulomb interaction.
The charge-symmetry breaking (CSB) term of nuclear interaction originates from the mass difference 
of protons and neutrons and 
the $ \pi^0 $-$ \eta $ and $ \rho^0 $-$ \omega $ meson-exchange processes,
and
the charge-independence breaking (CIB) term of nuclear interaction is mainly due to the mass difference between $ \pi^0 $ and $ \pi^{\pm} $~\cite{
  Coon1982Phys.Rev.C26_2402}.
These two terms are defined by 
\begin{subequations}
  \begin{align}
    v_{\urm{CSB}}
    & \equiv
      v_{nn} - v_{pp}, \\
    v_{\urm{CIB}}
    & \equiv
      v_{pn}
      -
      \frac{v_{nn} + v_{pp}}{2},
  \end{align}
\end{subequations}
respectively.
Effects of the ISB terms of the nuclear interaction on the nuclear properties have been discussed~\cite{
  Okamoto1964Phys.Lett.11_150,
  Nolen1969Annu.Rev.Nucl.Sci.19_471,
  Shlomo1978Rep.Prog.Phys.41_957,
  Hatsuda1991Phys.Rev.Lett.66_2851,
  Sagawa1995Phys.Lett.B353_7,
  Brown1998Phys.Rev.C58_220,
  Brown2000Phys.Lett.B483_49,
  Liang2009Phys.Rev.C79_064316,
  Kaneko2010Phys.Rev.C82_061301,
  Kaneko2012Phys.Rev.Lett.109_092504,
  Satula2012Phys.Rev.C86_054316,
  Wiringa2013Phys.Rev.C88_044333,
  Kaneko2013Phys.Rev.Lett.110_172505,
  Kaneko2014Phys.Rev.C89_031302,
  Kaneko2015Phys.Scr.T166_014011,
  Hardy2015Phys.Rev.C91_025501,
  Kaneko2017Phys.Lett.B773_521,
  Dong2018Phys.Rev.C97_021301,
  Kaneko2018Phys.Rev.C97_054326,
  Roca-Maza2018Phys.Rev.Lett.120_202501,
  Loc2019Phys.Rev.C99_014311,
  Dong2019Phys.Rev.C99_014319,
  Roca-Maza2020Phys.Rev.C101_014320,
  Novario2023Phys.Rev.Lett.130_032501},
as well as impact on neutron-star mass-radius relation~\cite{
  Selva2021Symmetry13_144}.
\par
The Coulomb interaction affects properties of the nuclear structure,
and breaks the isospin symmetry of the atomic nuclei~\cite{
  Shlomo1978Rep.Prog.Phys.41_957,
  Auerbach1983Phys.Rep.98_273}.
The ISB terms of nuclear interaction and Coulomb interaction are of different origins.
These effects are, in general, measured as a net effect,
while the Coulomb interaction plays a major role.
To disentangle these effects from the experimental data,
it is necessary to understand which quantities are sensitive to the ISB or Coulomb interaction.
Hence, sensitivity studies for the Coulomb and the ISB terms of nuclear interactions are indispensable.
\par
Here, a key issue to discuss such sensitivity studies is the accuracy of the calculation,
since the ISB terms of the nuclear interaction are only a tiny part of the whole.
To evaluate the contribution of the Coulomb interaction to nuclear properties, 
recently, a high-accuracy treatment of the Coulomb interaction for nuclear structure calculations was developed~\cite{
  Naito2018Phys.Rev.C97_044319,
  Naito2019Phys.Rev.C99_024309,
  Naito2020Phys.Rev.C101_064311}.
In this series of works, the density gradient effect was considered
for the Coulomb exchange energy density functional (EDF) 
using the generalized gradient approximation (GGA).
On top of that, the proton and neutron electric form factors were taken into account self-consistently,
and the vacuum polarization for the Coulomb interaction was considered.
\par
The ISB terms of the nuclear interactions
have also been included in the Skyrme EDF~\cite{
  Vautherin1972Phys.Rev.C5_626}
in Refs.~\cite{
  Roca-Maza2018Phys.Rev.Lett.120_202501,
  Baczyk2018Phys.Lett.B778_178,
  Baczyk2019J.Phys.G46_03LT01,
  Baczyk2021Phys.Rev.C103_054320}.
We discussed a possibility to determine the CSB strength of the Skyrme interaction
referring to \textit{ab initio} calculations~\cite{
  Naito2022Phys.Rev.C105_L021304}.
\par
This paper aims to a complete sensitivity study of the nuclear EoS and the neutron-skin thickness to the Coulomb and ISB terms.
In the previous Letter~\cite{
  Naito:2022hyb},
we discussed
the effect of ISB terms on the charge-radii difference of mirror nuclei $ \Delta R_{\urm{ch}} $ 
and, accordingly, on estimating the density dependence of the symmetry energy, $ L $,
using $ \Delta R_{\urm{ch}} $.
In this paper,
we will discuss 
similar analyses for different quantities related to isospin symmetry breaking:
the neutron-skin thickness and the mass differences of mirror nuclei.
The ISB effect on the difference between the calculated charge radius of $ \nuc{Ca}{40}{} $ and that of $ \nuc{Ca}{48}{} $,
where it was claimed that such difference is related to the symmetry energy of an employed EDF~\cite{
  Perera2021Phys.Rev.C104_064313}, 
will also be discussed.
\par
This paper is organized as follows:
In Sec.~\ref{sec:matter}, effects of the ISB terms on nuclear matter properties will be discussed.
In Sec.~\ref{sec:theoretical}, the theoretical framework will be shown.
In Sec.~\ref{sec:calculation}, the sensitivity study of the neutron-skin thickness $ \Delta R_{np} $,
the difference between the charge radius of $ \nuc{Ca}{40}{} $ and that of $ \nuc{Ca}{48}{} $,
and mass differences of mirror nuclei will be investigated.
The ISB effect on the correlation between the neutron-skin thickness and the density dependence of the symmetry energy~\cite{
  Myers1969Ann.Phys.55_395,
  Roca-Maza2011Phys.Rev.Lett.106_252501,
  Reinhard2022Phys.Rev.C105_L021301}
will also be discussed.
In Sec.~\ref{sec:conclusion}, we will summarize this paper.
%
%
\section{Isospin Symmetry Breaking Interaction and Nuclear Equation of State}
\label{sec:matter}
\par
In this section, the ISB contributions to nuclear matter properties are discussed.
In order to discuss it, first, the energy density of the Skyrme-ISB interaction is shown.
Although only the leading-order ($ t_0 $-like) Skyrme-ISB interaction is considered in our numerical calculations,
we will show the momentum-dependent ($ t_1 $ and $ t_2 $-like) Skyrme-ISB contributions for the EDF here.
\par
The Skyrme CSB and CIB interactions are denoted by 
\begin{widetext}
  \begin{subequations}
    \label{eq:Skyrme_ISB}
    \begin{align}
      v_{\urm{Sky}}^{\urm{CSB}} \left( \ve{r} \right)
      & =
        \left\{
        s_0
        \left( 1 + y_0 P_{\sigma} \right)
        \delta \left( \ve{r} \right)
        +
        \frac{s_1}{2}
        \left( 1 + y_1 P_{\sigma} \right)
        \left[
        \ve{k}^{\dagger 2} \delta \left( \ve{r} \right)
        +
        \delta \left( \ve{r} \right) \ve{k}^2
        \right]
        +
        s_2
        \left( 1 + y_2 P_{\sigma} \right)
        \ve{k}^{\dagger}
        \cdot
        \delta \left( \ve{r} \right)
        \ve{k}
        \right\}
        \frac{\tau_{z1} + \tau_{z2}}{4},
        \label{eq:Skyrme_CSB} \\
      v_{\urm{Sky}}^{\urm{CIB}} \left( \ve{r} \right)
      & =
        \left\{
        u_0
        \left( 1 + z_0 P_{\sigma} \right)
        \delta \left( \ve{r} \right)
        +
        \frac{u_1}{2}
        \left( 1 + z_1 P_{\sigma} \right)
        \left[
        \ve{k}^{\dagger 2} \delta \left( \ve{r} \right)
        +
        \delta \left( \ve{r} \right) \ve{k}^2
        \right]
        +
        u_2
        \left( 1 + z_2 P_{\sigma} \right)
        \ve{k}^{\dagger}
        \cdot
        \delta \left( \ve{r} \right)
        \ve{k}
        \right\}
        \left(
        a_1 \ve{\tau}_1 \cdot \ve{\tau}_2
        +
        a_2 \tau_{z1} \tau_{z2}
        \right),
        \label{eq:Skyrme_CIB}
    \end{align}
  \end{subequations}
  respectively,
  in analogy with the isospin-symmetric Skyrme interaction~\cite{
    Skyrme1958Nucl.Phys.9_615,
    Vautherin1972Phys.Rev.C5_626}
  \begin{align}
    v_{\urm{Sky}}^{\urm{IS}} \left( \ve{r} \right)
    & = 
      t_0
      \left( 1 + x_0 P_{\sigma} \right)
      \delta \left( \ve{r} \right)
      +
      \frac{t_1}{2}
      \left( 1 + x_1 P_{\sigma} \right)
      \left[
      \ve{k}^{\dagger 2} \delta \left( \ve{r} \right)
      +
      \delta \left( \ve{r} \right) \ve{k}^2
      \right]
      +
      t_2
      \left( 1 + x_2 P_{\sigma} \right)
      \ve{k}^{\dagger}
      \cdot
      \delta \left( \ve{r} \right)
      \ve{k}
      \notag \\
    & \quad
      +
      \frac{t_3}{6}
      \left( 1 + x_3 P_{\sigma} \right)
      \delta \left( \ve{r} \right)
      \left[ 
      \rho \left( \ve{R} \right)
      \right]^{\alpha}
      +
      i W_0
      \ve{\sigma}
      \cdot 
      \ve{k}^{\dagger}
      \times
      \delta \left( \ve{r} \right)
      \ve{k}, 
      \label{eq:Skyrme_IS} 
  \end{align}
\end{widetext}
where
$ \ve{r} = \ve{r}_1 - \ve{r}_2 $
and
$ \ve{R} = \left( \ve{r}_1 + \ve{r}_2 \right) / 2 $.
See Ref.~\cite{
  Roca-Maza2018Prog.Part.Nucl.Phys.101_96}
for the standard definitions of the other symbols.
\par
It is worthwhile to discuss the form of the CIB operator.
Three types of the CIB operator---the simple form
$ \tau_{z1} \tau_{z2} $,
the isotensor form $ T_{12} = \ve{\tau}_1 \cdot \ve{\tau}_2 - 3 \tau_{z1} \tau_{z2} $,
and
the general form
$ a_1 \ve{\tau}_1 \cdot \ve{\tau}_2 + a_2 \tau_{z1} \tau_{z2} $---are widely used~\cite{
  Sagawa1995Phys.Lett.B353_7,
  Miller2006Annu.Rev.Nucl.Part.Sci.56_253,
  Roca-Maza2018Phys.Rev.Lett.120_202501,
  Baczyk2018Phys.Lett.B778_178}.
However, there is no criterion to fix
values of $ a_1 $ and $ a_2 $ in $ a_1 \ve{\tau}_1 \cdot \ve{\tau}_2 + a_2 \tau_{z1} \tau_{z2} $ from any fundamental theory.
For example, as shown in Appendix~\ref{sec:form_CIB},
according to the one-pion exchange nuclear interaction,
a relation $ a_1 = - a_2 $ can be derived;
however, the one-pion exchange interaction gives only a part of the CIB interaction.
The difference between the isotensor and the one-pion exchange forms can be absorbed in the isospin symmetric part,
i.e., the term $ \ve{\tau}_1 \cdot \ve{\tau}_2 $ itself is isospin-symmetric.
In order to keep generality, here, the ``general'' form of CIB operator,
$ a_1 \ve{\tau}_1 \cdot \ve{\tau}_2 + a_2 \tau_{z1} \tau_{z2} $,
is adopted.
\par
It will be shown that if one uses $ a_1 = - a_2 $ for the CIB operator
and does not assume the formalism of the proton-neutron mixed density functional theory~\cite{
  Stoitsov2005Comput.Phys.Commun.167_43,
  Sato2013Phys.Rev.C88_061301,
  Sheikh2014Phys.Rev.C89_054317},
the CIB contributions to the energy density vanishes.~\footnote{
  This is true even if one takes a Gogny interaction.}
For other cases, whatever form of CIB operator is used,
the CIB contribution to the nuclear matter does not vanish.
Therefore, whichever form of the CIB operator other than $ a_1 = - a_2 $ is used, there is neither disadvantage nor advantage.
Note that the ISB contributions to the energy density in the formalism of the proton-neutron mixed density functional theory is given in Ref.~\cite{
  Dobaczewski2021J.Phys.G48_102001}.
\subsection{ISB nuclear energy density}
\par
Although the ISB nuclear energy density has been shown in Ref.~\cite{
  Sagawa2019Eur.Phys.J.A55_227},
it is convenient to show it here again to discuss effects of ISB terms on nuclear matter properties.
The nuclear energy density for the isospin symmetric part is shown in, for example, Refs.~\cite{
  Vautherin1972Phys.Rev.C5_626,
  Reinhard1999Phys.Rev.C60_014316,
  Stoitsov2005Comput.Phys.Commun.167_43,
  Danielewicz2009Nucl.Phys.A818_36}.
Here, we do not consider the proton-neutron mixed density, i.e.,
$ \rho_{pn} \left( \ve{r} \right) $ and $ \rho_{np} \left( \ve{r} \right) $ are assumed to be zero.
\par
Using the expectation values of the CSB and CIB operators,
expanding the wave function on a basis where the $ \ve{\tau} $ and $ \tau_z $ are good quantum numbers,
one obtains
\begin{widetext}
  \begin{subequations}
    \begin{align}
      \brakket{pp}{\ve{\tau}_1 \cdot \ve{\tau}_2}{pp}
      =
      \brakket{nn}{\ve{\tau}_1 \cdot \ve{\tau}_2}{nn}
      & =
        1, \\
      \brakket{pn}{\ve{\tau}_1 \cdot \ve{\tau}_2}{pn}
      =
      \brakket{np}{\ve{\tau}_1 \cdot \ve{\tau}_2}{np}
      & =
        -1, \\
      \brakket{pn}{\ve{\tau}_1 \cdot \ve{\tau}_2}{np}
      & =
        2.
    \end{align}
  \end{subequations}
  Accordingly, we get
  \begin{subequations}
    \label{eq:Skyrme_EDF}
    \begin{align}
      \ca{E}_{\urm{CSB}}^{\urm{H}}
        & = 
          \frac{s_0}{4}
          \left( 1 + \frac{y_0}{2} \right)
          \left( \rho_n^2 - \rho_p^2 \right)
          +
          \frac{1}{8}
          \left[
          s_1 \left( 1 + \frac{y_1}{2} \right)
          +
          s_2 \left( 1 + \frac{y_2}{2} \right)
          \right]
          \left( \rho_n t_n - \rho_p t_p \right)
          \notag \\
        & \quad
          -
          \frac{1}{32}
          \left[
          3 s_1 \left( 1 + \frac{y_1}{2} \right)
          -
          s_2 \left( 1 + \frac{y_2}{2} \right)
          \right]
          \left( \rho_n \laplace \rho_n - \rho_p \laplace \rho_p \right)
          -
          \frac{1}{32}
          \left(
          s_1 y_1
          +
          s_2 y_2 
          \right)
          \left( \ve{J}_n^2 - \ve{J}_p^2 \right), 
          \label{eq:CSB_H} \\
      \ca{E}_{\urm{CSB}}^{\urm{x}}
        & = 
          -
          \frac{s_0}{4}
          \left( \frac{1}{2} + y_0  \right)
          \left( \rho_n^2 - \rho_p^2 \right)
          -
          \frac{1}{8}
          \left[
          s_1 
          \left( \frac{1}{2} + y_1 \right)
          -
          s_2 
          \left( \frac{1}{2} + y_2 \right)
          \right]
          \left( \rho_n t_n - \rho_p t_p \right)
          \notag \\
        & \quad
          +
          \frac{1}{32}
          \left[
          3 s_1 \left( \frac{1}{2} + y_1 \right)
          +
          s_2 \left( \frac{1}{2} + y_2 \right)
          \right]
          \left( \rho_n \laplace \rho_n - \rho_p \laplace \rho_p \right)
          +
          \frac{1}{32}
          \left( s_1 - s_2 \right)
          \left( \ve{J}_n^2 - \ve{J}_p^2 \right), 
          \label{eq:CSB_x} \\
      \ca{E}_{\urm{CIB}}^{\urm{H}}
        & =
          \left( a_1 + a_2 \right)
          \left\{
          \frac{u_0}{2}
          \left( 1 + \frac{z_0}{2} \right)
          \left( \rho_n - \rho_p \right)^2
          +
          \frac{1}{4}
          \left[
          u_1 \left( 1 + \frac{z_1}{2} \right)
          +
          u_2 \left( 1 + \frac{z_2}{2} \right)
          \right]
          \left( \rho_n - \rho_p \right)
          \left( t_n - t_p \right)
          \right.
          \notag \\
        & \qquad
          \left.
          -
          \frac{1}{16}
          \left[
          3 u_1 \left( 1 + \frac{z_1}{2} \right)
          -
          u_2 \left( 1 + \frac{z_2}{2} \right)
          \right]
          \left( \rho_n - \rho_p \right)
          \left( \laplace \rho_n - \laplace \rho_p \right)
          -
          \frac{1}{16}
          \left(
          u_1 z_1 
          +
          u_2 z_2 
          \right)
          \left( \ve{J}_n - \ve{J}_p \right)^2
          \right\}
          \label{eq:CIB_H} \\
      \ca{E}_{\urm{CIB}}^{\urm{x}}
        & =
          \left( a_1 + a_2 \right)
          \left\{
          -
          \frac{u_0}{2}
          \left( \frac{1}{2} + z_0 \right)
          \left( \rho_n^2 + \rho_p^2 \right)
          -
          \frac{1}{4}
          \left[
          u_1 \left( \frac{1}{2} + z_1 \right)
          -
          u_2 \left( \frac{1}{2} + z_2 \right)
          \right]
          \left( \rho_n t_n + \rho_p t_p \right)
          \right.
          \notag \\
        & \qquad
          \left.
          +
          \frac{1}{16}
          \left[
          3 u_1 \left( \frac{1}{2} + z_1 \right)
          +
          u_2 \left( \frac{1}{2} + z_2 \right)
          \right]
          \left( \rho_n \laplace \rho_n + \rho_p \laplace \rho_p \right)
          +
          \frac{1}{16}
          \left( u_1 - u_2 \right)
          \left( \ve{J}_n^2 + \ve{J}_p^2 \right)
          \right\},   
          \label{eq:CIB_x}
    \end{align}
  \end{subequations}
  where
  $ t_{\tau} = \sum_j \left| \nabla \phi_{j \tau} \right|^2 $
  and
  $ \ve{J}_{\tau} = \sum_j \phi_{j \tau} \ve{\sigma} \times \nabla \phi_{j \tau} $ 
  are
  the kinetic energy and spin-orbit densities for nucleon of species $ \tau $,
  and $ \ca{E}^{\urm{H}} $ and $ \ca{E}^{\urm{x}} $ correspond to the Hartree and exchange contributions, respectively.
  \subsection{Nuclear equation of state with ISB terms} 
  \par
  The nuclear equation of state can be calculated as
  \begin{subequations}
    \label{eq:EoS}
    \begin{align}
      \frac{E_{\urm{Skyrme}}}{A} \left( \rho, \beta \right)
      & =
        \epsilon_{\urm{Skyrme}} \left( \rho, \beta \right)
        \notag \\
      & =
        \epsilon_0 \left( \rho \right)
        +
        \epsilon_1 \left( \rho \right)
        \beta
        +
        \epsilon_2 \left( \rho \right)
        \beta^2
        +
        O \left( \beta^3 \right), \\
      \epsilon_0 \left( \rho \right)
      & =
        \frac{3}{5}
        \frac{\hbar^2}{2m}
        \left( \frac{3 \pi^2}{2} \right)^{2/3}
        \rho^{2/3}
        +
        \frac{1}{8}
        \left[
        3 t_0
        -
        \left( a_1 + a_2 \right) u_0 \left( 1 + 2 z_0 \right)
        \right]
        \rho
        \notag \\
      & \quad
        +
        \frac{3}{80}
        \left( \frac{3 \pi^2}{2} \right)^{2/3}
        \left\{
        3 t_1
        +
        t_2 \left( 5 + 4 x_2 \right)
        -
        \left( a_1 + a_2 \right)
        \left[
        u_1 \left( 1 + 2 z_1 \right)
        -
        u_2 \left( 1 + 2 z_2 \right)
        \right]
        \right\}
        \rho^{5/3}
        +
        \frac{t_3}{16}
        \rho^{\alpha + 1}, \\
      \epsilon_1 \left( \beta \right)
      & =
        \frac{s_0}{8}
        \left( 1 - y_0 \right)
        \rho
        +
        \frac{1}{20}
        \left( \frac{3 \pi^2}{2} \right)^{2/3}
        \left[ s_1 \left( 1 - y_1 \right) + 3 s_2 \left( 1 + y_2 \right) \right]
        \rho^{5/3}, \\
      \epsilon_2 \left( \beta \right)
      & =
        \frac{1}{3}
        \frac{\hbar^2}{2m}
        \left( \frac{3 \pi^2}{2} \right)^{2/3}
        \rho^{2/3}
        -
        \frac{1}{8}
        \left[
        t_0 \left( 1 + 2 x_0 \right)
        -
        3 \left( a_1 + a_2 \right) u_0
        \right]
        \rho
        \notag \\
      & \quad
        -
        \frac{1}{24}
        \left( \frac{3 \pi^2}{2} \right)^{2/3}
        \left\{
        3 t_1 x_1
        -
        t_2 \left( 4 + 5 x_2 \right)
        -
        \left( a_1 + a_2 \right)
        \left[
        u_1 \left( 4 - z_1 \right)
        +
        u_2 \left( 8 + 7 z_2 \right)
        \right]
        \right\}
        \rho^{5/3}
        -
        \frac{t_3}{48}
        \left( 1 + 2 x_3 \right)
        \rho^{\alpha + 1},
    \end{align}
  \end{subequations}
\end{widetext}
where
$ \rho = \rho_n + \rho_p $
and 
$ \beta = \left( \rho_n - \rho_p \right) / \rho $.
It is obviously found that
the CIB term contributes to the isoscalar term and $ \beta^2 $ term,
while the CSB term generates $ \beta $ term.
Here, $ E_{\urm{Skyrme}} $ is the Skyrme EDF which also includes the CSB and CIB contributions [Eq.~\eqref{eq:Skyrme_EDF}],
as well as the ordinary isospin symmetric part.
\par
If one does not consider the CSB term,
$ \epsilon_1 \equiv 0 $ holds;
accordingly, the symmetry energy $ \epsilon_{\urm{sym}} $ can be simply defined by
$ \epsilon_{\urm{sym}} \left( \rho \right) = \epsilon_2 \left( \rho \right) $,
which is the usual definition.
Once the CSB term is introduced,
$ \epsilon_1 $ term appears and there can be several possible definitions of $ \epsilon_{\urm{sym}} $:
$ \epsilon_{\urm{sym}} $ is
defined by
$ \epsilon_{\urm{sym}} \left( \rho \right)
=
\left.
  \frac{1}{2}
  \frac{\partial^2 \epsilon \left( \rho, \beta \right)}{\partial \beta^2}
\right|_{\beta = 0} $ 
or
by $ \epsilon_{\urm{sym}} \left( \rho \right) = \epsilon \left( \rho, 1 \right) - \epsilon \left( \rho, 0 \right) $.
As discussed in Ref.~\cite{
  Naito:2022hyb},
the latter definition 
\begin{align}
  \epsilon_{\urm{sym}} \left( \rho \right)
  & =
    \epsilon \left( \rho, 1 \right)
    -
    \epsilon \left( \rho, 0 \right)
    \notag \\
  & \simeq
    \epsilon_1 \left( \rho \right)
    +
    \epsilon_2 \left( \rho \right)
    \label{eq:def_sym}
\end{align}
leads to the straightforward extension of the relation between the pressure of neutron matter
at the saturation density,
$ P \left( \rho_{\urm{sat}}, 1 \right) $,
and the density dependence of symmetry energy, $ L $,
as $ P \left( \rho_{\urm{sat}}, 1 \right) = L \rho_{\urm{sat}} / 3 $.
The approximation of Eq.~\eqref{eq:def_sym}
is justified around the saturation density
since the terms higher order than $ \beta^2 $ are small.
\par
EoS parameters $ \epsilon_{\urm{sat}} $, $ K_{\infty} $, $ J $, $ L $, and $ K_{\urm{sym}} $ and their extensions to ISB terms are defined by~\cite{
  Mondal2018Int.J.Mod.Phys.E27_1850078}
\begin{widetext}
  \begin{subequations}
    \begin{align}
      \epsilon_0 \left( \rho \right)
      & =
        \epsilon_{\urm{sat}}
        +
        \epsilon_{\urm{sat}}^{\urm{CIB}}
        +
        \frac{1}{2}
        \left(
        K_{\infty}
        +
        K_{\infty}^{\urm{CIB}}
        \right)
        \left(
        \frac{\rho - \rho_{\urm{sat}}}{3 \rho_{\urm{sat}}}
        \right)^2
        +
        \ldots, \\
      \epsilon_{\urm{sym}} \left( \rho \right)
      & =
        \left(
        J
        +
        J^{\urm{CIB}}
        +
        J^{\urm{CSB}}
        \right)
        +
        \left(
        L
        +
        L^{\urm{CIB}}
        +
        L^{\urm{CSB}}
        \right)
        \left(
        \frac{\rho - \rho_{\urm{sat}}}{3 \rho_{\urm{sat}}}
        \right)
        +
        \frac{1}{2}
        \left(
        K_{\urm{sym}}
        +
        K_{\urm{sym}}^{\urm{CIB}}
        +
        K_{\urm{sym}}^{\urm{CSB}}
        \right)
        \left(
        \frac{\rho - \rho_{\urm{sat}}}{3 \rho_{\urm{sat}}}
        \right)^2
        +
        \ldots,
    \end{align}
  \end{subequations}
  where the CSB contribution to $ \epsilon_{\urm{sat}} $ and $ K_{\infty} $ are zero.
  These EoS parameters read
  \begin{subequations}
    \begin{align}
      \epsilon_{\urm{sat}}
      & =
        \frac{3}{5} 
        \frac{\hbar^2}{2m}
        \left( \frac{3 \pi^2}{2} \right)^{2/3}
        \rho_{\urm{sat}}^{2/3}
        +
        \frac{3}{8}
        t_0
        \rho_{\urm{sat}}
        +
        \frac{3}{80}
        \left( \frac{3 \pi^2}{2} \right)^{2/3}
        \left[
        3 t_1
        +
        t_2 \left( 5 + 4 x_2 \right)
        \right]
        \rho_{\urm{sat}}^{5/3}
        +
        \frac{t_3}{16}
        \rho_{\urm{sat}}^{\alpha + 1}, \\
      \epsilon_{\urm{sat}}^{\urm{CIB}}
      & =
        -
        \frac{1}{8}
        \left( a_1 + a_2 \right) u_0 \left( 1 + 2 z_0 \right)
        \rho_{\urm{sat}}
        -
        \frac{3}{80}
        \left( \frac{3 \pi^2}{2} \right)^{2/3}
        \left( a_1 + a_2 \right)
        \left[
        u_1 \left( 1 + 2 z_1 \right)
        -
        u_2 \left( 1 + 2 z_2 \right)
        \right]
        \rho_{\urm{sat}}^{5/3}, \\
      K_{\infty}
      & = 
        -
        \frac{6}{5}
        \frac{\hbar^2}{2m}
        \left( \frac{3 \pi^2}{2} \right)^{2/3}
        \rho_{\urm{sat}}^{2/3}
        +
        \frac{3}{8}
        \left( \frac{3 \pi^2}{2} \right)^{2/3}
        \left[
        3 t_1
        +
        t_2 \left( 5 + 4 x_2 \right)
        \right]
        \rho_{\urm{sat}}^{5/3}
        +
        \frac{9}{16}
        t_3
        \alpha \left( \alpha + 1 \right)
        \rho_{\urm{sat}}^{\alpha + 1}, \\
      K_{\infty}^{\urm{CIB}}
      & =
        -
        \frac{3}{8}
        \left( \frac{3 \pi^2}{2} \right)^{2/3}
        \left( a_1 + a_2 \right)
        \left[
        u_1 \left( 1 + 2 z_1 \right)
        -
        u_2 \left( 1 + 2 z_2 \right)
        \right]
        \rho_{\urm{sat}}^{5/3}, \\
      J
      & = 
        \frac{1}{3}
        \frac{\hbar^2}{2m}
        \left( \frac{3 \pi^2}{2} \right)^{2/3}
        \rho_{\urm{sat}}^{2/3}
        -
        \frac{t_0}{8}
        \left( 1 + 2 x_0 \right)
        \rho_{\urm{sat}}
        -
        \frac{1}{24}
        \left( \frac{3 \pi^2}{2} \right)^{2/3}
        \left[
        3 t_1 x_1
        -
        t_2 \left( 4 + 5 x_2 \right)
        \right]
        \rho_{\urm{sat}}^{5/3}
        -
        \frac{t_3}{48}
        \left( 1 + 2 x_3 \right)
        \rho_{\urm{sat}}^{\alpha + 1}, \\
      J^{\urm{CIB}}
      & = 
        \frac{3}{8}
        \left( a_1 + a_2 \right) u_0
        \rho_{\urm{sat}}
        +
        \frac{1}{24}
        \left( \frac{3 \pi^2}{2} \right)^{2/3}
        \left( a_1 + a_2 \right)
        \left[
        u_1 \left( 4 - z_1 \right)
        +
        u_2 \left( 8 + 7 z_2 \right)
        \right]
        \rho_{\urm{sat}}^{5/3}, \\
      J^{\urm{CSB}}
      & =
        \frac{s_0}{8}
        \left( 1 - y_0 \right)
        \rho_{\urm{sat}}
        +
        \frac{1}{20}
        \left( \frac{3 \pi^2}{2} \right)^{2/3}
        \left[ s_1 \left( 1 - y_1 \right) + 3 s_2 \left( 1 + y_2 \right) \right]
        \rho_{\urm{sat}}^{5/3}, \\
      L
      & = 
        \frac{2}{3}
        \frac{\hbar^2}{2m}
        \left( \frac{3 \pi^2}{2} \right)^{2/3}
        \rho_{\urm{sat}}^{2/3}
        -
        \frac{3}{8}
        t_0 \left( 1 + 2 x_0 \right)
        \rho_{\urm{sat}}
        -
        \frac{5}{24}
        \left( \frac{3 \pi^2}{2} \right)^{2/3}
        \left[
        3 t_1 x_1 
        -
        t_2 \left( 4 + 5 x_2 \right)
        \right]
        \rho_{\urm{sat}}^{5/3}
        -
        \frac{t_3}{16}
        \left( 1 + 2 x_3 \right)
        \left( \alpha + 1 \right)
        \rho_{\urm{sat}}^{\alpha + 1}, \\
      L^{\urm{CIB}}
      & = 
        \frac{9}{8}
        \left( a_1 + a_2 \right) u_0
        \rho_{\urm{sat}}
        +
        \frac{5}{24}
        \left( \frac{3 \pi^2}{2} \right)^{2/3}
        \left( a_1 + a_2 \right)
        \left[
        u_1 \left( 4 - z_1 \right)
        +
        u_2 \left( 8 + 7 z_2 \right)
        \right]
        \rho_{\urm{sat}}^{5/3}, \\
      L^{\urm{CSB}}
      & =
        \frac{3}{8}
        s_0 \left( 1 - y_0 \right)
        \rho_{\urm{sat}}
        +
        \frac{1}{4}
        \left( \frac{3 \pi^2}{2} \right)^{2/3}
        \left[ s_1 \left( 1 - y_1 \right) + 3 s_2 \left( 1 + y_2 \right) \right]
        \rho_{\urm{sat}}^{5/3}, \\
      K_{\urm{sym}}
      & = 
        -
        \frac{2}{3}
        \frac{\hbar^2}{2m}
        \left( \frac{3 \pi^2}{2} \right)^{2/3}
        \rho_{\urm{sat}}^{2/3}
        -
        \frac{5}{12}
        \left( \frac{3 \pi^2}{2} \right)^{2/3}
        \left[
        3 t_1 x_1
        -
        t_2 \left( 4 + 5 x_2 \right)
        \right]
        \rho_{\urm{sat}}^{5/3}
        -
        \frac{3}{16}
        t_3 \left( 1 + 2 x_3 \right)
        \alpha \left( \alpha + 1 \right)
        \rho_{\urm{sat}}^{\alpha + 1}, \\
      K_{\urm{sym}}^{\urm{CIB}}
      & = 
        \frac{5}{12}
        \left( \frac{3 \pi^2}{2} \right)^{2/3}
        \left( a_1 + a_2 \right)
        \left[
        u_1 \left( 4 - z_1 \right)
        +
        u_2 \left( 8 + 7 z_2 \right)
        \right]
        \rho_{\urm{sat}}^{5/3}, \\
      K_{\urm{sym}}^{\urm{CSB}}
      & =
        \frac{1}{2}
        \left( \frac{3 \pi^2}{2} \right)^{2/3}
        \left[ s_1 \left( 1 - y_1 \right) + 3 s_2 \left( 1 + y_2 \right) \right]
        \rho_{\urm{sat}}^{5/3}, 
    \end{align}
  \end{subequations}
  respectively.
  \par
  The pressure of nuclear matter reads
  \begin{align}
    & P \left( \rho, \beta \right)
      \notag \\
    & = 
      \rho^2
      \frac{\partial \epsilon_{\urm{Skyrme}} \left( \rho, \beta \right)}{\partial \rho}
      \notag \\
    & \simeq 
      \left\{
      \frac{2}{5}
      \frac{\hbar^2}{2m}
      \left( \frac{3 \pi^2}{2} \right)^{2/3}
      \rho^{5/3}
      +
      \frac{1}{8}
      \left[
      3 t_0
      -
      \left( a_1 + a_2 \right) u_0 \left( 1 + 2 z_0 \right)
      \right]
      \rho^2
      \right.
      \notag \\
    & \qquad
      \left.
      +
      \frac{1}{16}
      \left( \frac{3 \pi^2}{2} \right)^{2/3}
      \left\{
      3 t_1
      +
      t_2 \left( 5 + 4 x_2 \right)
      -
      \left( a_1 + a_2 \right)
      \left[
      u_1 \left( 1 + 2 z_1 \right)
      -
      u_2 \left( 1 + 2 z_2 \right)
      \right]
      \right\}
      \rho^{8/3}
      +
      \frac{t_3}{16}
      \left( \alpha + 1 \right)
      \rho^{\alpha + 2}
      \right\}
      \notag \\
    & \quad
      +
      \left\{
      \frac{s_0}{8}
      \left( 1 - y_0 \right)
      \rho^2
      +
      \frac{1}{12}
      \left( \frac{3 \pi^2}{2} \right)^{2/3}
      \left[ s_1 \left( 1 - y_1 \right) + 3 s_2 \left( 1 + y_2 \right) \right]
      \rho^{8/3}
      \right\}
      \beta
      \notag \\
    & \quad
      +
      \left\{
      \frac{2}{9}
      \frac{\hbar^2}{2m}
      \left( \frac{3 \pi^2}{2} \right)^{2/3}
      \rho^{5/3}
      -
      \frac{1}{8}
      \left[
      t_0 \left( 1 + 2 x_0 \right)
      -
      3 \left( a_1 + a_2 \right) u_0
      \right]
      \rho^2
      \right.
      \notag \\
    & \qquad
      \left.
      -
      \frac{5}{72}
      \left( \frac{3 \pi^2}{2} \right)^{2/3}
      \left\{
      3 t_1 x_1
      -
      t_2 \left( 4 + 5 x_2 \right)
      -
      \left( a_1 + a_2 \right)
      \left[
      u_1 \left( 4 - z_1 \right)
      +
      u_2 \left( 8 + 7 z_2 \right)
      \right]
      \right\}
      \rho^{8/3}
      -
      \frac{t_3}{48}
      \left( \alpha + 1 \right)
      \left( 1 + 2 x_3 \right)
      \rho^{\alpha + 2} 
      \right\}
      \beta^2,
      \label{eq:pressure}
  \end{align}
  where the higher order terms than $ \beta^2 $ are neglected.
  The pressure of the pure neutron matter at the saturation density reads
  \begin{align}
    & P \left( \rho_{\urm{sat}}, 1 \right)
      \notag \\
    & \simeq 
      \left\{
      \frac{s_0}{8}
      \left( 1 - y_0 \right)
      \rho_{\urm{sat}}^2
      +
      \frac{1}{12}
      \left( \frac{3 \pi^2}{2} \right)^{2/3}
      \left[ s_1 \left( 1 - y_1 \right) + 3 s_2 \left( 1 + y_2 \right) \right]
      \rho_{\urm{sat}}^{8/3}
      \right\}
      \notag \\
    & \quad
      +
      \left\{
      \frac{2}{9}
      \frac{\hbar^2}{2m}
      \left( \frac{3 \pi^2}{2} \right)^{2/3}
      \rho_{\urm{sat}}^{5/3}
      -
      \frac{1}{8}
      \left[
      t_0 \left( 1 + 2 x_0 \right)
      -
      3 \left( a_1 + a_2 \right) u_0
      \right]
      \rho_{\urm{sat}}^2
      \right.
      \notag \\
    & \qquad
      \left.
      -
      \frac{5}{72}
      \left( \frac{3 \pi^2}{2} \right)^{2/3}
      \left\{
      3 t_1 x_1
      -
      t_2 \left( 4 + 5 x_2 \right)
      -
      \left( a_1 + a_2 \right)
      \left[
      u_1 \left( 4 - z_1 \right)
      +
      u_2 \left( 8 + 7 z_2 \right)
      \right]
      \right\}
      \rho_{\urm{sat}}^{8/3}
      -
      \frac{t_3}{48}
      \left( \alpha + 1 \right)
      \left( 1 + 2 x_3 \right)
      \rho_{\urm{sat}}^{\alpha + 2} 
      \right\}
      \notag \\
    & = 
      \frac{L + L^{\urm{CIB}} + L^{\urm{CSB}}}{3}
      \rho_{\urm{sat}}.
  \end{align}
  \par
  The saturation density $ \rho_{\urm{sat}} $, which is defined by 
  $ P \left( \rho_{\urm{sat}}, 0 \right) = 0 $,
  satisfies
  \begin{align}
    & \frac{2}{5}
      \frac{\hbar^2}{2m}
      \left( \frac{3 \pi^2}{2} \right)^{2/3}
      \rho_{\urm{sat}}^{5/3}
      +
      \frac{1}{8}
      \left[
      3 t_0
      -
      \left( a_1 + a_2 \right) u_0 \left( 1 + 2 z_0 \right)
      \right]
      \rho_{\urm{sat}}^2
      \notag \\
    & +
      \frac{1}{16}
      \left( \frac{3 \pi^2}{2} \right)^{2/3}
      \left\{
      3 t_1
      +
      t_2 \left( 5 + 4 x_2 \right)
      -
      \left( a_1 + a_2 \right)
      \left[
      u_1 \left( 1 + 2 z_1 \right)
      -
      u_2 \left( 1 + 2 z_2 \right)
      \right]
      \right\}
      \rho_{\urm{sat}}^{8/3}
      +
      \frac{t_3}{16}
      \left( \alpha + 1 \right)
      \rho_{\urm{sat}}^{\alpha + 2}
      \simeq
      0,
  \end{align}
\end{widetext}
where the approximation comes from Eq.~\eqref{eq:pressure} and is correct up to $ \beta^2 $.
This equation implies that the saturation density $ \rho_{\urm{sat}} $ itself is changed due to the CIB term.
\par
Effects of ISB terms to the saturation density and EoS parameters are discussed in Sec.~\ref{sec:calc_matter}.
%
%
\section{EDFs of ISB and Coulomb Interactions}
\label{sec:theoretical}
\par
To calculate the density and total energy, the self-consistent nuclear density functional theory~\cite{
  Hohenberg1964Phys.Rev.136_B864,
  Kohn1965Phys.Rev.140_A1133,
  Kohn1999Rev.Mod.Phys.71_1253,
  Bender2003Rev.Mod.Phys.75_121,
  Meng2006Prog.Part.Nucl.Phys.57_470,
  Liang2015Phys.Rep.570_1}
is used.
In nuclear density functional theory including the case of Skyrme EDFs,
the ground-state energy is written as
\begin{align}
  E_{\urm{gs}} 
  & = 
    T_0
    +
    E_{\urm{IS}} \left[ \rho_p, \rho_n \right]
    \notag \\
  & \quad
    +
    E_{\urm{CSB}} \left[ \rho_p, \rho_n \right]
    +
    E_{\urm{CIB}} \left[ \rho_p, \rho_n \right]
    +
    E_{\urm{Coul}} \left[ \rho_{\urm{ch}} \right],
    \label{eq:Egs}
\end{align}
where $ T_0 $, $ E_{\urm{IS}} $, $ E_{\urm{CSB}} $, $ E_{\urm{CIB}} $, and $ E_{\urm{Coul}} $ 
are the Kohn-Sham kinetic energy,
the isospin symmetric, CSB, CIB, and the Coulomb EDFs, respectively.
The proton and neutron density distribution are denoted by $ \rho_p $ and $ \rho_n $, respectively.
The definition of the charge density distribution $ \rho_{\urm{ch}} $ will be discussed in the next subsection.
\subsection{Nuclear part}
\par
In SHF, the isospin symmetric nuclear term $ E_{\urm{IS}} $ is the standard Skyrme EDF~\cite{
  Vautherin1972Phys.Rev.C5_626,
  Reinhard1999Phys.Rev.C60_014316}.
In this paper, we mainly use the SAMi EDF~\cite{
  Roca-Maza2012Phys.Rev.C86_031306}
and 
the SAMi-J EDF family~\cite{
  Roca-Maza2013Phys.Rev.C87_034301}.
In addition, we also use the SAMi-ISB EDF~\cite{
  Roca-Maza2018Phys.Rev.Lett.120_202501}.
Note that the SAMi-ISB EDF includes $ E_{\urm{CSB}} $ and $ E_{\urm{CIB}} $ as well;
we call the SAMi-ISB EDF without ISB terms,
as the ``SAMi-noISB'' EDF to avoid any confusion
with the SAMi EDF without ISB terms.
The $ E_{\urm{IS}} $ of SAMi-noISB EDF is different from that of the original SAMi EDF,
since $ E_{\urm{IS}} $, $ E_{\urm{CSB}} $, and $ E_{\urm{CIB}} $ are optimized altogether simultaneously,
although the same protocol was adopted.
In the original papers of these EDFs~\cite{
  Roca-Maza2012Phys.Rev.C86_031306,
  Roca-Maza2018Phys.Rev.Lett.120_202501},
only limited digits are shown.
However, since we will discuss details of numerical results,
more digits are demanded to achieve higher accuracy.
The precise values of the parameters of SAMi and SAMi-ISB EDFs are shown
in Appendix~\ref{sec:skyrme}.
\par
For the ISB terms of nuclear part, $ E_{\urm{CSB}} $ and $ E_{\urm{CIB}} $,
we adopt the SAMi-ISB EDF,
whose forms are
the leading-order Skyrme-ISB interaction,
i.e., $ s_1 = s_2 = u_1 = u_2 = 0 $ in Eqs.~\eqref{eq:Skyrme_ISB},
with $ a_1 = 0 $ and $ a_2 = 1/2 $.
The CSB and CIB EDFs, respectively, read
\begin{align}
  E_{\urm{CSB}} \left[ \rho_p, \rho_n \right]
  & = 
    \frac{s_0 \left( 1 - y_0 \right)}{8}
    \int
    \left\{
    \left[ \rho_n \left( \ve{r} \right) \right]^2
    -
    \left[ \rho_p \left( \ve{r} \right) \right]^2
    \right\}
    \, d \ve{r}, 
    \label{eq:CSB} \\
  E_{\urm{CIB}} \left[ \rho_p, \rho_n \right]
  & = 
    \frac{u_0 \left( 1 - z_0 \right)}{8}
    \int
    \left\{
    \left[ \rho_n \left( \ve{r} \right) \right]^2
    +
    \left[ \rho_p \left( \ve{r} \right) \right]^2
    \right\}
    \, d \ve{r}
    \notag \\
  & \quad
    -
    \frac{u_0 \left( 2 + z_0 \right)}{4}
    \int
    \rho_n \left( \ve{r} \right) 
    \rho_p \left( \ve{r} \right) 
    \, d \ve{r}
    \label{eq:CIB}
\end{align}
with $ y_0 = z_0 = -1 $.
\subsection{Coulomb part}
\par
The Coulomb part $ E_{\urm{Coul}} $ is, in general, divided into four terms,
the Coulomb Hartree term $ E_{\urm{CH}} $,
the Coulomb exchange term $ E_{\urm{Cx}} $,
the vacuum polarization term $ E_{\urm{VP}} $,
and the electromagnetic spin-orbit term $ E_{\urm{EMSO}} $~\cite{
  Naito2020Phys.Rev.C101_064311}.
Note that many-body effects of the Coulomb interaction, namely, the Coulomb correlation~\cite{
  Bulgac1996Nucl.Phys.A601_103,
  Bulgac1999Eur.Phys.J.A5_247,
  Bulgac1999Phys.Lett.B469_1,
  Naito2018Phys.Rev.C97_044319},
are not considered in this paper and left for future perspectives.
We start from the Hartree-Fock-Slater approximation~\cite{
  Dirac1930Proc.Camb.Phil.Soc.26_376,
  Slater1951Phys.Rev.81_385},
i.e., the Coulomb LDA exchange EDF for $ E_{\urm{Cx}} $
with $ E_{\urm{VP}} \equiv 0 $ and $ E_{\urm{EMSO}} \equiv 0 $,
together with the point-particle approximation $ \rho_{\urm{ch}} \equiv \rho_p $.
On top of the Hartree-Fock-Slater approximation,
in order to see effects of the Coulomb interaction,
the precise treatment of the Coulomb interaction---the GGA,
the proton finite-size effect,
the neutron finite-size effect,
and 
the vacuum polarization $ E_{\urm{VP}} $---is introduced step by step as proposed in Ref.~\cite{
  Naito2020Phys.Rev.C101_064311}.
We use abbreviations
``NoEx,'' ``LDA,'' ``GGA,'' ``$ p $-fin,'' ``$ pn $-fin,'' and ``All''
for
\begin{subequations}
  \label{eq:Coul_def}
  \begin{align}
    E_{\urm{Coul}}^{\urm{NoEx}}
    & = 
      E_{\urm{CH}} \left[ \rho_p \right],
      \label{eq:Coul_def_H} \\
    E_{\urm{Coul}}^{\urm{LDA}}
    & = 
      E_{\urm{CH}} \left[ \rho_p \right]
      +
      E_{\urm{Cx}}^{\urm{LDA}} \left[ \rho_p \right], 
      \label{eq:Coul_def_LDA} \\
    E_{\urm{Coul}}^{\urm{GGA}}
    & = 
      E_{\urm{CH}} \left[ \rho_p \right]
      +
      E_{\urm{Cx}}^{\urm{GGA}} \left[ \rho_p \right], 
      \label{eq:Coul_def_GGA} \\
    E_{\urm{Coul}}^{\urm{$ p $-fin}}
    & = 
      E_{\urm{CH}} \left[ \rho_{\urm{ch}}^{\urm{$ p $-fin}} \right]
      +
      E_{\urm{Cx}}^{\urm{GGA}} \left[ \rho_{\urm{ch}}^{\urm{$ p $-fin}} \right], 
      \label{eq:Coul_def_pfin} \\
    E_{\urm{Coul}}^{\urm{$ pn $-fin}}
    & = 
      E_{\urm{CH}} \left[ \rho_{\urm{ch}}^{\urm{$ pn $-fin}} \right]
      +
      E_{\urm{Cx}}^{\urm{GGA}} \left[ \rho_{\urm{ch}}^{\urm{$ pn $-fin}} \right], 
      \label{eq:Coul_def_pnfin} \\
    E_{\urm{Coul}}^{\urm{All}}
    & = 
      E_{\urm{CH}} \left[ \rho_{\urm{ch}}^{\urm{$ pn $-fin}} \right]
      +
      E_{\urm{Cx}}^{\urm{GGA}} \left[ \rho_{\urm{ch}}^{\urm{$ pn $-fin}} \right]
      +
      E_{\urm{VP}} \left[ \rho_p \right],
      \label{eq:Coul_def_all} 
  \end{align}
\end{subequations}
respectively.
Here, 
$ E_{\urm{Cx}}^{\urm{LDA}} $ and $ E_{\urm{Cx}}^{\urm{GGA}} $ are the Coulomb exchange EDFs in the LDA and GGA,
\begin{subequations}
  \begin{align}
    E_{\urm{Cx}}^{\urm{LDA}} \left[ \rho \right]
    & =
      -
      \frac{3e^2}{4}
      \left( \frac{3}{\pi} \right)^{1/3}
      \int
      \left[
      \rho \left( \ve{r} \right)
      \right]^{4/3}
      \, d \ve{r}, \\
    E_{\urm{Cx}}^{\urm{GGA}} \left[ \rho \right]
    & =
      -
      \frac{3e^2}{4}
      \left( \frac{3}{\pi} \right)^{1/3}
      \int
      F \left( s \left( \ve{r} \right) \right)
      \left[
      \rho \left( \ve{r} \right)
      \right]^{4/3}
      \, d \ve{r}, 
  \end{align}
\end{subequations}
respectively.
In this paper, we use the modified Perdew-Burke-Ernzerhof GGA enhancement factor~\cite{
  Perdew1996Phys.Rev.Lett.77_3865}
\begin{subequations}
  \begin{align}
    F \left( s \right)
    & =
      1 + \kappa
      -
      \frac{\kappa}{1 + \lambda \mu s^2 / \kappa}, \\
    s
    & =
      \frac{\left| \nabla \rho \right|}{2 k_{\urm{F}} \rho}, \\
    k_{\urm{F}}
    & =
      \left(
      3 \pi^2 \rho
      \right)^{1/3}, \\
    \mu
    & =
      0.21951, \\
    \kappa
    & =
      0.804,
  \end{align}
\end{subequations}
with $ \lambda = 1.25 $~\cite{
  Naito2019Phys.Rev.C99_024309},
which is determined to reproduce the exact-Fock energy at the level of the point-particle approximation.
Here, $ \rho_{\urm{ch}}^{\urm{$ p $-fin}} $ and $ \rho_{\urm{ch}}^{\urm{$ pn $-fin}} $ are
charge densities in which only proton finite size and both proton and neutron finite size are considered, respectively.
They are defined in the momentum space as 
\begin{subequations}
  \label{eq:charge}
  \begin{align}
    \tilde{\rho}_{\urm{ch}}^{\urm{$ p $-fin}} \left( q \right)
    & =
      \tilde{G}_{\urm{E} p} \left( q^2 \right)
      \tilde{\rho}_p \left( q \right) , \\
    \tilde{\rho}_{\urm{ch}}^{\urm{$ pn $-fin}} \left( q \right)
    & =
      \tilde{G}_{\urm{E} p} \left( q^2 \right)
      \tilde{\rho}_p \left( q \right)
      +
      \tilde{G}_{\urm{E} n} \left( q^2 \right)
      \tilde{\rho}_n \left( q \right)
      \notag \\
    & =
      \tilde{\rho}_{\urm{ch}} \left( q \right) ,
  \end{align}
\end{subequations}
respectively,
where $ \tilde{\rho} \left( q \right) $ is the Fourier transform of the density in the coordinate representation $ \rho \left( r \right) $.
In this paper, we only consider the electric form factors of nucleons,
$ \tilde{G}_{\urm{E} \tau} $
($ \tau = p $, $ n $),
and use the form factors obtained by Friedrich and Walcher~\cite{
  Friedrich2003Eur.Phys.J.A17_607}.
We will use $ \rho_{\urm{ch}} $ to calculate the Coulomb energy only:
only the electric form factors are considered in Eq.~\eqref{eq:charge}
and
effects of the magnetic form factors are considered perturbatively as the electromagnetic spin-orbit interaction.
Since the vacuum polarization is weak compared to the Coulomb Hartree and exchange terms,
the finite-size effect on the vacuum polarization is not considered~\cite{
  Naito2018Phys.Rev.C97_044319}.
On top of ``All'', the electromagnetic spin-orbit term $ E_{\urm{EMSO}} $ is considered perturbatively,
which is abbreviated as ``$ \text{All} + \text{EMSO} $.''
Since $ E_{\urm{EMSO}} $ is considered at first-order perturbation theory, it does not affect the density distribution, i.e., charge radius or $ \Delta R_{np} $.
\subsection{Calculation setup}
\par
All the terms shown above have been implemented in the calculation code \textsc{skyrme\_rpa}~\cite{
  Colo2013Comput.Phys.Commun.184_142}.
The spherical symmetry is assumed and the pairing correlation is neglected in the calculation,
since we focus on only the doubly-magic nuclei.
A meshed box of $ 0.1 \, \mathrm{fm} \times 150 $ is used.
%
%
\section{Numerical Results and Discussion}
\label{sec:calculation}
\par
This section is devoted to show numerical results:
ISB effects on nuclear matter properties
and 
the sensitivity study to the Coulomb interaction and the ISB strength dependence of the following physical observables---the neutron-skin thickness,
the difference between the charge radius of $ \nuc{Ca}{40}{} $ and that of $ \nuc{Ca}{48}{} $, 
and mass differences of mirror nuclei of $ \nuc{Ca}{48}{} $-$ \nuc{Ni}{48}{} $ isobars.
For the first check, we select $ \nuc{O}{16}{} $, $ \nuc{Ca}{40}{} $, $ \nuc{Ca}{48}{} $, $ \nuc{Ni}{48}{} $, and $ \nuc{Pb}{208}{} $ as examples to study the neutron-skin thickness.
\subsection{Nuclear matter properties}
\label{sec:calc_matter}
\par
The parameter sets of SAMi and SAMi-ISB EDFs are optimized by using the same protocol.
To see more precise effects of CSB and CIB nuclear matter properties,
we switch on and off CSB and CIB terms of SAMi-ISB in Table~\ref{tab:matter}.
The saturation density $ \rho_{\urm{sat}} $ and EoS parameters
$ \epsilon_{\urm{sat}} $, $ J $, and $ L $
defined in Sec.~\ref{sec:matter}
are calculated by
SAMi-noISB, SAMi-CIB, SAMi-CSB, and SAMi-ISB EDFs
shown in Table~\ref{tab:matter}.
To see the effect of refitting
of SAMi and SAMi-noISB, 
i.e., the effect of difference of Skyrme parameters,
$ t_0 $-$ t_3 $, $ x_0 $--$ x_3 $, $ W_0 $, $ W'_0 $, and $ \alpha $,
results of the original SAMi are also listed.
Here, SAMi-noISB, SAMi-CIB, and SAMi-CSB, respectively, refer to
SAMi-ISB without any ISB terms,
only with CIB term,
and
only with CSB term.
Summary of their abbreviations is also shown in Table~\ref{tab:matter}.
\par
The CIB term makes $ \rho_{\urm{sat}} $ smaller and $ \epsilon_{\urm{sat}} $ larger,
but their effects are, respectively,
less than $ 0.002 \, \mathrm{fm}^{-3} $ and $ 0.2 \, \mathrm{MeV} $, which are negligible.
The refitting effect, i.e., difference between $ \rho_{\urm{sat}} $ and $ \epsilon_{\urm{sat}} $ obtained by SAMi and those by SAMi-noISB,
is also quite tiny.
\par
The CIB term makes 
$ J + J^{\urm{CIB}} + J^{\urm{CSB}} $ and $ L + L^{\urm{CIB}} + L^{\urm{CSB}} $
larger, respectively,
by $ 0.6 \, \mathrm{MeV} $ and $ 2.3 \, \mathrm{MeV} $
and
the CSB term makes them smaller, respectively,
by $ 1.1 \, \mathrm{MeV} $ and $ 3.2 \, \mathrm{MeV} $.
Effects of these two terms almost cancel each other,
and, eventually, 
$ J + J^{\urm{CIB}} + J^{\urm{CSB}} $ and $ L + L^{\urm{CIB}} + L^{\urm{CSB}} $
obtained by SAMi-noISB and SAMi-ISB are quite similar.
In contrast, the refitting effect on
$ J $ and $ L $
are,
respectively, $ 2.7 \, \mathrm{MeV} $ and $ 6.5 \, \mathrm{MeV} $.
The refitting effect on $ L $ may not be negligible.
\begin{table*}[tb]
  \centering
  \caption{
    The saturation density $ \rho_{\urm{sat}} $ and EoS parameters
    $ \epsilon_{\urm{sat}} $, $ J $, and $ L $
    calculated by
    SAMi-noISB, SAMi-CIB, SAMi-CSB, and SAMi-ISB EDFs.
    To see the effect of refitting, those by SAMi EDF are also shown.}
  \label{tab:matter}
  \begin{ruledtabular}
    \begin{tabular}{llddddd}
      \multicolumn{2}{l}{EDF}
      & \multicolumn{1}{c}{SAMi}
      & \multicolumn{1}{c}{SAMi-noISB}
      & \multicolumn{1}{c}{SAMi-CIB}
      & \multicolumn{1}{c}{SAMi-CSB}
      & \multicolumn{1}{c}{SAMi-ISB} \\
      \hline
      \multicolumn{2}{l}{$ E_{\urm{IS}} $}
      & \multicolumn{1}{c}{SAMi}
      & \multicolumn{1}{c}{SAMi-noISB}
      & \multicolumn{1}{c}{SAMi-noISB}
      & \multicolumn{1}{c}{SAMi-noISB}
      & \multicolumn{1}{c}{SAMi-noISB} \\
      \multicolumn{2}{l}{$ E_{\urm{CIB}} $}
      & \multicolumn{1}{c}{No}
      & \multicolumn{1}{c}{No}
      & \multicolumn{1}{c}{Yes}
      & \multicolumn{1}{c}{No}
      & \multicolumn{1}{c}{Yes} \\
      \multicolumn{2}{l}{$ E_{\urm{CSB}} $}
      & \multicolumn{1}{c}{No}
      & \multicolumn{1}{c}{No}
      & \multicolumn{1}{c}{No}
      & \multicolumn{1}{c}{Yes}
      & \multicolumn{1}{c}{Yes} \\
      \hline
      $ \rho_{\urm{sat}} $
      & ($ \mathrm{fm}^{-3} $)
      &   0.1587
      &   0.1613
      &   0.1597
      &   0.1613
      &   0.1597 \\
      $ \epsilon_{\urm{sat}} + \epsilon_{\urm{sat}}^{\urm{CIB}} $ 
      & ($ \mathrm{MeV} $)
      & -15.9271 
      & -16.0288
      & -15.7700
      & -16.0288
      & -15.7700 \\
      $ J + J^{\urm{CIB}} + J^{\urm{CSB}} $
      & ($ \mathrm{MeV} $)
      &  28.1256
      &  30.8274
      &  31.4337
      &  29.7667
      &  30.3835 \\
      $ L + L^{\urm{CIB}} + L^{\urm{CSB}} $
      & ($ \mathrm{MeV} $)
      &  43.5582
      &  50.0953
      &  52.3624
      &  46.9132
      &  49.2118 \\
    \end{tabular}
  \end{ruledtabular}
\end{table*}
\subsection{Neutron-skin thickness}
\label{sec:calc_nskin}
\subsubsection{Coulomb term and $ L $ parameter}
\label{sec:calc_LvsR}
\par
It has been shown that the neutron-skin thickness $ \Delta R_{np} $ is related to the density dependence of the symmetry energy of nuclear matter $ L $~\cite{
  Myers1969Ann.Phys.55_395,
  Brown2000Phys.Rev.Lett.85_5296,
  Chen2005Phys.Rev.C72_064309,
  Roca-Maza2011Phys.Rev.Lett.106_252501,
  Reinhard2022Phys.Rev.C105_L021301}.
Firstly, we show the sensitivity study of the $ L $ parameter on the Coulomb part of the EDF.
In this calculation, the ISB terms are not considered, i.e., 
$ E_{\urm{CSB}} \equiv 0 $ and $ E_{\urm{CIB}} \equiv 0 $.
For the Coulomb part $ E_{\urm{Coul}} $,
we adopt 
NoEx, LDA, GGA, $ p $-fin, $ pn $-fin, and All,
which are defined in Eqs.~\eqref{eq:Coul_def}.
\par
All the calculations are performed with the SAMi EDF and the SAMi-J EDF family.
The parameter sets of the SAMi-J family are determined by the same criteria of the SAMi EDF
with a fixed symmetry energy $ J $.
Accordingly, each SAMi-J EDF has different symmetry parameter $ L $,
as shown in Table~\ref{tab:SAMi_J}.
The neutron-skin thickness $ \Delta R_{np} $ for various $ L $ can be calculated using SAMi and SAMi-J EDFs,
and the data are fitted to
\begin{equation}
  \label{eq:LvsR}
  \Delta R_{np} \equiv R_n - R_p = a + bL,
\end{equation}
i.e., the same as Ref.~\cite{
  Roca-Maza2011Phys.Rev.Lett.106_252501},
where $ R_n $ and $ R_p $ are the root-mean-square radii of the neutron and proton density distributions,
respectively.
\begin{table}[tb]
  \centering
  \caption{The saturation density $ \rho_{\urm{sat}} $,
    the symmetry energy $ J $,
    and its slope $ L $ of the SAMi EDF~\cite{
      Roca-Maza2012Phys.Rev.C86_031306}
    and the SAMi-J family~\cite{
      Roca-Maza2013Phys.Rev.C87_034301}.}
  \label{tab:SAMi_J}
  \begin{ruledtabular}
    \begin{tabular}{lddd}
      \multicolumn{1}{c}{EDF} & \multicolumn{1}{c}{$ \rho_{\urm{sat}} $ ($ \mathrm{fm}^{-3} $)} & \multicolumn{1}{c}{$ J $ ($ \mathrm{MeV} $)} & \multicolumn{1}{c}{$ L $ ($ \mathrm{MeV} $)} \\
      \hline
      SAMi-J27 & 0.1595 & 27 &  30.0001 \\
      SAMi-J28 & 0.1587 & 28 &  39.7416 \\
      SAMi-J29 & 0.1579 & 29 &  51.6040 \\
      SAMi-J30 & 0.1571 & 30 &  63.1784 \\
      SAMi-J31 & 0.1563 & 31 &  74.3683 \\
      SAMi-J32 & 0.1555 & 32 &  85.1014 \\
      SAMi-J33 & 0.1548 & 33 &  95.4072 \\
      SAMi-J34 & 0.1542 & 34 & 105.3074 \\
      SAMi-J35 & 0.1537 & 35 & 114.9543 \\
      \hline
      SAMi     & 0.1587 & 28.1256 &  43.5582 \\
    \end{tabular}
  \end{ruledtabular}
\end{table}
\par
\begin{figure}[tb]
  \centering
  \includegraphics[width=1.0\linewidth]{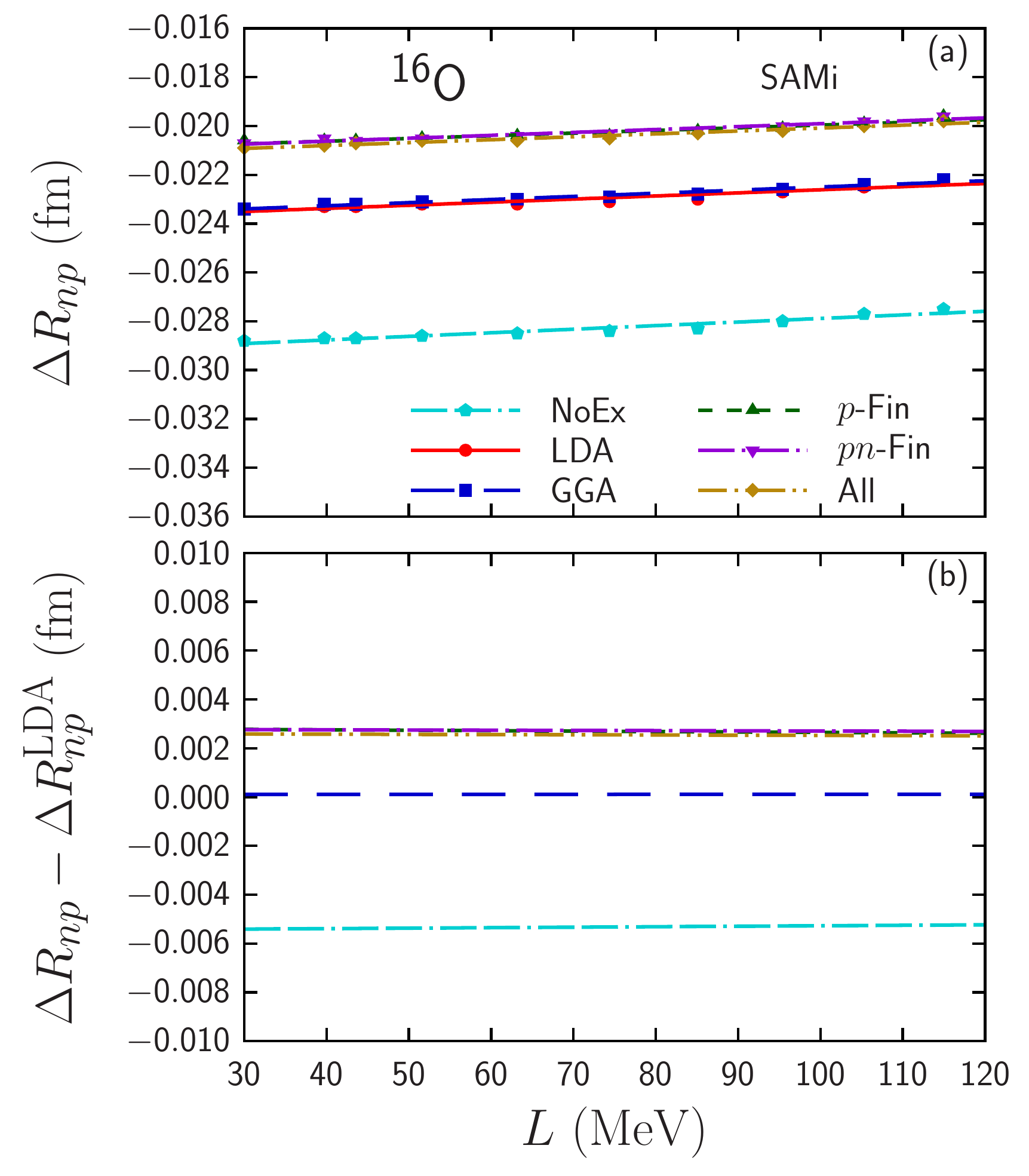}
  \caption{
    (a)~Neutron-skin thickness $ \Delta R_{np} $ as a function of the symmetry parameter $ L $ for $ \nuc{O}{16}{} $.
    The dash-dash-dotted, solid, long-dashed, dashed, dash-dotted, and dash-dot-dotted lines
    with pentagon, circle, square, upper-triangle, down-triangle, and diamond symbols
    show the results
    without Coulomb exchange (NoEx),
    of Coulomb LDA and Coulomb GGA in point-particle approximation (LDA and GGA),
    of Coulomb GGA with proton and proton-neutron finite-size effects ($ p $-fin and $ pn $-fin),
    and of Coulomb GGA with proton-neutron finite-size effects and the vacuum polarization (All),
    respectively.
    (b)~Change of $ \Delta R_{np} $ from that of the Coulomb LDA EDF.}
  \label{fig:LvsR_008_016}
\end{figure}
\begin{figure}[tb]
  \centering
  \includegraphics[width=1.0\linewidth]{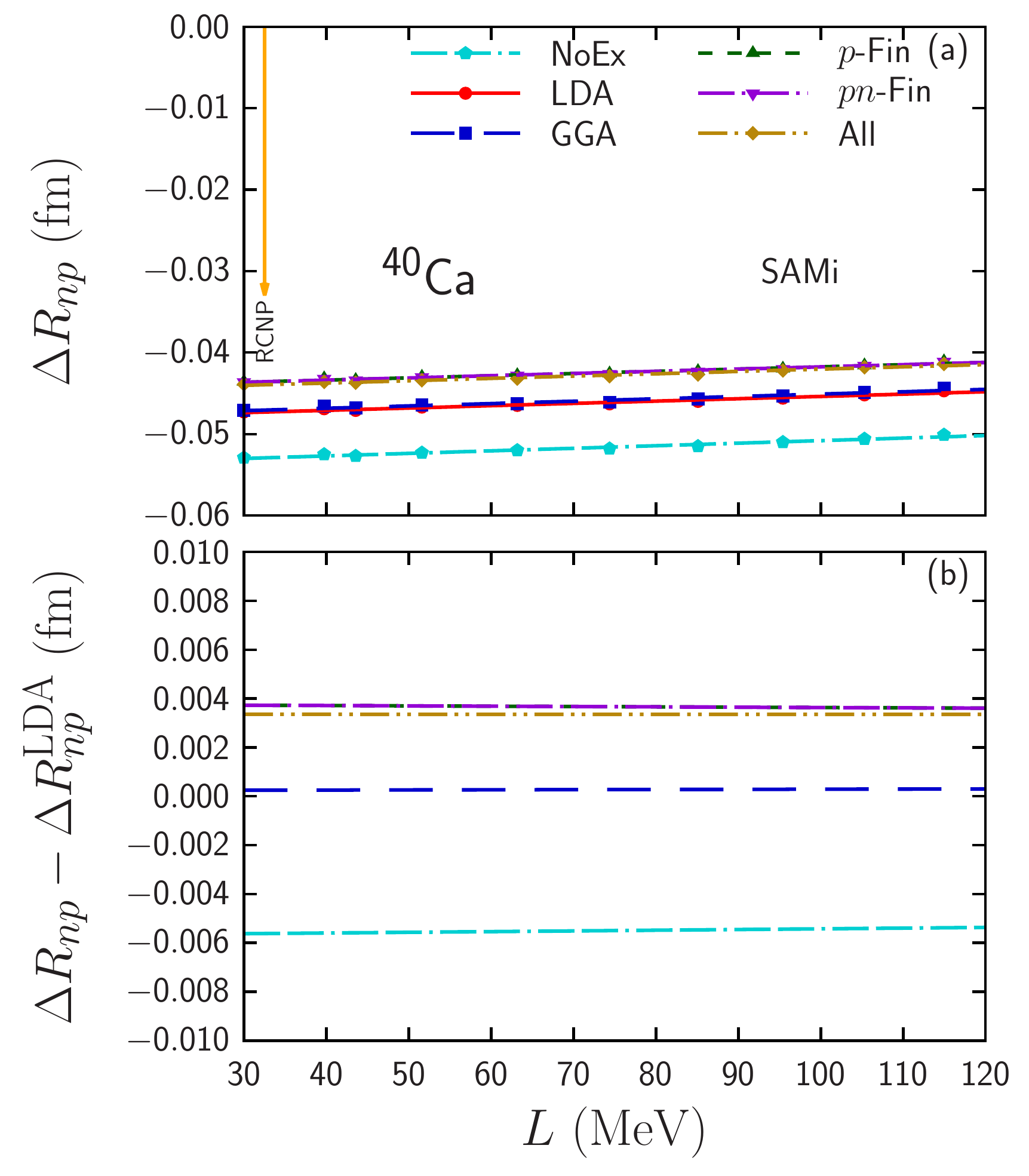}
  \caption{
    The same as Fig.~\ref{fig:LvsR_008_016} but for $ \nuc{Ca}{40}{} $.
    The experimental value of $ \Delta R_{np} = -0.010_{-0.023}^{+0.022} \, \mathrm{fm} $~\cite{
      Zenihiro_2018}
    is shown as a vertical line.}
  \label{fig:LvsR_020_040}
\end{figure}
\begin{figure}[tb]
  \centering
  \includegraphics[width=1.0\linewidth]{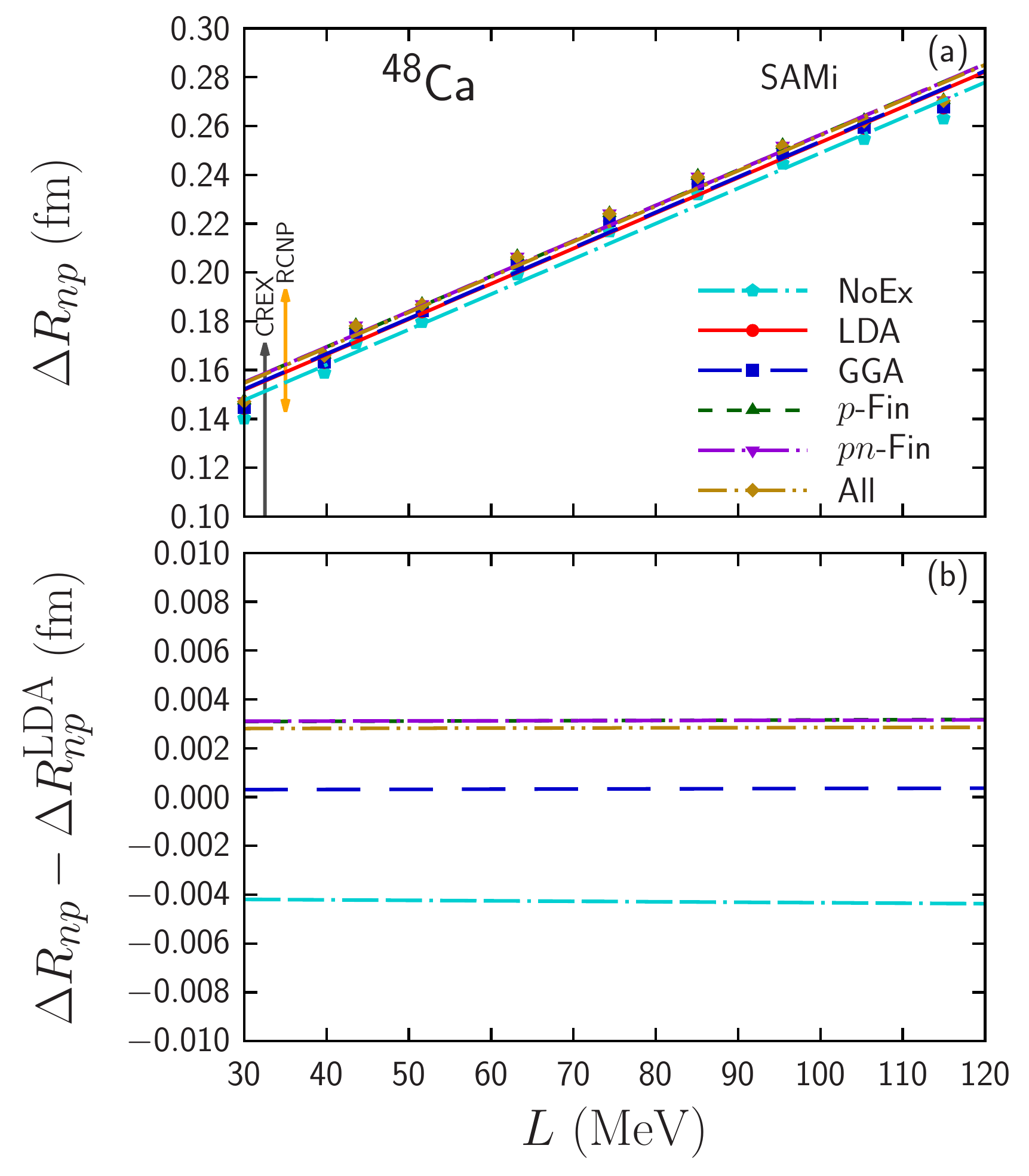}
  \caption{
    The same as Fig.~\ref{fig:LvsR_008_016} but for $ \nuc{Ca}{48}{} $.
    The experimental value of $ \Delta R_{np} = 0.168_{-0.028}^{+0.025} \, \mathrm{fm} $ (RCNP)~\cite{
      Zenihiro_2018}
    and
    $ 0.121 \pm 0.050 \, \mathrm{fm} $ (CREX)~\cite{
      CREX:2022kgg}
    are shown as vertical lines.}
  \label{fig:LvsR_020_048}
\end{figure}
\begin{figure}[tb]
  \centering
  \includegraphics[width=1.0\linewidth]{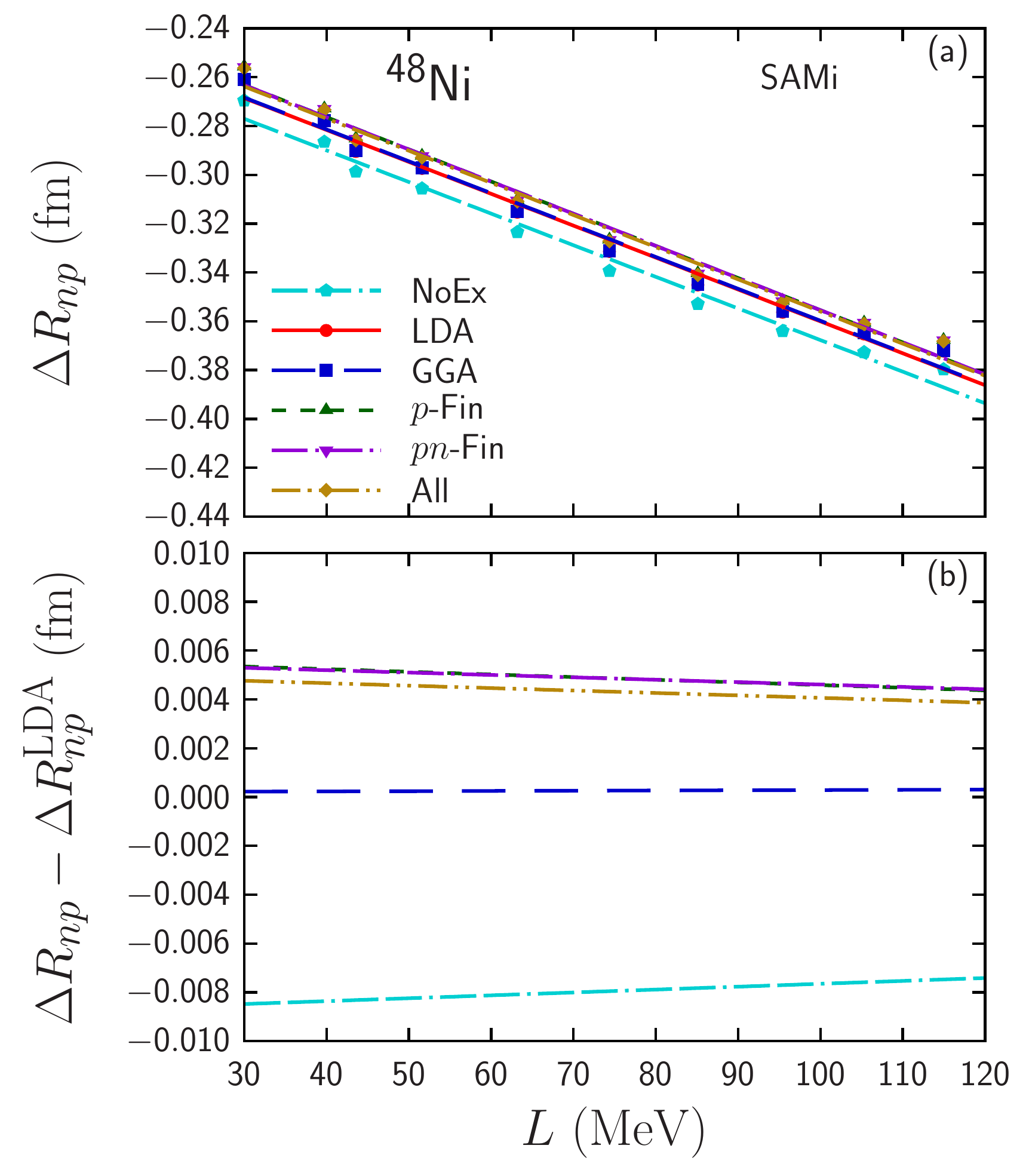}
  \caption{
    The same as Fig.~\ref{fig:LvsR_008_016} but for $ \nuc{Ni}{48}{} $.}
  \label{fig:LvsR_028_048}
\end{figure}
\begin{figure}[tb]
  \centering
  \includegraphics[width=1.0\linewidth]{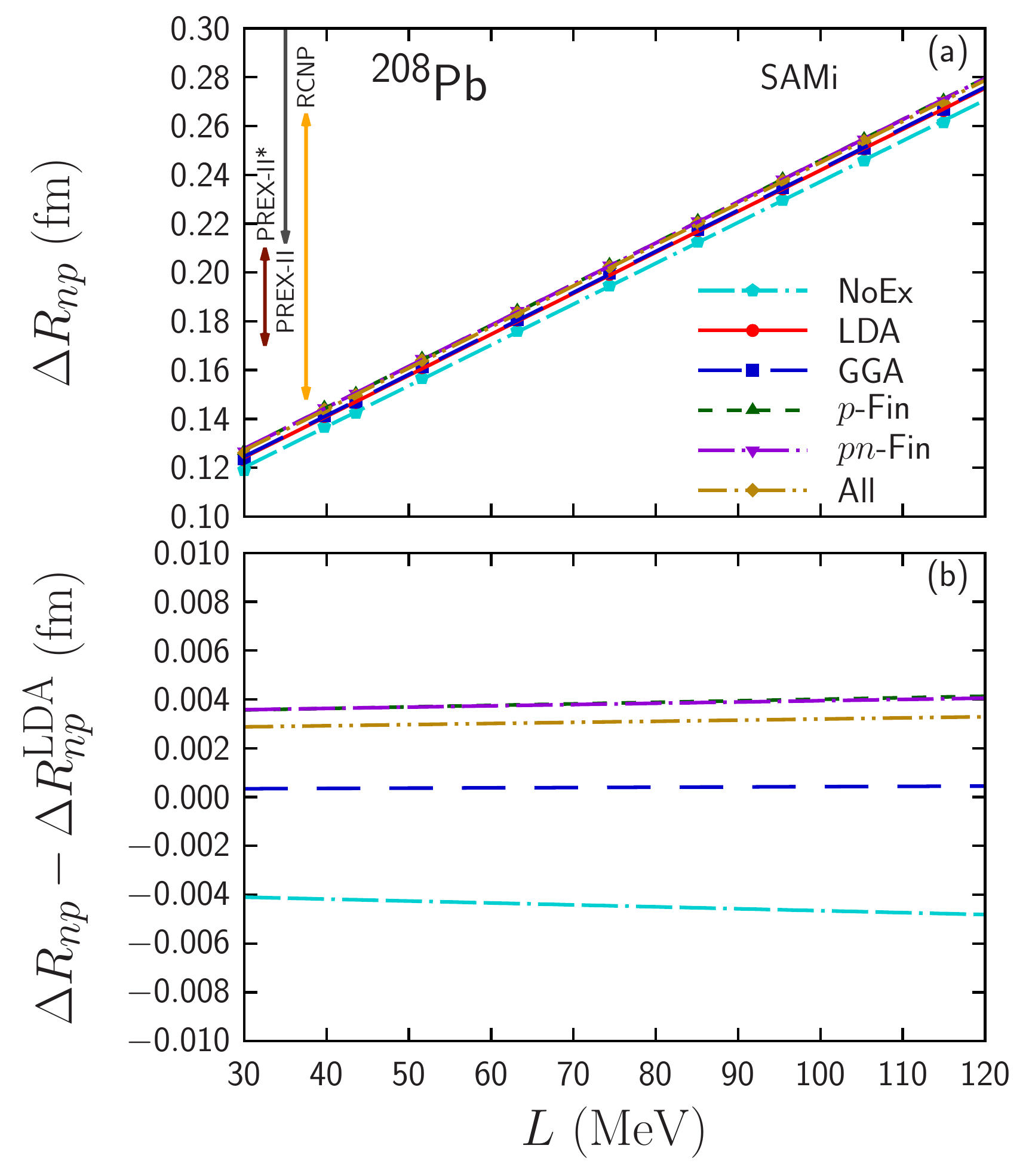}
  \caption{
    The same as Fig.~\ref{fig:LvsR_008_016} but for $ \nuc{Pb}{208}{} $.
    The experimental values of
    $ \Delta R_{np} = 0.211_{-0.063}^{+0.054} \, \mathrm{fm} $ (RCNP)~\cite{
      Zenihiro2010Phys.Rev.C82_044611}
    and 
    $ 0.283 \pm 0.071 \, \mathrm{fm} $ (PREX-II)~\cite{
      Adhikari2021Phys.Rev.Lett.126_172502}
    are shown as vertical lines.
    Reanalyzed data of PREX-II
    $ \Delta R_{np} = 0.190 \pm 0.020 \, \mathrm{fm} $ (PREX-II*)~\cite{
      Reinhard2021Phys.Rev.Lett.127_232501}
    is also shown as a vertical line.}
  \label{fig:LvsR_082_208}
\end{figure}

Panels (a) of Figs.~\ref{fig:LvsR_008_016}--\ref{fig:LvsR_082_208} show the neutron-skin thickness $ \Delta R_{np} $ as a function of the symmetry parameter $ L $ for $ \nuc{O}{16}{} $, $ \nuc{Ca}{40}{} $, $ \nuc{Ca}{48}{} $, $ \nuc{Ni}{48}{} $, and $ \nuc{Pb}{208}{} $, respectively.
The pentagon, circle, square, upper-triangle, down-triangle, and diamond symbols
show the results of NoEx, LDA, GGA, $ p $-fin, $ pn $-fin, and All, respectively.
Using the data, $ a $ and $ b $ in Eq.~\eqref{eq:LvsR} are determined as shown in Table~\ref{tab:LvsR}.
Panels (b) of Figs.~\ref{fig:LvsR_008_016}--\ref{fig:LvsR_082_208} show the difference between
the neutron-skin thickness $ \Delta R_{np} $ calculated by these Coulomb EDFs and
that by the Coulomb LDA EDF.
As will be discussed later, the treatment of the Coulomb interaction scarcely affects the neutron-skin thickness $ \Delta R_{np} $.
The experimental values of $ \Delta R_{np} $ for
$ \nuc{Ca}{40}{} $
($ \Delta R_{np} = -0.010_{-0.023}^{+0.022} \, \mathrm{fm} $~\cite{
  Zenihiro_2018}),
$ \nuc{Ca}{48}{} $
($ 0.168_{-0.028}^{+0.025} \, \mathrm{fm} $~\cite{
  Zenihiro_2018}
and
$ 0.121 \pm 0.050 \, \mathrm{fm} $~\cite{
  CREX:2022kgg}),
and
$ \nuc{Pb}{208}{} $
($ \Delta R_{np} = 0.211_{-0.063}^{+0.054} \, \mathrm{fm} $~\cite{
  Zenihiro2010Phys.Rev.C82_044611},
$ 0.283 \pm 0.071 \, \mathrm{fm} $~\cite{
  Adhikari2021Phys.Rev.Lett.126_172502},
and the reanalyzed data of PREX-II experiment
$ \Delta R_{np} = 0.190 \pm 0.020 \, \mathrm{fm} $~\cite{
  Reinhard2021Phys.Rev.Lett.127_232501})
are also shown as vertical lines in Figs.~\ref{fig:LvsR_020_040}, \ref{fig:LvsR_020_048}, and \ref{fig:LvsR_082_208}, respectively.
\begin{table}[tb]
  \centering
  \caption{
    Parameters $ a $ and $ b $ in Eq.~\eqref{eq:LvsR}.
    See the text for more detail.}
  \label{tab:LvsR}
  \begin{ruledtabular}
    \begin{tabular}{rldd}
      \multicolumn{1}{c}{Nuclei} & \multicolumn{1}{c}{Coulomb} & \multicolumn{1}{c}{$ a $ ($ \mathrm{fm} $)} & \multicolumn{1}{c}{$ b $ ($ \mathrm{fm} \, \mathrm{MeV}^{-1} $)} \\ \hline
      $ \nuc{O}{16}{} $
                                 & NoEx       & -0.029353 &  0.00001469 \\
                                 & LDA        & -0.023886 &  0.00001274 \\
                                 & GGA        & -0.023782 &  0.00001283 \\
                                 & $ p $-fin  & -0.021061 &  0.00001097 \\
                                 & $ pn $-fin & -0.021098 &  0.00001192 \\
                                 & All        & -0.021278 &  0.00001192 \\
      \hline
      $ \nuc{Ca}{40}{} $
                                 & NoEx       & -0.053942 &  0.00003132 \\
                                 & LDA        & -0.048232 &  0.00002847 \\
                                 & GGA        & -0.048001 &  0.00002903 \\
                                 & $ p $-fin  & -0.044467 &  0.00002712 \\
                                 & $ pn $-fin & -0.044467 &  0.00002712 \\
                                 & All        & -0.044879 &  0.00002843 \\
      \hline
      $ \nuc{Ca}{48}{} $
                                 & NoEx       &  0.104330 &  0.00144605 \\
                                 & LDA        &  0.108463 &  0.00144801 \\
                                 & GGA        &  0.108747 &  0.00144866 \\
                                 & $ p $-fin  &  0.111525 &  0.00144897 \\
                                 & $ pn $-fin &  0.111554 &  0.00144857 \\
                                 & All        &  0.111254 &  0.00144857 \\
      \hline
      $ \nuc{Ni}{48}{} $
                                 & NoEx       & -0.238189 & -0.00129505 \\
                                 & LDA        & -0.229357 & -0.00130689 \\
                                 & GGA        & -0.229163 & -0.00130595 \\
                                 & $ p $-fin  & -0.223682 & -0.00131776 \\
                                 & $ pn $-fin & -0.223767 & -0.00131670 \\
                                 & All        & -0.224293 & -0.00131690 \\
      \hline
      $ \nuc{Pb}{208}{} $
                                 & NoEx       &  0.069984 &  0.00167210 \\
                                 & LDA        &  0.073847 &  0.00168003 \\
                                 & GGA        &  0.074152 &  0.00168124 \\
                                 & $ p $-fin  &  0.077225 &  0.00168631 \\
                                 & $ pn $-fin &  0.077268 &  0.00168528 \\
                                 & All        &  0.076585 &  0.00168461 \\
    \end{tabular}
  \end{ruledtabular}
\end{table}
\par
First, let us compare the $ L $ dependence of $ \Delta R_{np} $ among all the calculated nuclei.
In $ N = Z $ nuclei, the neutron-skin thickness $ \Delta R_{np} $ is almost independent of the slope parameter $ L $,
because $ b $ is small [$ O \left( 10^{-5} \right) \, \mathrm{MeV} \, \mathrm{fm}^{-1} $].
In contrast, in $ N > Z $ nuclei, the neutron-skin thickness $ \Delta R_{np} $ has a strong $ L $ dependence as mentioned in Refs.~\cite{
  Roca-Maza2011Phys.Rev.Lett.106_252501,
  Reinhard2022Phys.Rev.C105_L021301}
with $ b \simeq O \left( 10^{-3} \right) \, \mathrm{MeV} \, \mathrm{fm}^{-1} $.
Moreover, as seen in $ \nuc{Ca}{48}{} $ and $ \nuc{Ni}{48}{} $,
the values $ b $ of the mirror nuclei have almost the same absolute value but opposite signs.
In contrast, the absolute value $ a $ for $ \nuc{Ni}{48}{} $ is almost twice of that for $ \nuc{Ca}{48}{} $,
which is quite a significant difference
considering the isospin symmetry of the nuclear interaction.
In these calculations in Figs.~\ref{fig:LvsR_008_016}--\ref{fig:LvsR_082_208},
the ISB terms of the nuclear interaction are not considered,
and the effects of the ISB terms of the nuclear interaction are left for the next section.
Note that negative values of $ b $ in $ \nuc{O}{16}{} $, $ \nuc{Ca}{40}{} $, and $ \nuc{Ni}{48}{} $ indicate proton skins, instead of neutron skins.
The detailed discussion of the origin of the correlation between $ \Delta R_{np} $ and $ L $ can be found in several papers in the literature, e.g., in Refs.~\cite{
  Myers1969Ann.Phys.55_395,
  Baldo2016Prog.Part.Nucl.Phys.91_203}.
Note that the mean-field calculation sometimes underestimates the isospin impurity~\cite{
  Hamamoto1993Phys.Rev.C48_R960,
  Sagawa1996Phys.Rev.C53_2163,
  Sagawa2022Phys.Lett.B829_137072}.
\par
Next, we recognize a clear dependence on the treatment of the Coulomb interaction
in the value $ a $, i.e., in the absolute value of $ \Delta R_{np} $.
Even if the treatment of the Coulomb interaction is changed,
the $ L $ dependence of $ \Delta R_{np} $ is almost unchanged,
i.e., $ b $ is almost constant.
This is due to the fact that, in the atomic nuclei,
the nuclear interaction $ E_{\urm{IS}} $ dominates
and the gross structure of $ \rho_p $ and $ \rho_n $ are determined by $ E_{\urm{IS}} $.
The subdominant Coulomb interaction mainly affects proton distribution, and thus $ R_p $,
but its effect on $ R_p $ is less than $ 0.01 \, \mathrm{fm} $ order.
That is, the Coulomb interaction changes the absolute value, and thus $ a $,
but it hardly changes the slope $ b $.
This is the same as the charge-radii difference of mirror nuclei~\cite{
  Naito:2022hyb}.
The fact that the slope parameter $ b $ is hardly changed by the treatment of the Coulomb interaction
can be also understood referring to Ref.~\cite{
  Centelles2010Phys.Rev.C82_054314}.
Treatment of the Coulomb interaction mainly changes the Coulomb potential in the surface region~\cite{
  Naito2019Phys.Rev.C99_024309},
while the correlation between $ L $ and $ \Delta R_{np} $ of a nucleus with a well-developed bulk like $ \nuc{Pb}{208}{} $ is slightly affected by the surface effect~\cite{
  Centelles2010Phys.Rev.C82_054314}.
\par
Let us see in more detail the effects of the Coulomb interaction.
If the Coulomb exchange term is neglected,
$ \Delta R_{np} $ becomes smaller.
Since the Coulomb exchange term $ E_{\urm{Cx}} $ is effectively attractive,
the Coulomb effect gets stronger if $ E_{\urm{Cx}} $ is neglected.
Then, $ \rho_p $ extends, i.e., $ R_p $ becomes larger, while $ \rho_n $ scarcely changes.
As discussed in Ref.~\cite{
  Naito2019Phys.Rev.C99_024309},
the GGA EDF scarcely changes $ \rho_p $ nor $ \rho_n $;
hence, $ \Delta R_{np} $ is also scarcely changed.
The proton finite-size effect makes the Coulomb interaction between protons weaker
\cite{
  Naito2020Phys.Rev.C101_064311}.
Accordingly, $ \rho_p $ shrinks, i.e., $ R_p $ becomes smaller, 
while $ \rho_n $ remains almost the same,
so that $ \Delta R_{np} $ becomes larger.
In contrast, the vacuum polarization makes the Coulomb interaction between protons stronger
and thus $ \Delta R_{np} $ becomes smaller.
\par
To understand the effect of neutron finite size to $ \Delta R_{np} $,
we introduce here the effective charge formalism to implement the finite-size effect.
For simplicity, we consider only the Hartree term,
while it can be straightforwardly extended to the Coulomb exchange term.
If the effective charges of the protons and neutrons, $ e_{\urm{eff} p} $ and $ e_{\urm{eff} n} $, are introduced~\footnote{
  The effective charges may have $ \ve{r} $, $ Z $, or $ N $ dependences,
  while such dependences are not considered here for simplicity.},
the charge density distributions are written as
\begin{equation}
  \label{eq:eff_ch}
  \rho_{\urm{ch}} \left( \ve{r} \right)
  \simeq
  e_{\urm{eff} p} 
  \rho_p \left( \ve{r} \right)
  +
  e_{\urm{eff} n} 
  \rho_n \left( \ve{r} \right).
\end{equation}
Here, $ e_{\urm{eff} n} $ is assumed to be negative since the mean-square radius of neutron charge distribution $ r_{\urm{E} n}^2 $ is negative.
Since $ e_{\urm{eff} n} $ is negative, the proton-neutron Coulomb interaction is attractive.
At the same time, $ \rho_{\urm{ch}} $ shrinks when the neutron finite-size effect is considered,
because of $ e_{\urm{eff} n} < 0 $.
Therefore, the behavior is rather complicated, and eventually, the intercept $ a $ is almost unchanged,
because the neutron finite-size effect is weak.
\par
At last, we discuss how much the Coulomb interaction affects the estimation of $ L $ value from $ \Delta R_{np} $ in $ \nuc{Pb}{208}{} $.
Here, uncertainties due to the linear fitting is not considered.
For instance, if $ \Delta R_{np} = 0.20 \, \mathrm{fm} $ is assumed, 
adopting the present estimations with the Coulomb LDA EDF,
the slope parameter is estimated as
$ L = 75 \, \mathrm{MeV} $,
while
it is estimated as $ 73 \, \mathrm{MeV} $ with the ``All'' Coulomb EDF.
This difference is much smaller than the experimental error or uncertainty due to the linear fitting.
The treatment of the Coulomb interaction does not impact much on the extraction of $ L $ from the experimental result for $ \Delta R_{np} $.
\subsubsection{Comparison between Coulomb and ISB interactions}
\label{sec:calc_CvsR}
\par
\begin{figure}[tb]
  \centering
  \includegraphics[width=1.0\linewidth]{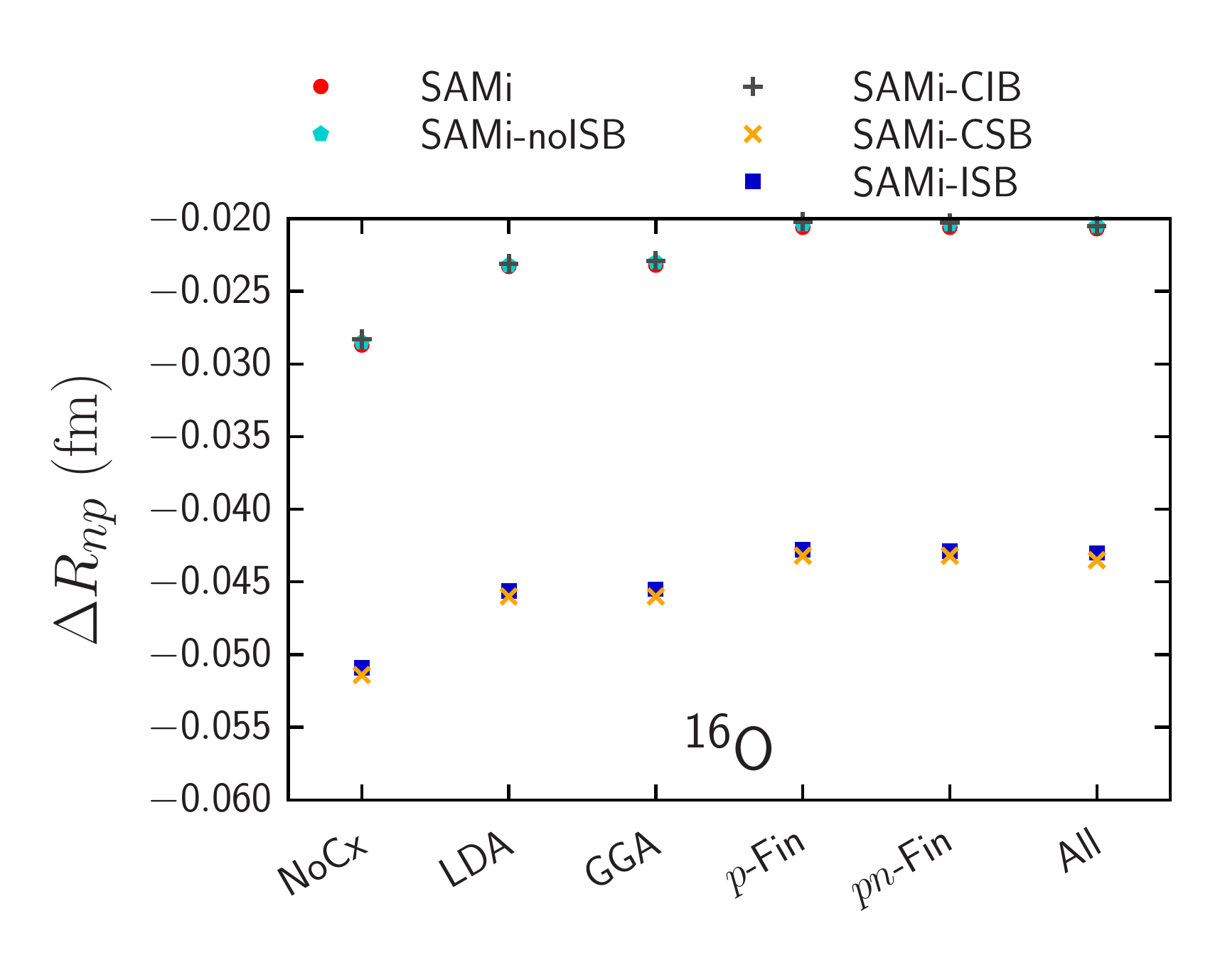}
  \caption{Neutron-skin thickness of $ \nuc{O}{16}{} $ 
    calculated with several Coulomb treatment with
    SAMi (circle),
    SAMi-noISB (pentagon),
    SAMi-CIB (plus),
    SAMi-CSB (cross),
    and SAMi-ISB (square) EDFs.}
  \label{fig:CoulvsRnp_008_016}
\end{figure}
\begin{figure}[tb]
  \centering
  \includegraphics[width=1.0\linewidth]{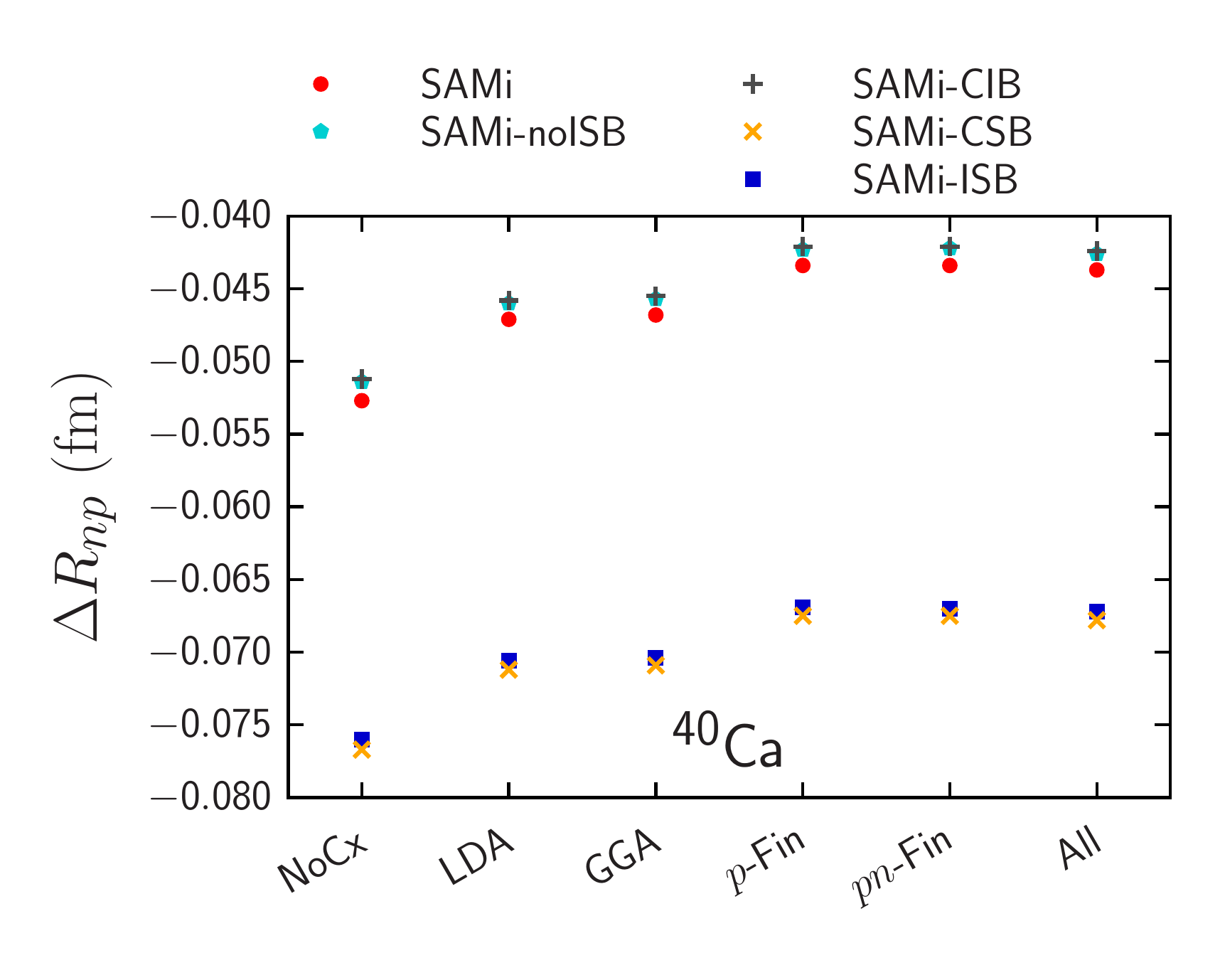}
  \caption{
    The same as Fig.~\ref{fig:CoulvsRnp_008_016} but for $ \nuc{Ca}{40}{} $.}
  \label{fig:CoulvsRnp_020_040}
\end{figure}
\begin{figure}[tb]
  \centering
  \includegraphics[width=1.0\linewidth]{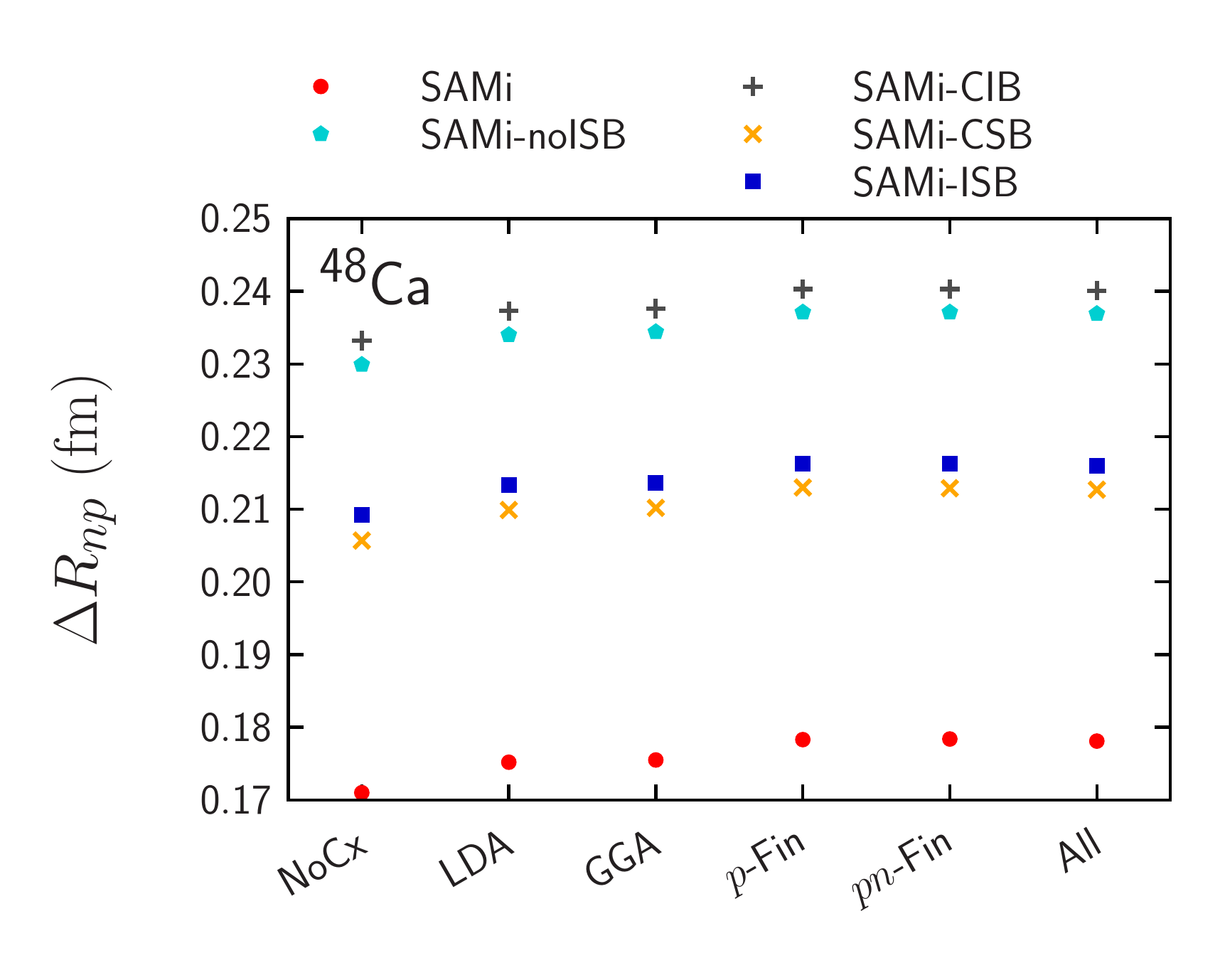}
  \caption{
    The same as Fig.~\ref{fig:CoulvsRnp_008_016} but for $ \nuc{Ca}{48}{} $.}
  \label{fig:CoulvsRnp_020_048}
\end{figure}
\begin{figure}[tb]
  \centering
  \includegraphics[width=1.0\linewidth]{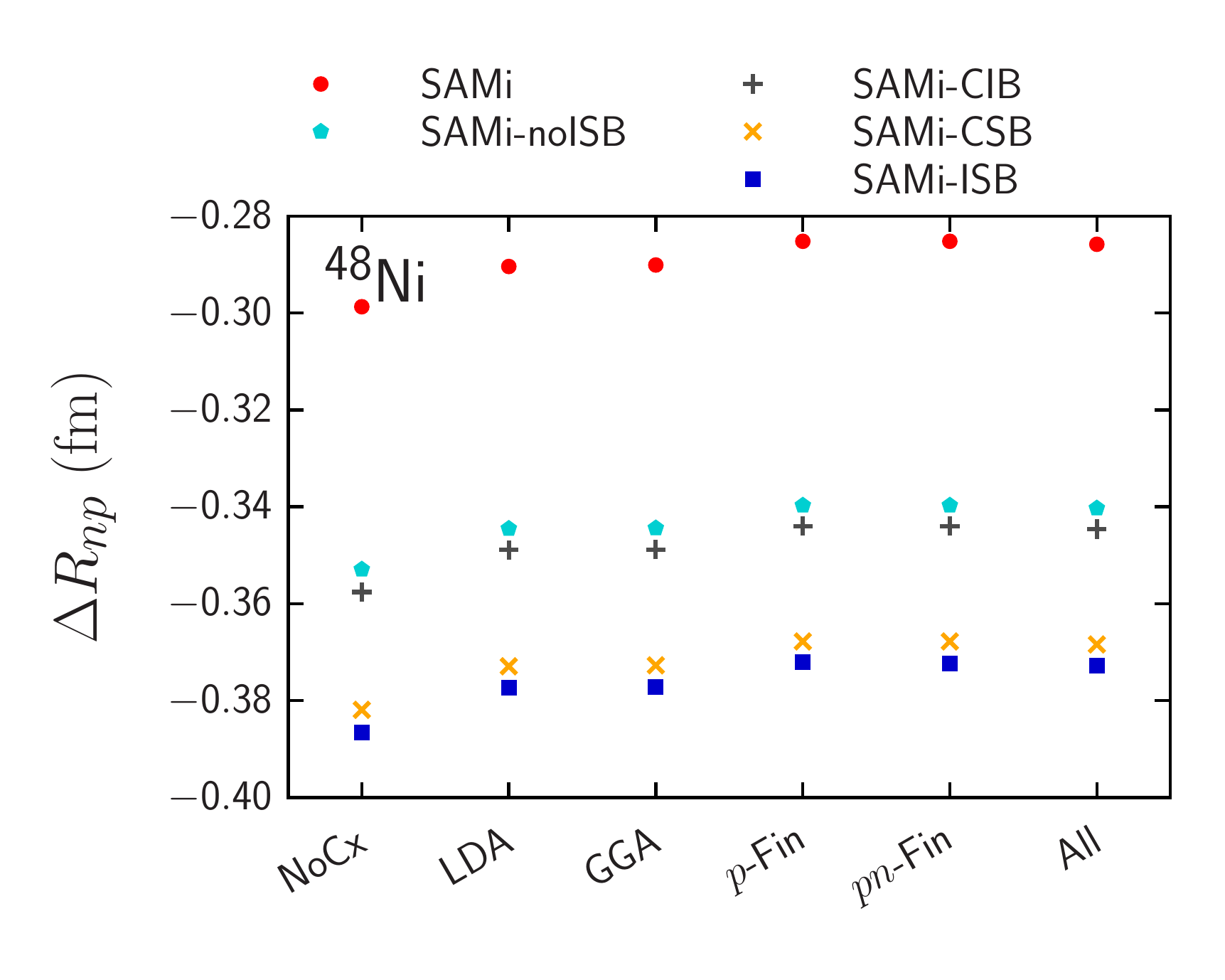}
  \caption{
    The same as Fig.~\ref{fig:CoulvsRnp_008_016} but for $ \nuc{Ni}{48}{} $.}
  \label{fig:CoulvsRnp_028_048}
\end{figure}
\begin{figure}[tb]
  \centering
  \includegraphics[width=1.0\linewidth]{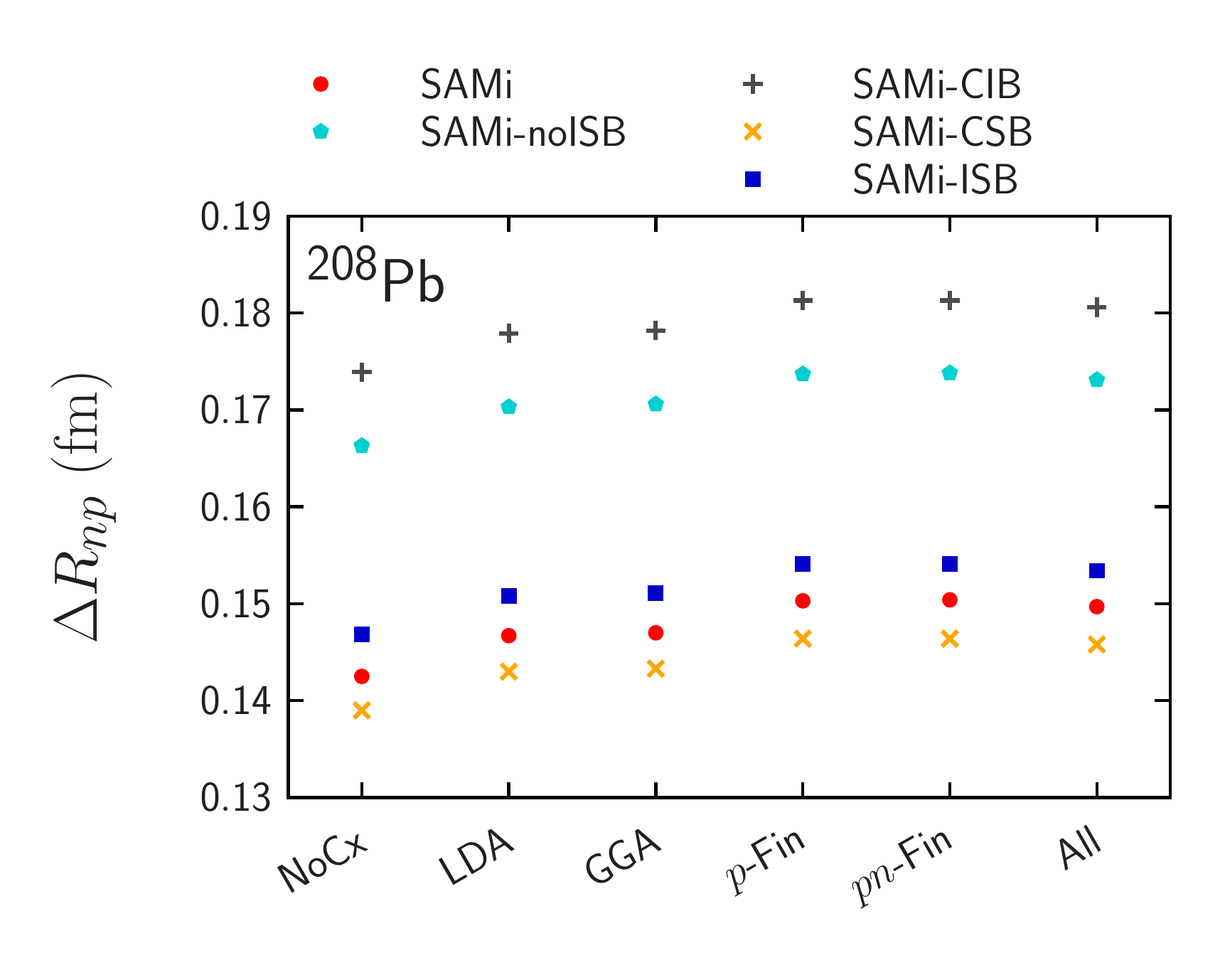}
  \caption{
    The same as Fig.~\ref{fig:CoulvsRnp_008_016} but for $ \nuc{Pb}{208}{} $.}
  \label{fig:CoulvsRnp_082_208}
\end{figure}

Next, we show the dependence
of the treatment of the Coulomb interaction and the ISB interaction  on $ \Delta R_{np} $
in Figs.~\ref{fig:CoulvsRnp_008_016}--\ref{fig:CoulvsRnp_082_208}.
Circles, pentagons, pluses, crosses, and squares show calculated results with
SAMi, SAMi-noISB, SAMi-CIB, SAMi-CSB, and SAMi-ISB EDFs, respectively.
Their values in $ \nuc{Ca}{40}{} $, $ \nuc{Ca}{48}{} $, and $ \nuc{Pb}{208}{} $ are also shown in 
Table~\ref{tab:R_Coul}.
\par
Both the CSB and CIB terms contribute to $ \Delta R_{np} $:
The former decreases $ \Delta R_{np} $ for both $ N = Z $ and $ N \ne Z $ nuclei;
the latter slightly decreases $ \Delta R_{np} $ in $ N < Z $ nuclei, 
whereas it slightly increases for $ N > Z $ nuclei.
The CIB term does not affect $ \Delta R_{np} $ in $ N = Z $ nuclei.
The values of the CSB and CIB contributions can be found in Table~\ref{tab:R_Coul} and Figs.~\ref{fig:CoulvsRnp_008_016}--\ref{fig:CoulvsRnp_082_208}.
\par
It is shown that SAMi and SAMi-noISB give almost the same $ \Delta R_{np} $ in $ N = Z $ nuclei,
while SAMi-noISB gives larger (smaller) $ \Delta R_{np} $ than SAMi in $ N > Z $ ($ N < Z $) nuclei.
In $ N = Z $ nuclei, if neither the ISB interaction nor the Coulomb interaction is considered,
$ \Delta R_{np} $ is equal to zero due to the isospin symmetry.
Thus, $ \Delta R_{np} $ of the SAMi and SAMi-noISB are constructed only due to the Coulomb interaction,
and thus these two EDFs give the similar $ \Delta R_{np} $ in $ N = Z $ nuclei.
In contrast, $ N \ne Z $ nuclei, difference between the SAMi and the SAMi-noISB reflects the property of these EDFs,
especially their $ L $ values
(SAMi: $ L = 44 \, \mathrm{MeV} $, SAMi-noISB: $ L = 50 \, \mathrm{MeV} $).
\par
Comparing how $ \Delta R_{np} $ is changed as the Coulomb interaction is treated precisely step-by-step,
one can find that such changes are universal among the results calculated with the SAMi, SAMi-noISB, SAMi-CIB, SAMi-CSB, and SAMi-ISB.
Thus,
one can conclude that
the effect of the model dependence associated with the treatment of the Coulomb interaction is unchanged
[$ O \left( 0.001 \right) \, \mathrm{fm} $],
which is comparable with the effect of the CIB term on $ \Delta R_{np} $ for $ N \ne Z $ nuclei.
\par
References~\cite{
  Brown1998Phys.Rev.C58_220,
  Brown2000Phys.Lett.B483_49,
  Goriely2008Phys.Rev.C77_031301}
claimed that the ISB interaction and the Coulomb exchange give a non-trivial cancellation.
If the claim in Refs.~\cite{
  Brown1998Phys.Rev.C58_220,
  Brown2000Phys.Lett.B483_49,
  Goriely2008Phys.Rev.C77_031301}
is valid,
$ \Delta R_{np} $ calculated
with the SAMi EDF without the Coulomb exchange term
and
that
with the SAMi-ISB EDF with the Coulomb LDA exchange (or all the Coulomb) term
should be identical.
However, as seen in Table~\ref{tab:R_Coul} and Figs.~\ref{fig:CoulvsRnp_008_016}--\ref{fig:CoulvsRnp_082_208},
$ \Delta R_{np} $ for $ \nuc{Ca}{40}{} $, $ \nuc{Ca}{48}{} $, and $ \nuc{Pb}{208}{} $ calculated with the SAMi without the Coulomb exchange term
are
$ -0.0527 $,
$  0.1710 $, and
$  0.1425 \, \mathrm{fm} $, respectively,
whereas
those with the SAMi-ISB and the Coulomb exchange (all Coulomb)
are
$ -0.0706 $, 
$  0.2133 $, and
$  0.1508 \, \mathrm{fm} $
($ -0.0672 $, $  0.2160 $, and $  0.1534 \, \mathrm{fm} $);
these two are still non-negligibly different, except $ \nuc{Pb}{208}{} $.
The case of $ \nuc{Pb}{208}{} $ may be accidental.
Thus, such treatment is not fully acceptable to discuss $ \Delta R_{np} $ quantitatively.
\begingroup
\squeezetable
\begin{table*}[tb]
  \centering
  \caption{
    Neutron-skin thickness $ \Delta R_{np} $ of
    $ \nuc{Ca}{40}{} $, $ \nuc{Ca}{48}{} $, and $ \nuc{Pb}{208}{} $
    calculated with SAMi-ISB EDF
    without Coulomb exchange term (NoCx),
    with Coulomb LDA (LDA),
    or with full Coulomb treatment (All)
    without ISB term, only with CSB term, only with CIB term, and with all ISB terms.
    For comparison, $ \Delta R_{np} $ calculated  with the SAMi EDF is also shown.
    All the values are in $ \mathrm{fm} $.}
  \label{tab:R_Coul}
  \begin{ruledtabular}
    \begin{tabular}{llddddddddd}
      \multicolumn{1}{c}{$ E_{\urm{IS}} $} & \multicolumn{1}{c}{ISB} & \multicolumn{3}{c}{$ \nuc{Ca}{40}{} $} & \multicolumn{3}{c}{$ \nuc{Ca}{48}{} $} & \multicolumn{3}{c}{$ \nuc{Pb}{208}{} $} \\
                                           & & \multicolumn{1}{c}{NoCx} & \multicolumn{1}{c}{LDA} & \multicolumn{1}{c}{All} & \multicolumn{1}{c}{NoCx} & \multicolumn{1}{c}{LDA} & \multicolumn{1}{c}{All} & \multicolumn{1}{c}{NoCx} & \multicolumn{1}{c}{LDA} & \multicolumn{1}{c}{All} \\
      \hline
      SAMi       & No ISB   & -0.0527 & -0.0471 & -0.0437 & 0.1710 & 0.1752 & 0.1781 & 0.1425 & 0.1467 & 0.1497 \\
      \hline
      SAMi-noISB & No ISB   & -0.0514 & -0.0460 & -0.0426 & 0.2299 & 0.2340 & 0.2369 & 0.1663 & 0.1703 & 0.1731 \\
      SAMi-noISB & Only CIB & -0.0512 & -0.0458 & -0.0424 & 0.2332 & 0.2373 & 0.2401 & 0.1739 & 0.1779 & 0.1806 \\
      SAMi-noISB & Only CSB & -0.0767 & -0.0712 & -0.0678 & 0.2057 & 0.2099 & 0.2127 & 0.1390 & 0.1430 & 0.1458 \\
      SAMi-noISB & All ISB  & -0.0760 & -0.0706 & -0.0672 & 0.2092 & 0.2133 & 0.2160 & 0.1468 & 0.1508 & 0.1534 \\
    \end{tabular}
  \end{ruledtabular}
\end{table*}
\endgroup
\subsubsection{ISB term and neutron-skin thickness}
\label{sec:calc_ISBvsR}
\par
At last, we study how much the ISB strength correlates with the neutron-skin thickness $ \Delta R_{np} $.
In this calculation, the SAMi-noISB EDF is used for $ E_{\urm{IS}} $.
The CSB strength $ -s_0 $ in $ E_{\urm{CSB}} $ is gradually changed from $ 0 \, \mathrm{MeV} \, \mathrm{fm}^3 $ to $ 50 \, \mathrm{MeV} \, \mathrm{fm}^3 $,
while the CIB strength $ u_0 $ in $ E_{\urm{CIB}} $ is kept $ 0 \, \mathrm{MeV} \, \mathrm{fm}^3 $,
or
the CIB strength $ u_0 $ in $ E_{\urm{CIB}} $ is gradually changed from $ 0 \, \mathrm{MeV} \, \mathrm{fm}^3 $ to $ 50 \, \mathrm{MeV} \, \mathrm{fm}^3 $,
while $ E_{\urm{CSB}} = 0 \, \mathrm{MeV} \, \mathrm{fm}^3 $ is kept.
Note that this calculation with
$ u_0 = 25.8 \, \mathrm{MeV} \, \mathrm{fm}^3 $ and
$ -s_0 = 26.3 \, \mathrm{MeV} \, \mathrm{fm}^3 $ exactly corresponds to the SAMi-ISB EDF.
The Coulomb LDA EDF [Eq.~\eqref{eq:Coul_def_LDA}] is used for the calculation.
\par
Using these data, $ u_0 $ or $ - s_0 $ dependence of $ \Delta R_{np} $ is parametrized as 
\begin{subequations}
  \label{eq:ISBvsR}
  \begin{align}
    \Delta R_{np}
    & =
      c + d \left( - s_0 \right),
      \label{eq:ISBvsR_CSB} \\
    \Delta R_{np}
    & =
      e + f u_0.
      \label{eq:ISBvsR_CIB}
  \end{align}
\end{subequations}
\par
\begin{figure}[tb]
  \centering
  \includegraphics[width=1.0\linewidth]{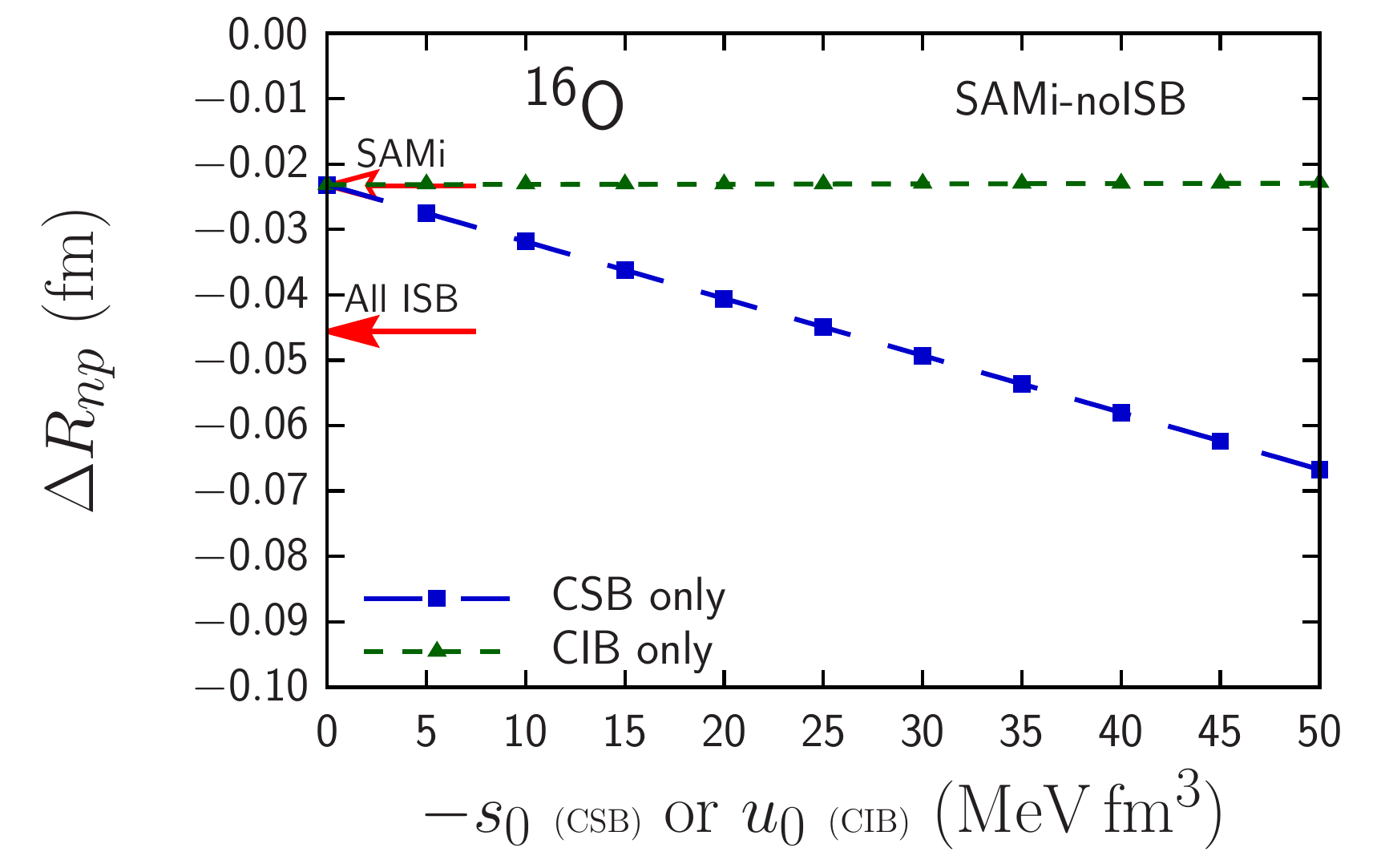}
  \caption{
    Neutron-skin thickness $ \Delta R_{np} $ as functions of the CSB and CIB strength $ -s_0 $ and $ u_0 $, respectively, for $ \nuc{O}{16}{} $.
    The long-dashed and dashed lines 
    show results only with CSB and only with CIB, respectively.
    Filled and empty arrows show $ \Delta R_{np} $ calculated by the full SAMi-ISB
    and the SAMi EDFs, respectively.}
  \label{fig:ISBvsR_008_016}
\end{figure}
\begin{figure}[tb]
  \centering
  \includegraphics[width=1.0\linewidth]{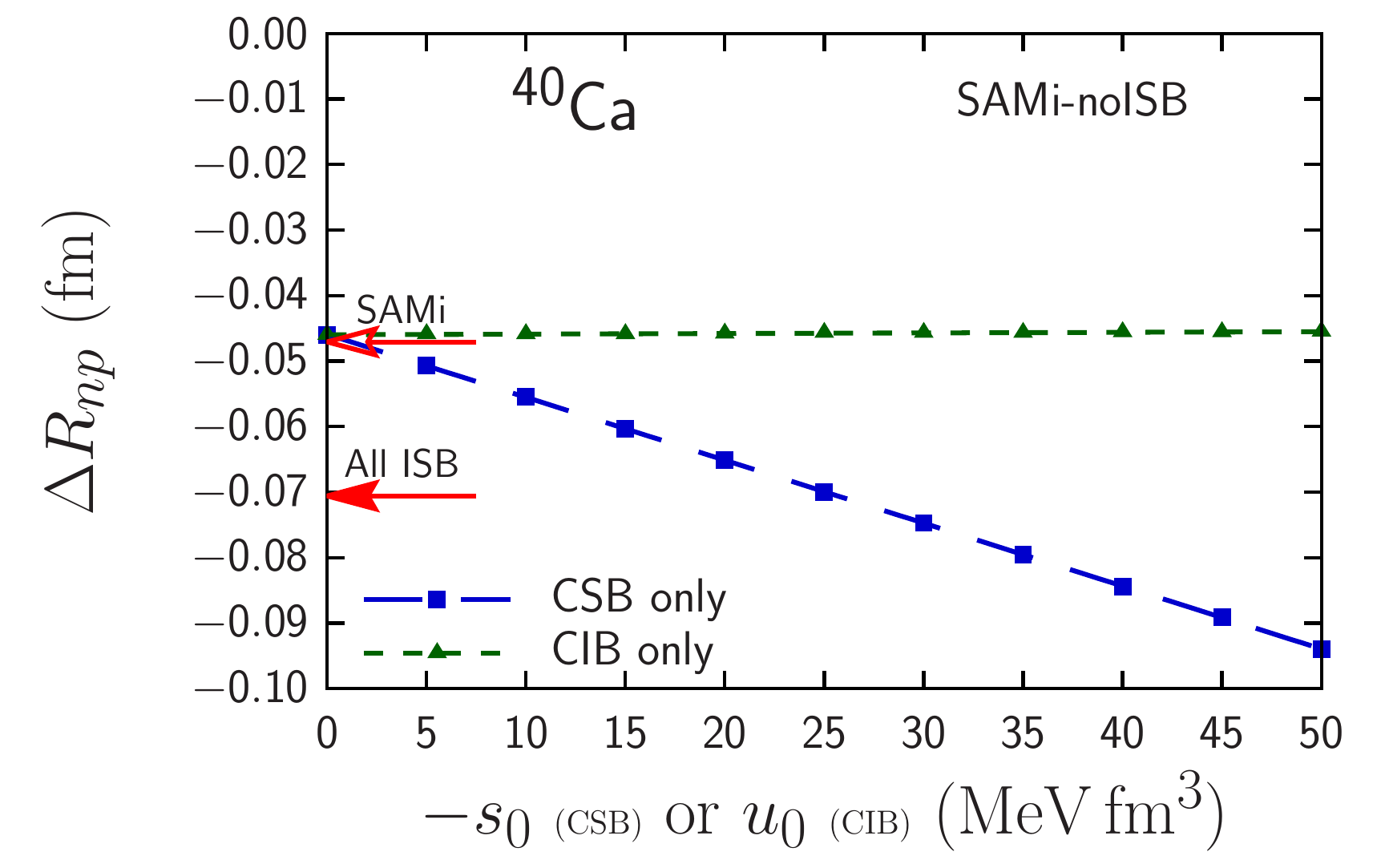}
  \caption{
    The same as Fig.~\ref{fig:ISBvsR_008_016} but for $ \nuc{Ca}{40}{} $.}
  \label{fig:ISBvsR_020_040}
\end{figure}
\begin{figure}[tb]
  \centering
  \includegraphics[width=1.0\linewidth]{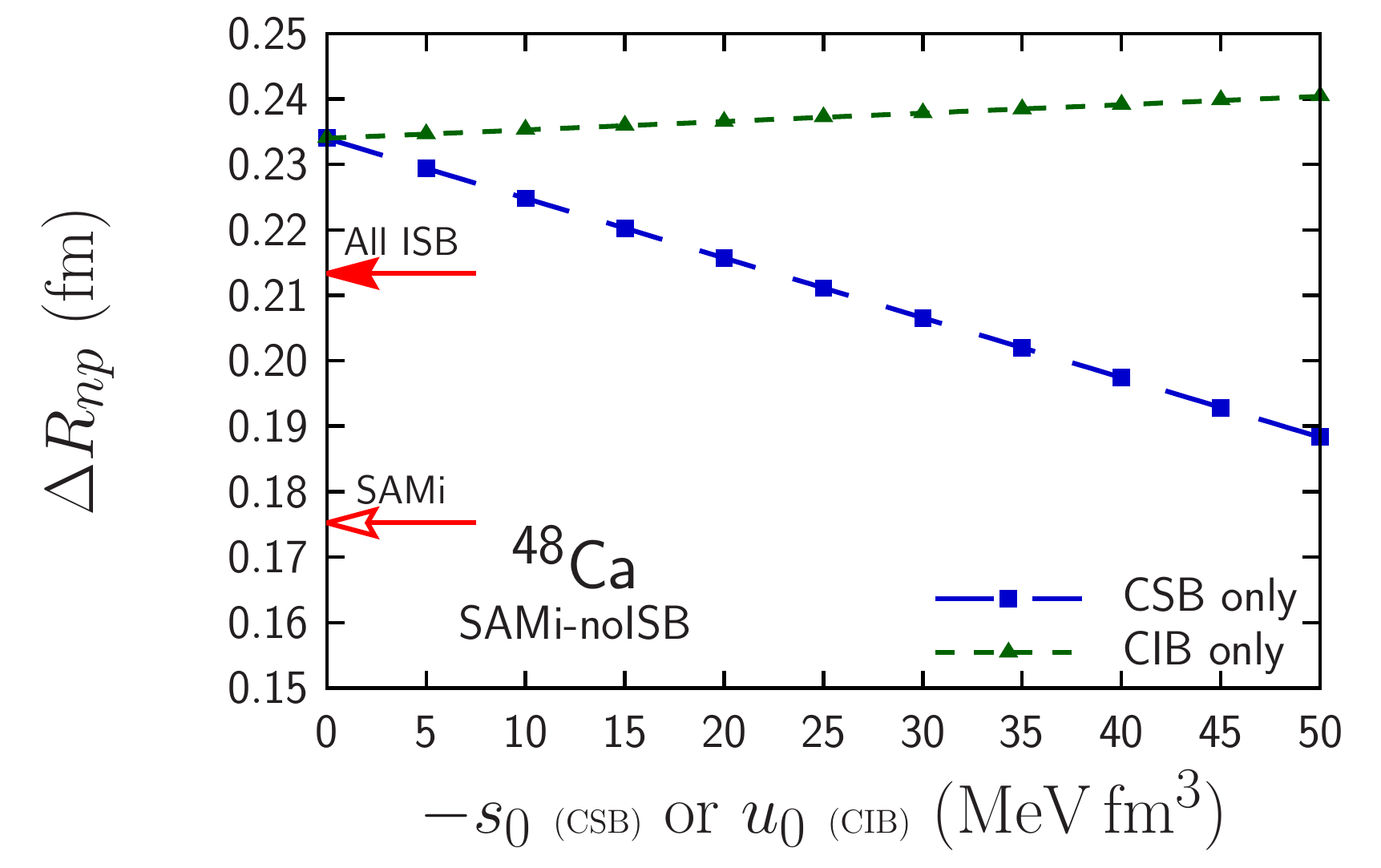}
  \caption{
    The same as Fig.~\ref{fig:ISBvsR_008_016} but for $ \nuc{Ca}{48}{} $.}
  \label{fig:ISBvsR_020_048}
\end{figure}
\begin{figure}[tb]
  \centering
  \includegraphics[width=1.0\linewidth]{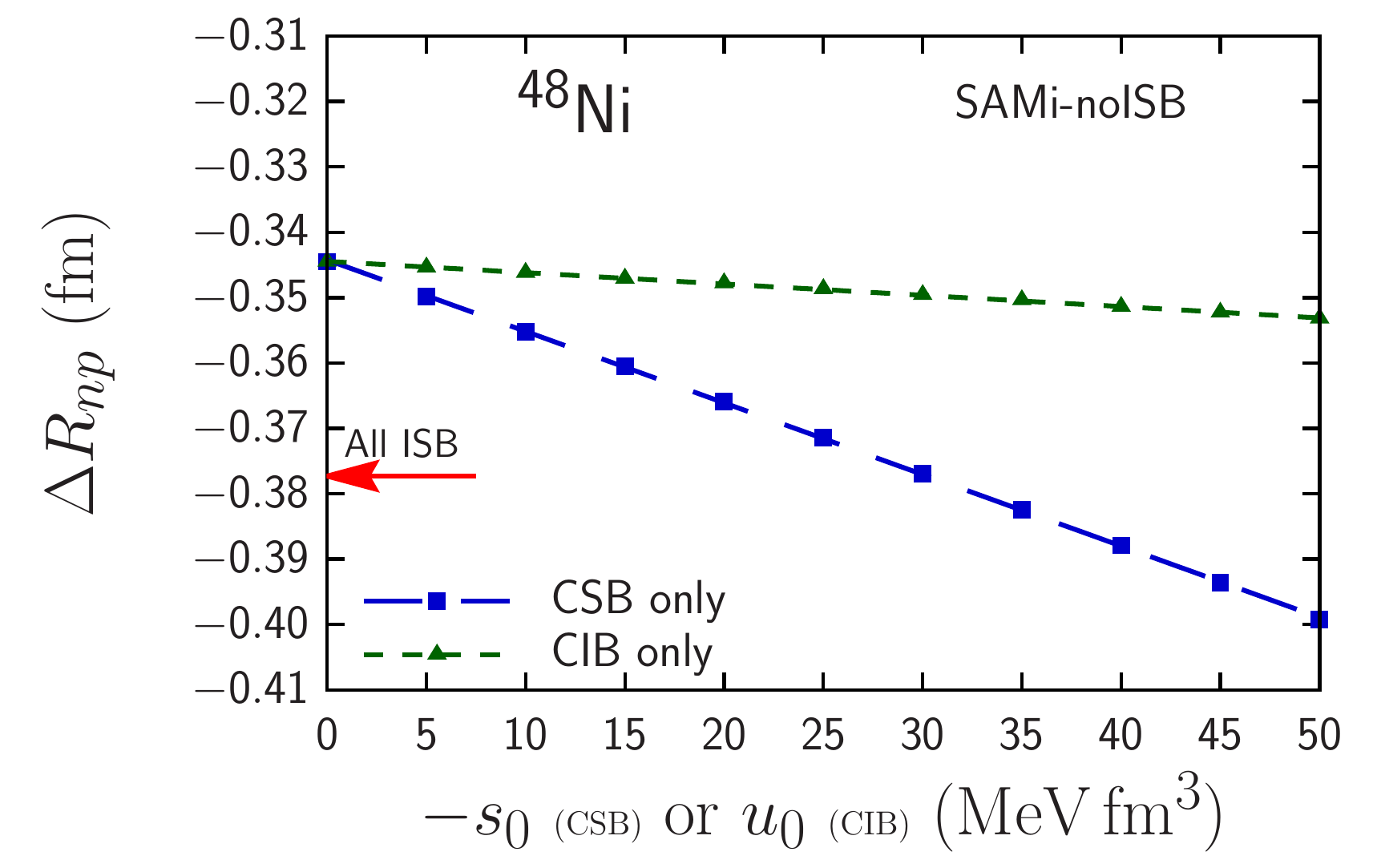}
  \caption{
    The same as Fig.~\ref{fig:ISBvsR_008_016} but for $ \nuc{Ni}{48}{} $.}
  \label{fig:ISBvsR_028_048}
\end{figure}
\begin{figure}[tb]
  \centering
  \includegraphics[width=1.0\linewidth]{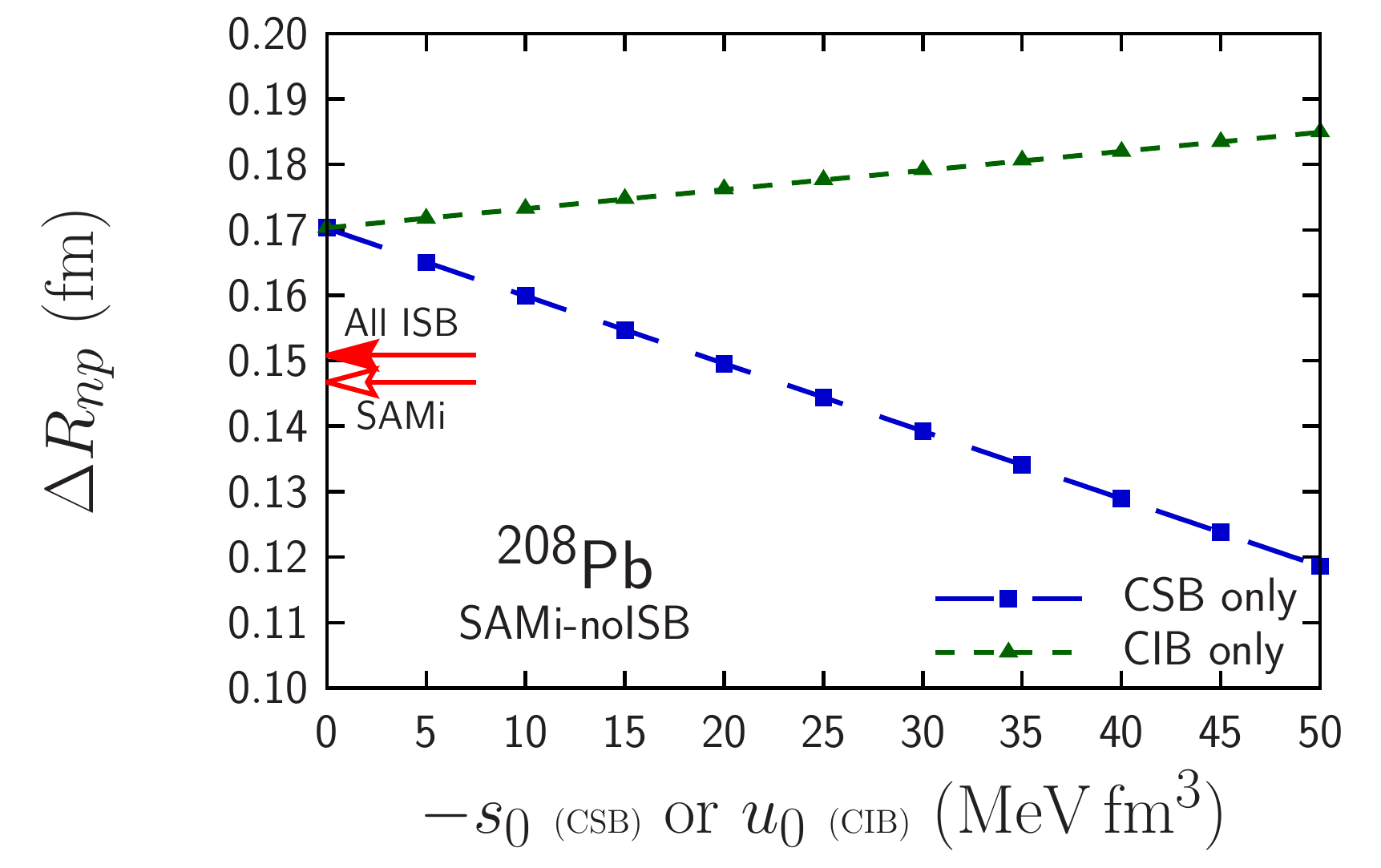}
  \caption{
    The same as Fig.~\ref{fig:ISBvsR_008_016} but for $ \nuc{Pb}{208}{} $.}
  \label{fig:ISBvsR_082_208}
\end{figure}

Figures~\ref{fig:ISBvsR_008_016}--\ref{fig:ISBvsR_082_208} show the neutron-skin thickness $ \Delta R_{np} $ as functions of the ISB strength $ -s_0 $ and $ u_0 $ for $ \nuc{O}{16}{} $, $ \nuc{Ca}{40}{} $, $ \nuc{Ca}{48}{} $, $ \nuc{Ni}{48}{} $, and $ \nuc{Pb}{208}{} $.
Since the refitting effect of the Skyrme EDF is non-negligible,
the experimental data are not shown in the figures.
The long-dashed and dashed lines 
show the results only with CSB and only with CIB, respectively.
Filled and empty arrows show $ \Delta R_{np} $ calculated by the full SAMi-ISB and the original SAMi EDFs, respectively.
Using the data, $ c $ and $ d $ in Eq.~\eqref{eq:ISBvsR_CSB}
and $ e $ and $ f $ in Eq.~\eqref{eq:ISBvsR_CIB} are determined as shown in Table~\ref{tab:ISBvsR}.
\begin{table}[tb]
  \centering
  \caption{
    Parameters $ c $, $ d $, $ e $, and $ f $ in Eqs.~\eqref{eq:ISBvsR}.
    See text for more detail.}
  \label{tab:ISBvsR}
  \begin{ruledtabular}
    \begin{tabular}{rldd}
      \multicolumn{1}{c}{Nuclei} & \multicolumn{1}{c}{ISB} & \multicolumn{1}{c}{$ c $ or $ e $ ($ \mathrm{fm} $)} & \multicolumn{1}{c}{$ d $ or $ f $ ($ \mathrm{MeV}^{-1} \, \mathrm{fm}^{-2} $)} \\ \hline
      $ \nuc{O}{16}{} $
                                 & CSB & -0.023145 & -0.00087127 \\
                                 & CIB & -0.023168 &  0.00000455 \\
      \hline
      $ \nuc{Ca}{40}{} $
                                 & CSB & -0.045923 & -0.00096055 \\
                                 & CIB & -0.045986 &  0.00000927 \\
      \hline
      $ \nuc{Ca}{48}{} $
                                 & CSB &  0.233945 & -0.00091309 \\
                                 & CIB &  0.233977 &  0.00012818 \\
      \hline
      $ \nuc{Ni}{48}{} $
                                 & CSB & -0.344227 & -0.00109382 \\
                                 & CIB & -0.344459 & -0.00017291 \\
      \hline
      $ \nuc{Pb}{208}{} $
                                 & CSB &  0.170209 & -0.00103236 \\
                                 & CIB &  0.170300 &  0.00029200 \\
    \end{tabular}
  \end{ruledtabular}
\end{table}
\par
The CSB strength makes $ \Delta R_{np} $ smaller
and its slope $ d $ has almost the same value,
$ d \simeq -0.001 \, \mathrm{MeV}^{-1} \, \mathrm{fm}^{-2} $,
among all the calculated nuclei.
On the other hand, a larger CIB strength
changes $ \Delta R_{np} $ around $ 10 $--$ 20 \, \% $ of the CSB case in $ N \ne Z $ nuclei,
while around $ 0.5 $--$ 1 \, \% $ in $ N = Z $ nuclei.
Moreover, $ \Delta R_{np} $ becomes larger in $ N > Z $ nuclei and smaller in $ N < Z $ nuclei
as the CIB strength $ u_0 $ becomes larger.
The absolute value of the change of $ \Delta R_{np} $ is almost the same among the mirror nuclei.
Eventually, the CSB term gives the dominant contribution to $ \Delta R_{np} $,
and the ISB contribution to $ \Delta R_{np} $ is qualitatively universal in all the selected nuclei.
\par
We will consider the main reason of such behaviors~\footnote{
  The mechanism how the CSB interaction affects $ \Delta R_{np} $ was already discussed in the previous paper~\cite{
    Naito2022Phys.Rev.C105_L021304}.}.
The Skyrme-like zero-range CSB interaction for proton-proton, proton-neutron, and neutron-neutron are~\cite{
  Sagawa1995Phys.Lett.B353_7,
  Roca-Maza2018Phys.Rev.Lett.120_202501}
\begin{subequations}
  \label{eq:2body_CSB}
  \begin{align}
    v_{\urm{CSB}}^{pp} \left( \ve{r}_1, \ve{r}_2 \right)
    & =
      - \frac{s_0}{2} \left( 1 - P_{\sigma} \right)
      \delta \left( \ve{r}_1 - \ve{r}_2 \right), \\
    v_{\urm{CSB}}^{pn} \left( \ve{r}_1, \ve{r}_2 \right)
    & =
      0, \\
    v_{\urm{CSB}}^{nn} \left( \ve{r}_1, \ve{r}_2 \right)
    & =
      + \frac{s_0}{2} \left( 1 - P_{\sigma} \right)
      \delta \left( \ve{r}_1 - \ve{r}_2 \right), 
  \end{align}
\end{subequations}
respectively, with $ s_0 < 0 $.
Hence, as $ \left| s_0 \right| $ becomes larger,
the proton-proton repulsive interaction
and the neutron-neutron attractive interaction become stronger.
Accordingly, $ \rho_p $ expands and $ \rho_n $ shrinks.
Consequently, $ \Delta R_{np} $ becomes smaller as $ \left| s_0 \right| $ becomes larger.
\par
On the contrary, the Skyrme-like zero-range CIB interaction for proton-proton, proton-neutron, and neutron-neutron are~\cite{
  Sagawa1995Phys.Lett.B353_7,
  Roca-Maza2018Phys.Rev.Lett.120_202501}
\begin{subequations}
  \label{eq:2body_CIB}
  \begin{align}
    v_{\urm{CIB}}^{pp} \left( \ve{r}_1, \ve{r}_2 \right)
    & =
      + \frac{u_0}{2} \left( 1 - P_{\sigma} \right)
      \delta \left( \ve{r}_1 - \ve{r}_2 \right), \\
    v_{\urm{CIB}}^{pn} \left( \ve{r}_1, \ve{r}_2 \right)
    & =
      - \frac{u_0}{2} \left( 1 - P_{\sigma} \right)
      \delta \left( \ve{r}_1 - \ve{r}_2 \right), \\
    v_{\urm{CIB}}^{nn} \left( \ve{r}_1, \ve{r}_2 \right)
    & =
      + \frac{u_0}{2} \left( 1 - P_{\sigma} \right)
      \delta \left( \ve{r}_1 - \ve{r}_2 \right), 
  \end{align}
\end{subequations}
respectively, with $ u_0 > 0 $.
Hence, as $ u_0 $ becomes larger,
the proton-proton and neutron-neutron repulsive interactions and
the proton-neutron attractive interaction become stronger.
The effective CIB potential for protons and neutrons read 
\begin{subequations}
  \begin{align}
    V_{\urm{CIB}}^p \left( \ve{r} \right)
    & =
      \frac{\delta E_{\urm{CIB}} \left[ \rho_p, \rho_n \right]}{\delta \rho_p \left( \ve{r} \right)}
      \notag \\
    & =
      \frac{u_0}{2} \rho_p \left( \ve{r} \right)
      -
      \frac{u_0}{4} \rho_n \left( \ve{r} \right)
      \notag \\
    & =
      \frac{u_0}{4}
      \left[
      2 \rho_p \left( \ve{r} \right)
      -
      \rho_n \left( \ve{r} \right)
      \right], \\
    V_{\urm{CIB}}^n \left( \ve{r} \right)
    & =
      \frac{\delta E_{\urm{CIB}} \left[ \rho_p, \rho_n \right]}{\delta \rho_n \left( \ve{r} \right)}
      \notag \\
    & =
      \frac{u_0}{2} \rho_n \left( \ve{r} \right)
      -
      \frac{u_0}{4} \rho_p \left( \ve{r} \right)
      \notag \\
    & =
      \frac{u_0}{4}
      \left[
      2 \rho_n \left( \ve{r} \right)
      -
      \rho_p \left( \ve{r} \right)
      \right],
  \end{align}
\end{subequations}
respectively.
In $ N = Z $ nuclei, the proton and neutron density distributions are approximately the same,
$ \rho_p \left( \ve{r} \right) \simeq \rho_n \left( \ve{r} \right) $.
Therefore, $ V_{\urm{CIB}}^p \left( \ve{r} \right) \simeq V_{\urm{CIB}}^n \left( \ve{r} \right) $ also holds,
and thus the CIB effect on $ R_p $ is almost same as that on $ R_n $.
Hence, even though $ R_p $ and $ R_n $ are changed as the CIB strength $ u_0 $ is changed,
the neutron-skin thickness $ \Delta R_{np} $ is almost unchanged.
In $ N > Z $ nuclei, in general, $ \rho_n \left( \ve{r} \right) > \rho_p \left( \ve{r} \right) $ holds,
and thus the repulsive potential $ V_{\urm{CIB}}^p $ is weaker than $ V_{\urm{CIB}}^n $,
i.e., $ V_{\urm{CIB}}^p < V_{\urm{CIB}}^n $.
Hence, $ R_n $ extends more than $ R_p $, and accordingly, $ \Delta R_{np} $ increases.
Since $ V_{\urm{CIB}}^p $ is approximately equal to $ V_{\urm{CIB}}^n $ of the mirror nucleus,
the behaviors among the mirror nuclei have the same magnitude but with opposite signs.
\par
At last, we will discuss how large the CSB and CIB strengths affect the estimation of $ L $ value.
Here, $ \nuc{Pb}{208}{} $ is taken as an example.
As we did in Sec.~\ref{sec:calc_LvsR} and Ref.~\cite{
  Naito:2022hyb},
the $ L $-$ \Delta R_{np} $ correlation is derived by using the SAMi EDF and SAMi-J family.
Because the pressure of neutron matter is proportional to
$ L + L^{\urm{CIB}} + L^{\urm{CSB}} $,
the neutron-skin thickness is also expected to be correlated with $ L + L^{\urm{CIB}} + L^{\urm{CSB}} $.
If one does not consider CIB or CSB term,
$ L^{\urm{CIB}} $ and $ L^{\urm{CSB}} $ are equal to zero.
\par
\begin{figure}[tb]
  \centering
  \includegraphics[width=1.0\linewidth]{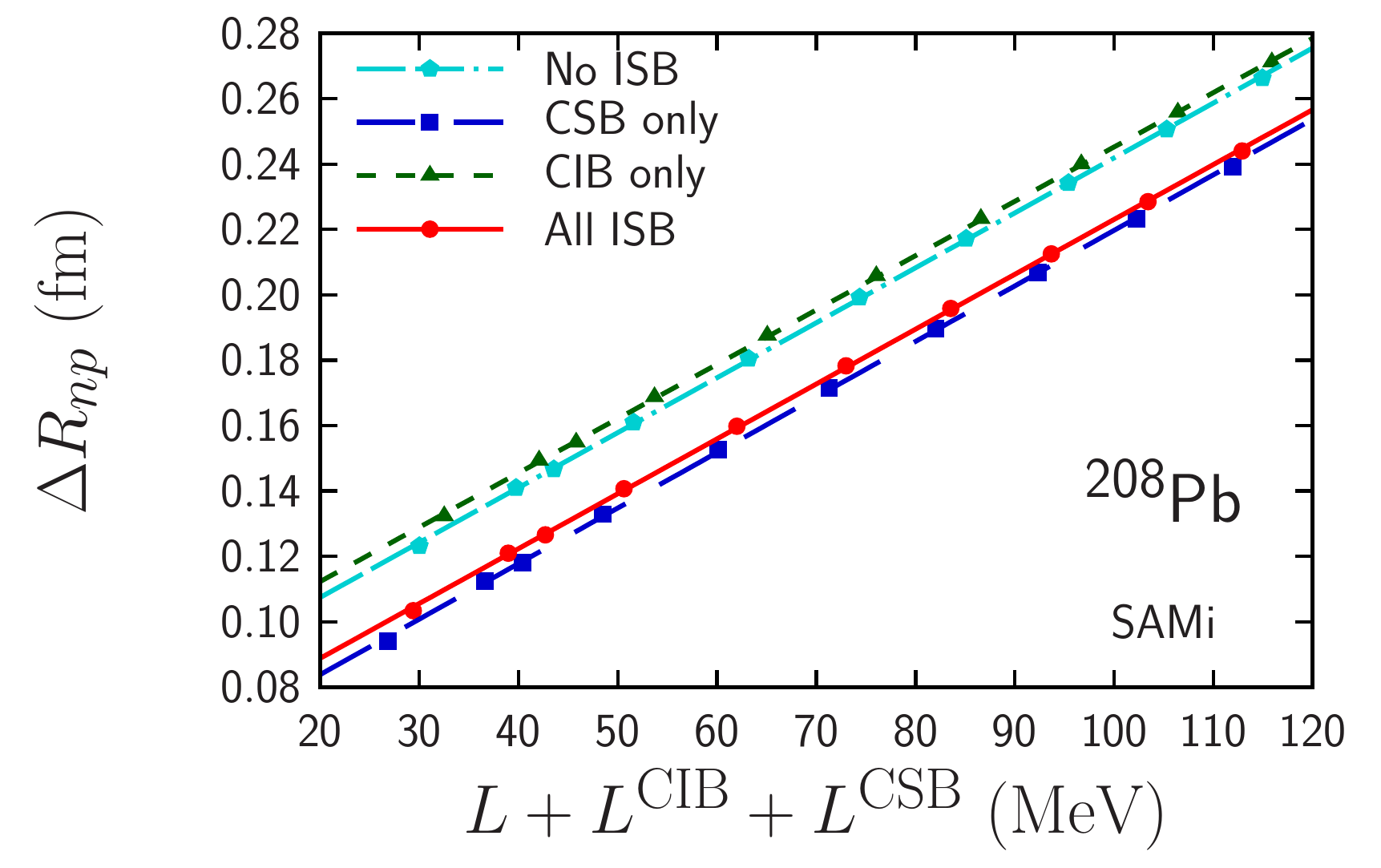}
  \caption{
    Correlation between $ L + L^{\urm{CIB}} + L^{\urm{CSB}} $ and $ \Delta R_{np} $ calculated
    without any ISB terms, only with CSB term, only with CIB term, and with the all ISB terms, respectively,
    shown in dash-dotted, long-dashed, dashed, and solid lines.
    The ISB terms of the SAMi-ISB EDF [Eqs.~\eqref{eq:CSB} and \eqref{eq:CIB}] are considered on top of the SAMi EDF and SAMi-J family.}
  \label{fig:Rnp_082_208}
\end{figure}

In order to see the ISB contribution,
the ISB terms of the SAMi-ISB EDF [Eqs.~\eqref{eq:CSB} and \eqref{eq:CIB}] are considered on top of the SAMi EDF and SAMi-J family.
The correlations without any ISB terms, only with CIB term, only with CSB term, and with both the CSB and CIB terms, denoted by all ISB, respectively, read
\begin{widetext}
  \begin{equation}
    \Delta R_{np}
    =
    \begin{cases}
      0.001680
      \left( L + L^{\urm{CIB}} + L^{\urm{CSB}} \right)
      +
      0.07385
      & \text{(without ISB terms)}, \\
      0.001661
      \left( L + L^{\urm{CIB}} + L^{\urm{CSB}} \right)
      +
      0.07907
      & \text{(only with CIB term)}, \\
      0.001698
      \left( L + L^{\urm{CIB}} + L^{\urm{CSB}} \right)
      +
      0.04987
      & \text{(only with CSB term)}, \\
      0.001678
      \left( L + L^{\urm{CIB}} + L^{\urm{CSB}} \right)
      +
      0.05526
      & \text{(with all ISB terms)},
    \end{cases}
  \end{equation}
\end{widetext}
and are shown in Fig.~\ref{fig:Rnp_082_208}.
If the same $ \Delta R_{np} $ is assumed,
the difference between obtained $ L + L^{\urm{CIB}} + L^{\urm{CSB}} $
without any ISB terms
and
that with all ISB terms
is $ 11.1 \, \mathrm{MeV} $.
Using $ L^{\urm{CIB}} = 2.3 \, \mathrm{MeV} $ and $ L^{\urm{CSB}} = -3.2 \, \mathrm{MeV} $,
one finds that $ L $ itself is changed by $ 12.0 \, \mathrm{MeV} $ once the ISB terms are considered.
Thus, the ISB contributions, in particular, the CSB one, to the $ L $ parameter may not be negligible.
In contrast to the case of the charge-radii difference of mirror nuclei $ \Delta R_{\urm{ch}} $,
the effect on $ L $ is smaller.
This is because
the CIB effect and the CSB one is opposite direction in $ \Delta R_{np} $ for $ N > Z $ nuclei,
while they are coherent in $ \Delta R_{\urm{ch}} $.
It should be noted that once $ E_{\urm{IS}} $ is refitted with considering the ISB terms,
the effect of ISB terms becomes rather mild.
Here, the uncertainty due to the fitting is not considered,
since both correlations between $ L $ and $ \Delta R_{np} $ and that between ISB strengths and $ \Delta R_{np} $
obtained in this paper 
are almost perfect (with $ r \approx 1.000 $),
and accordingly, the uncertainty due to the correlation is negligible.
\subsection{Charge radii difference between $ \nuc{Ca}{40}{} $ and $ \nuc{Ca}{48}{} $}
\label{sec:calc_Ca}
\par
\begin{figure}[tb]
  \centering
  \includegraphics[width=1.0\linewidth]{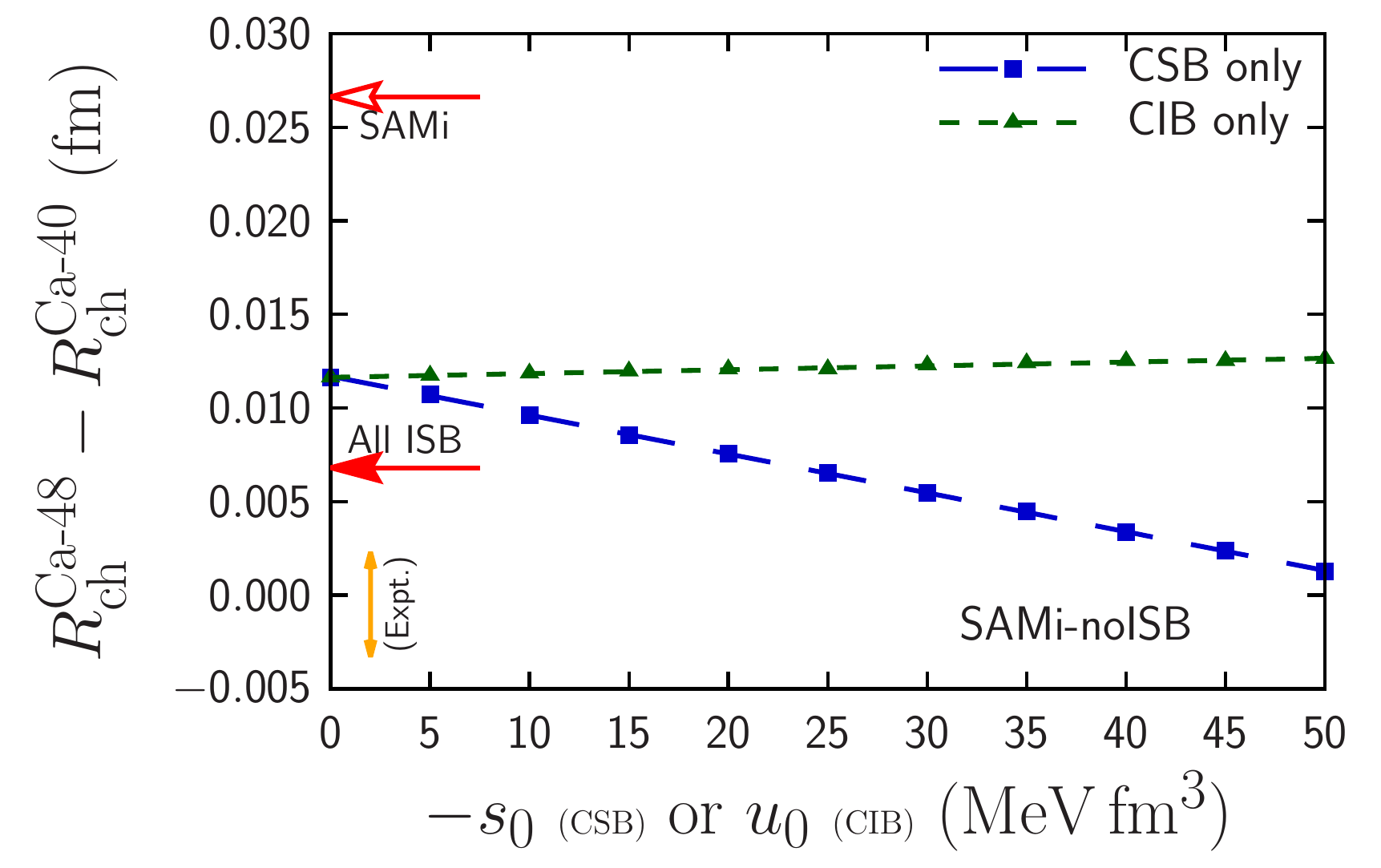}
  \caption{
    The CSB- and CIB-strength, $ -s_0 $ and $ u_0 $, dependences of
    the difference of the calculated charge radius of $ \nuc{Ca}{40}{} $ and that of $ \nuc{Ca}{48}{} $.
    Experimental value taken from Ref.~\cite{
      Angeli2013At.DataNucl.DataTables99_69}
    is shown as a vertical line.}
  \label{fig:Ca_ISB}
\end{figure}

Figure~\ref{fig:Ca_ISB} shows the CSB- and CIB-strength, $ -s_0 $ and $ u_0 $, dependences of
the difference of charge radii,
$ R_{\urm{ch}}^{\urm{Ca-48}} - R_{\urm{ch}}^{\urm{Ca-40}} $.
Here, the root-mean-square radius $ R_{\urm{ch}} $ is calculated by
\begin{equation}
  \label{eq:charge_radii}
  R_{\urm{ch}}^2
  =
  \int
  r^2
  \rho_{\urm{ch}} \left( \ve{r} \right)
  \, d \ve{r}
  +
  R_{\urm{so}}^2,
\end{equation}
where
$ \rho_{\urm{ch}} $ is the charge density distribution calculated by using Eq.~\eqref{eq:charge},
which only includes effects of nucleon electric form factors,
and
$ R_{\urm{so}}^2 $ is the spin-orbit (magnetic) contribution to charge radius~\cite{
  Horowitz2012Phys.Rev.C86_045503,
  Reinhard2021Phys.Rev.C103_054310,
  Naito2021Phys.Rev.C104_024316}.
The spin-orbit contribution can be calculated as
$ R_{\urm{so}}^2 = 0 \, \mathrm{fm}^2 $ and $ -0.101 \, \mathrm{fm}^2 $
for $ \nuc{Ca}{40}{} $ and $ \nuc{Ca}{48}{} $, respectively.
As we did in Sec.~\ref{sec:calc_ISBvsR},
on top of the SAMi-noISB EDF,
the CSB or CIB strength, $ -s_0 $ or $ u_0 $, is gradually changed from $ 0 \, \mathrm{MeV} \, \mathrm{fm}^3 $ to $ 50 \, \mathrm{MeV} \, \mathrm{fm}^3 $.
These data are fitted to
\begin{subequations}
  \begin{align}
    R_{\urm{ch}}^{\urm{Ca-48}} - R_{\urm{ch}}^{\urm{Ca-40}}
    & =
      0.011697 - 2.0767 \times 10^{-4} \left( - s_0 \right), \\
    R_{\urm{ch}}^{\urm{Ca-48}} - R_{\urm{ch}}^{\urm{Ca-40}}
    & =
      0.011637 + 2.0297 \times 10^{-5} u_0,
  \end{align}
\end{subequations}
respectively.
\par
Since the proton numbers of both nuclei are the same ($ Z = 20 $),
the difference of the proton radii, and thus the charge radii, is due to the proton-neutron interaction.
As shown in Eqs.~\eqref{eq:2body_CSB} and \eqref{eq:2body_CIB},
the attractive CIB interaction exists between protons and neutrons,
but
the CSB interaction does not.
Thus, one may think that the CIB strength $ u_0 $ influences the difference more than the CSB one.
Nonetheless, the figure shows a puzzling behavior;
the CSB strength affects the difference, while the CIB strength scarcely does.
\par
To understand such a behavior, let us consider the nuclear EoS.
As shown in Sec.~\ref{sec:matter},
the CSB and CIB interactions, respectively, give $ \beta $ and $ \beta^2 $ dependences on nuclear EoS
[see Eq.~\eqref{eq:EoS}].
In the case of $ \nuc{Ca}{48}{} $, the isospin asymmetry is approximately
$ \beta \approx \left( 28 - 20 \right) / 48 = 0.17 $.
Because $ \epsilon_1 $ is negative, whereas $ \epsilon_2^{\urm{IS}} \simeq J $ and $ 3 u_0 \rho / 16 $ are positive,
the slope of the CIB dependence is opposite to that of the CSB dependence.
Since $ 0 < \beta < 1 $ and
$ \left| \epsilon_1 \right| $ is almost the same value of the CIB contribution to $ \epsilon_2^{\urm{CIB}} $, 
the CSB contribution, which is the leading order with respect to $ \beta $, gives larger contribution.
\par
It should also be noted that different models of the Coulomb interaction give almost the same $ R_{\urm{ch}}^{\urm{Ca-48}} - R_{\urm{ch}}^{\urm{Ca-40}} $,
since both nuclei have the same proton numbers.
\subsection{Mass difference of mirror nuclei}
\label{sec:calc_mirror_mass}
\par
In this section, we discuss the model dependence of the mass difference of mirror nuclei.
The $ \nuc{Ca}{48}{} $-$ \nuc{Ni}{48}{} $ pair is selected as an example, 
$ \Delta E_{\urm{tot}} = E_{\urm{tot}}^{\urm{Ca-48}} - E_{\urm{tot}}^{\urm{Ni-48}} $.
\par
\begin{figure}[tb]
  \centering
  \includegraphics[width=1.0\linewidth]{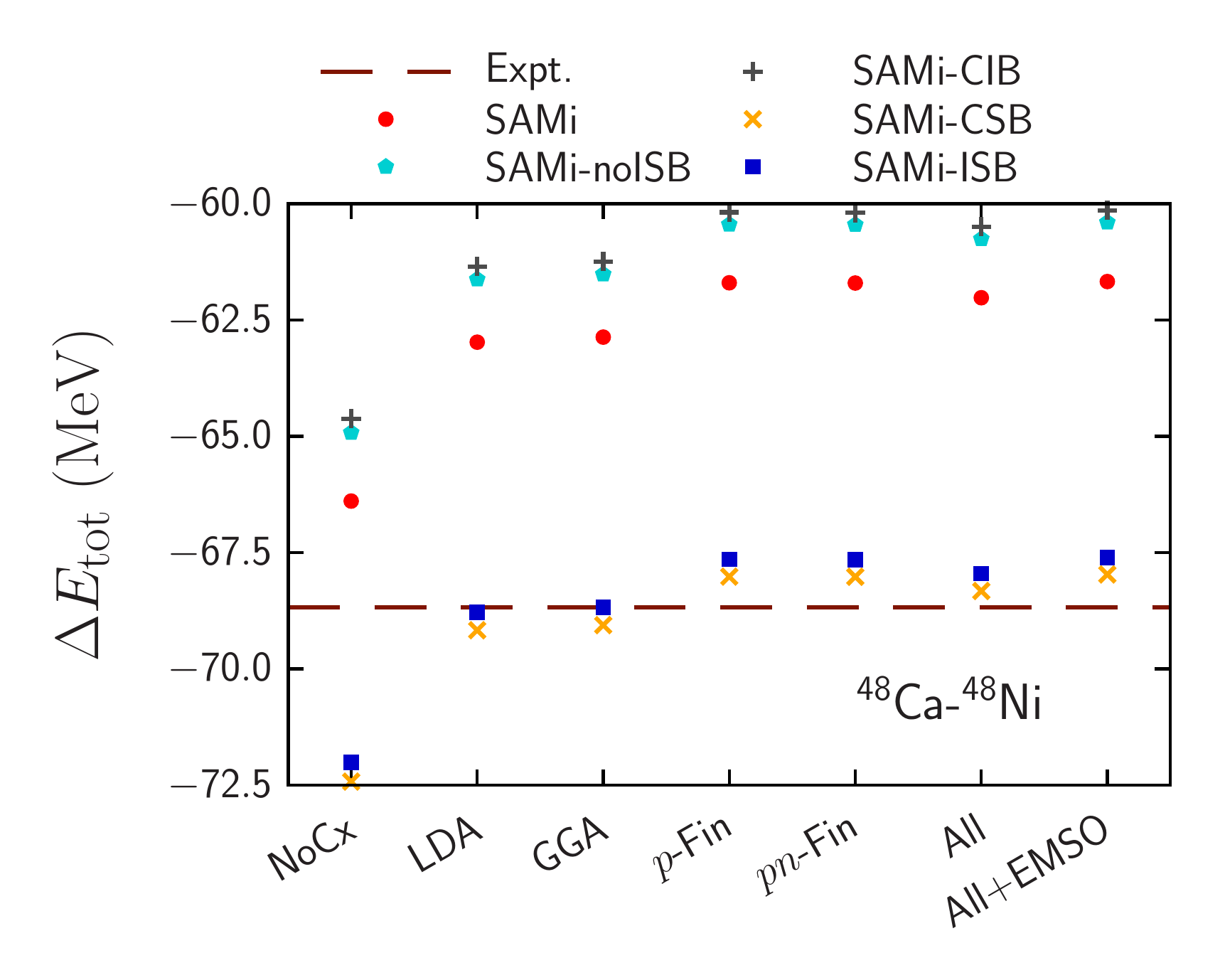}
  \caption{
    The mass difference of mirror nuclei $ \nuc{Ca}{48}{} $-$ \nuc{Ni}{48}{} $ 
    calculated with several Coulomb treatment with
    SAMi (circle),
    SAMi-noISB (pentagon),
    SAMi-CIB (plus),
    SAMi-CSB (cross),
    and SAMi-ISB (square) EDFs.
    Experimental data is taken from AME2020~\cite{
      Huang2021Chin.Phys.C45_030002}.}
  \label{fig:Coulvsmass_048}
\end{figure}

First, the dependence on the treatment of Coulomb interaction is shown in Fig.~\ref{fig:Coulvsmass_048}.
The difference of two estimated binding energies
($ 416.00120 \, \mathrm{MeV} $ for $ \nuc{Ca}{48}{} $ and $ 347.3 \, \mathrm{MeV} $ for $ \nuc{Ni}{48}{} $)
taken from AME2020~\cite{
  Huang2021Chin.Phys.C45_030002}
is shown as a horizontal line~\footnote{
  The binding energy of $ \nuc{Ni}{48}{} $ has not been measured yet,
  and this value is estimated value.
}.
\par
As discussed in Ref.~\cite{
  Naito2020Phys.Rev.C101_064311},
the SAMi-ISB EDF
reproduces the experimental value given by the AME2020.
The model dependence of Coulomb energy is about $ 1.2 \, \mathrm{MeV} $ and does not change this conclusion much.
Next, we focus on the comparison between the ISB and Coulomb effects.
First, effect of the precise treatment of Coulomb interaction is almost the same among all the tested $ E_{\urm{nucl}} $.
Comparing results by SAMi-noISB and SAMi-ISB, the ISB effects to $ \Delta E_{\urm{tot}} $ are around $ 7.2 \, \mathrm{MeV} $.
This is basically the effect of the CSB term,
whereas the effect of the CIB term on less than $ 0.3 \, \mathrm{MeV} $.
The difference between $ \Delta E_{\urm{tot}} $ calculated by using
SAMi and that by using SAMi-ISB
is 
around $ 5.8 \, \mathrm{MeV} $.
\par
Moreover, although the effects of the exchange term (or more precise treatment) of the Coulomb interaction are partially canceled with that of the ISB terms,
the difference between $ \Delta E_{\urm{tot}} $ calculated
with the SAMi EDF without Coulomb exchange term
and
that 
with the SAMi-ISB EDF with the Coulomb exchange (or all the Coulomb) term
is still non-negligibly different.
Thus, the non-trivial cancellation claimed in Refs.~\cite{
  Brown1998Phys.Rev.C58_220,
  Brown2000Phys.Lett.B483_49,
  Goriely2008Phys.Rev.C77_031301}
is not perfect.
\par
\begin{figure}[tb]
  \centering
  \includegraphics[width=1.0\linewidth]{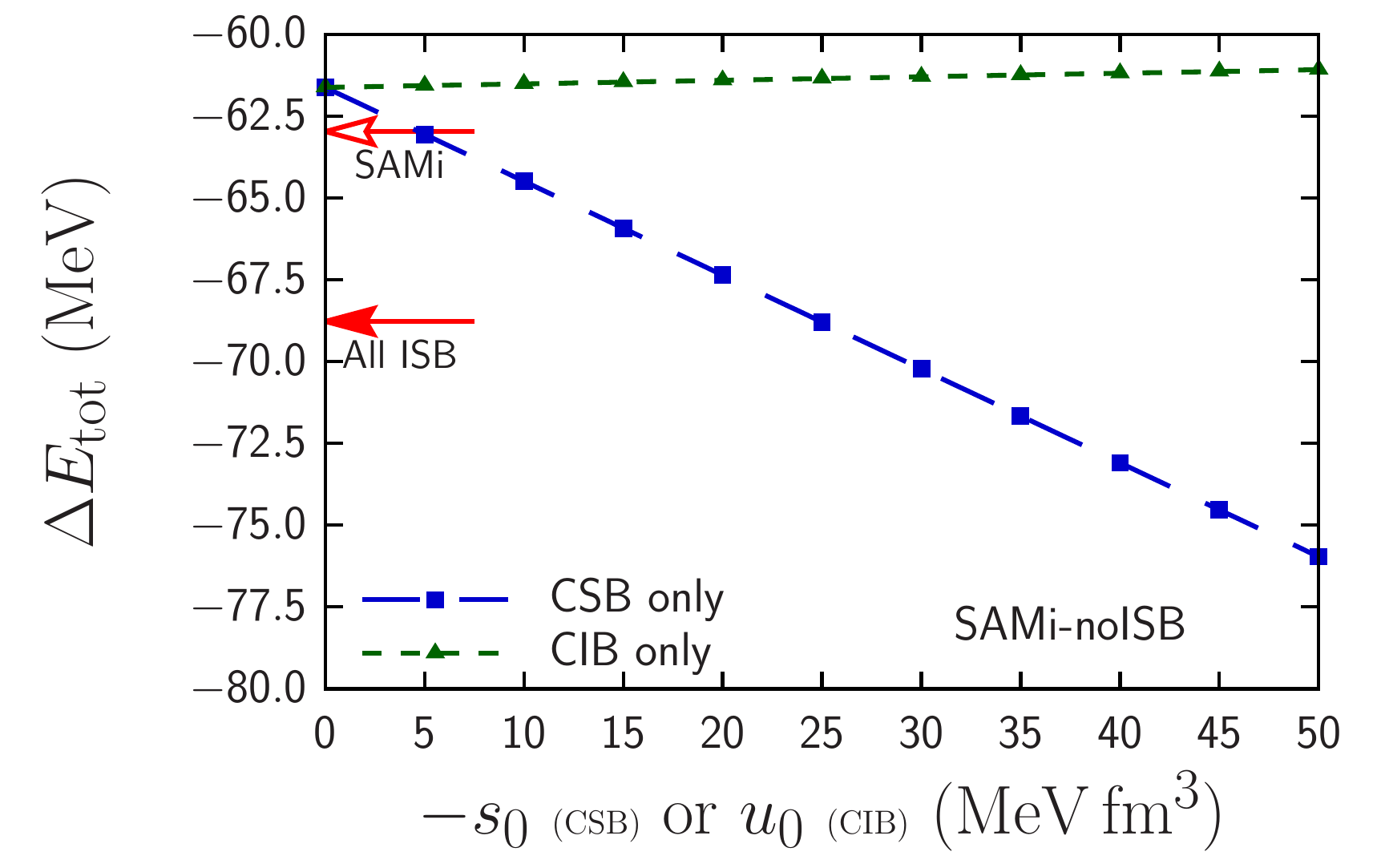}
  \caption{
    The mass difference of mirror nuclei $ \nuc{Ca}{48}{} $-$ \nuc{Ni}{48}{} $ 
    as functions of CSB (square) and CIB (up-triangle) strength $ - s_0 $ and $ u_0 $.
    Filled and empty arrows show $ \Delta R_{np} $ calculated by the full SAMi-ISB
    and the SAMi EDFs, respectively.}
  \label{fig:ISBvsmass_048}
\end{figure}

We discuss the effects of the CSB and CIB strength on $ \Delta E_{\urm{tot}} $.
As we did in Sec.~\ref{sec:calc_ISBvsR},
on top of the SAMi-noISB EDF,
the CSB or CIB strength, $ -s_0 $ or $ u_0 $, is gradually changed from $ 0 \, \mathrm{MeV} \, \mathrm{fm}^3 $ to $ 50 \, \mathrm{MeV} \, \mathrm{fm}^3 $.
The CSB- or CIB-strength dependence of $ \Delta E_{\urm{tot}} $ is shown in Fig.~\ref{fig:ISBvsmass_048}
in the squares and up-triangles, respectively.
These data are fitted to
\begin{subequations}
  \begin{align}
    \Delta E_{\urm{tot}}
    & =
      -61.625 - 0.28677 \left( - s_0 \right), \\
    \Delta E_{\urm{tot}}
    & =
      -61.625 + 0.01083 u_0, 
  \end{align}
\end{subequations}
respectively.
As seen in the figure, the mass difference of mirror nuclei $ \Delta E_{\urm{tot}} $ is sensitive to the CSB strength $ s_0 $.
In contrast, it is not sensitive to the CIB strength $ u_0 $,
although the absolute value of the total energy is changed.
This mechanism is discussed in details in Ref.~\cite{
  Naito2022Phys.Rev.C105_L021304}.
%
%
\section{Conclusion}
\label{sec:conclusion}
\par
In the previous Letter~\cite{
  Naito:2022hyb},
we had discussed the effects of the Coulomb and the isospin symmetry breaking (ISB) terms of nuclear interactions
on the charge-radii difference of mirror nuclei pair $ \nuc{Ca}{48}{} $ and $ \nuc{Ni}{48}{} $, $ \Delta R_{\urm{ch}} $.
We had found that the ISB terms of the nuclear interaction affect the estimation of the density dependence of the symmetry energy, $ L + L_{\urm{CSB}} + L_{\urm{CIB}} $, by about $ 6 $--$ 14 \, \mathrm{MeV} $,
using the correlation with $ \Delta R_{\urm{ch}} $.
In this paper,
we performed similar analyses,
i.e., the sensitivity checks of the model dependence of the Coulomb energy and the ISB nuclear interaction
to other properties related to the isospin symmetry breaking: 
the neutron-skin thickness $ \Delta R_{np} $ of several doubly-magic nuclei,
the difference of the charge radii between $ \nuc{Ca}{40}{} $ and $ \nuc{Ca}{48}{} $,
and the mass difference of mirror nuclei between $ \nuc{Ca}{48}{} $ and $ \nuc{Ni}{48}{} $.
\par
It is found that even if we treat the Coulomb interaction precisely,
its effect on $ \Delta R_{np} $ is less than $ 0.01 \, \mathrm{fm} $ in respect to $ L $.
This difference does not affect the extraction of $ L $ from the experimental $ \Delta R_{np} $.
\par
The ISB terms of the nuclear interaction is divided into two parts:
the charge-symmetry breaking (CSB) and the charge-independence breaking (CIB) ones.
The CSB interaction has a clear manifestation in the neutron-skin thickness and
the mass differences of mirror nuclei.
In contrast, the CIB interaction does not have a significant effect on neither the neutron-skin thickness of $ N = Z $ nuclei nor the mass difference of mirror nuclei.
The neutron-skin thickness of $ N \ne Z $ nuclei is affected by the CIB strength,
while its effect is small compared to the CSB and the Coulomb interaction in some cases.
Therefore, if one attempts to discuss effects of the CIB interaction on nuclear properties,
the Coulomb interaction must be treated precisely as well.
\par
The difference of the calculated charge radii of $ \nuc{Ca}{40}{} $ and $ \nuc{Ca}{48}{} $ is claimed to be related to the symmetry energy~\cite{
  Reinhard2017Phys.Rev.C95_064328,
  Perera2021Phys.Rev.C104_064313}.
In this paper, it was confirmed that the strength of the CSB interaction is correlated to such difference as well,
while the CIB one does not much.
\par
Among all the physical observables we tested, effects of the CIB interaction are smaller than those of the CSB one,
or even invisible on all the observables we discussed.
\par
Since both the CSB and the CIB interactions affect $ \Delta R_{np} $ appreciably,
the estimation of $ L + L_{\urm{CSB}} + L_{\urm{CIB}} $ is also affected as well
because of the strong correlation between two quantities.
For instance, if one assumes the CSB and CIB strengths as those used in the SAMi-ISB,
their effect to $ L $
estimated from the correlation with $ \Delta R_{np} $ 
is $ 12 \, \mathrm{MeV} $.
Therefore, in order to estimate $ L $ parameter using such experimental observables,
ISB contribution should be considered,
which has not been considered in the previous estimations of $ L $~\cite{
  Vinas2014Eur.Phys.J.A50_27,
  Sotani2022Prog.Theor.Exp.Phys.2022_041D01}.
Note that the CSB effect on $ \Delta R_{np} $ is opposite to the CIB one,
whereas they are coherent in $ \Delta R_{\urm{ch}} $.
Hence, the net ISB effect on $ \Delta R_{np} $
is slightly smaller than that on $ \Delta R_{\urm{ch}} $.
\par
It was claimed that the Coulomb exchange term and the ISB of the nuclear interaction are cancelled each other,
and accordingly the fitting of EDF without the Coulomb exchange term was performed, in some Skyrme EDF, such as the SKX EDF~\cite{
  Brown1998Phys.Rev.C58_220,
  Brown2000Phys.Lett.B483_49}.
However, we found that such treatment leads to non-negligible error for both $ \Delta R_{np} $ and $ \Delta E_{\urm{tot}} $
in comparison with proper inclusion of both the Coulomb exchange and ISB interactions.
\par
The magnitude of the ISB effect discussed above obviously depends on the strengths of CSB and CIB interactions.
Although the ISB terms affect most nuclear properties,
they should be taken into account properly, especially, for several nuclear properties,
for instance, $ \Delta R_{np} $, $ \Delta E_{\urm{tot}} $, and $ \Delta R_{np} $.
Thus, it is important to pin down their strengths precisely.
We had discussed the possibility of determining the CSB strength in comparison with the physical observables before~\cite{
  Naito2022Phys.Rev.C105_L021304},
while we found that phenomenological estimation of the CSB strength based on density functional theory
is $ 5 $--$ 10 $ times larger than the \textit{ab initio} estimation in the previous Letter~\cite{
  Naito:2022hyb}.
In this paper, we found that the CIB interaction is insensitive to $ \Delta R_{np} $, $ \Delta E_{\urm{tot}} $, and $ \Delta R_{\urm{ch}} $.
Hence, it is also important to find measurable quantities sensitive to the CIB strength.
%
%
%
\begin{acknowledgments}
  \par
  T.N.~and H.L.~would like to thank the RIKEN iTHEMS program
  and the RIKEN Pioneering Project: Evolution of Matter in the Universe.
  T.N.~acknowledges
  the JSPS Grant-in-Aid for Research Activity Start-up under Grant No.~22K20372,
  the JSPS Grant-in-Aid for Transformative Research Areas (A) under Grant No.~23H04526,
  the JSPS Grant-in-Aid for Scientific Research (B) under Grant No.~23H01845,
  the JSPS Grant-in-Aid for Scientific Research (C) under Grant No.~23K03426,
  the RIKEN Special Postdoctoral Researchers Program,
  and
  the Science and Technology Hub Collaborative Research Program from RIKEN Cluster for Science, Technology and Innovation Hub (RCSTI).
  G.C.~gratefully acknowledges the support and hospitality of YITP, Kyoto University, where part of this work has been carried out.
  H.L.~acknowledges the JSPS Grant-in-Aid for Early-Career Scientists under Grant No.~18K13549
  and the JSPS Grant-in-Aid for Scientific Research (S) under Grant No.~20H05648.
  H.S.~acknowledges the Grant-in-Aid for Scientific Research (C) under Grant No.~19K03858.
  The numerical calculations were performed on cluster computers at the RIKEN iTHEMS program.
\end{acknowledgments}
\appendix
%
%
\section{Isospin dependence of CIB interaction extracted from one-pion exchange interaction}
\label{sec:form_CIB}
\par
In this appendix, the isospin dependence of CIB interaction is discussed.
The main origin of the CIB interaction is mass difference of
the neutral pion $ \pi^0 $ and charged one $ \pi^{\pm} $.
Here, $ m_{\pi^0} $ and $ m_{\pi^{\pm}} $ denote mass of $ \pi^0 $ and $ \pi^{\pm} $, respectively.
The one-pion exchange potential reads~\cite{
  Yukawa1935Proc.Phys.Math.Soc.Jpn.Third17_48,
  Machleidt1987Phys.Rep.149_1}
\begin{equation}
  \label{eq:OPEP}
  V_{\urm{OPEP}} \left( m_{\pi},\ve{q} \right)
  \sim
  -
  \frac{\left( \ve{\sigma}_1 \cdot \ve{q} \right) \left( \ve{\sigma}_2 \cdot \ve{q} \right)}{m_{\pi}^2 + q^2}
  \ve{\tau}_1 \cdot \ve{\tau}_2 .
\end{equation}
Once the mass difference $ m_{\pi^0} \ne m_{\pi^{\pm}} $ is considered,
\begin{widetext}
  \begin{align}
    V_{\urm{OPEP}} \left( m_{\pi^0},\ve{q} \right)
    -
    V_{\urm{OPEP}} \left( m_{\pi^{\pm}},\ve{q} \right)
    & \sim
      -
      \frac{\left( \ve{\sigma}_1 \cdot \ve{q} \right) \left( \ve{\sigma}_2 \cdot \ve{q} \right)}{m_{\pi^0}^2 + q^2}
      \tau_{z1} \tau_{z2}
      +
      \frac{\left( \ve{\sigma}_1 \cdot \ve{q} \right) \left( \ve{\sigma}_2 \cdot \ve{q} \right)}{m_{\pi^{\pm}}^2 + q^2}
      \frac{\tau_1^{+} \tau_2^{-} + \tau_1^{-} \tau_2^{+}}{2}
      \notag \\
    & = 
      -
      \frac{\left( \ve{\sigma}_1 \cdot \ve{q} \right) \left( \ve{\sigma}_2 \cdot \ve{q} \right)}{m_{\pi^0}^2 + q^2}
      \ve{\tau}_1 \cdot \ve{\tau}_2 
      +
      \frac{\left( \ve{\sigma}_1 \cdot \ve{q} \right) \left( \ve{\sigma}_2 \cdot \ve{q} \right)}{m_{\pi^0}^2 + q^2}
      \frac{\Delta m_{\pi}^2}{m_{\pi^0}^2 - \Delta m_{\pi}^2 + q^2}
      \frac{\tau_1^{+} \tau_2^{-} + \tau_1^{-} \tau_2^{+}}{2}
      \notag \\
    & = 
      -
      \frac{\left( \ve{\sigma}_1 \cdot \ve{q} \right) \left( \ve{\sigma}_2 \cdot \ve{q} \right)}{m_{\pi^0}^2 + q^2}
      \ve{\tau}_1 \cdot \ve{\tau}_2 
      +
      \frac{\left( \ve{\sigma}_1 \cdot \ve{q} \right) \left( \ve{\sigma}_2 \cdot \ve{q} \right)}{m_{\pi^0}^2 + q^2}
      \frac{\Delta m_{\pi}^2}{m_{\pi^0}^2 - \Delta m_{\pi}^2 + q^2}
      \left(
      \ve{\tau}_1 \cdot \ve{\tau}_2 - \tau_{z1} \tau_{z2}
      \right),
      \label{eq:CIB_deriv}
  \end{align}
\end{widetext}
where
$ \tau_j^{\pm} = \tau_{xj} \pm i \tau_{yj} $ ($ j = 1 $, $ 2 $)
and 
$ \Delta m_{\pi}^2 = m_{\pi^0}^2 - m_{\pi^{\pm}}^2 $.
The first term of Eq.~\eqref{eq:CIB_deriv} corresponds to the isospin symmetric nuclear interaction,
while the second term corresponds to the charge independence breaking one.
%
%
\section{Precise values of Skyrme parameters}
\label{sec:skyrme}
\par
The precise values of the parameters of SAMi and SAMi-ISB EDFs are shown
in Tables~\ref{tab:param_SAMi} and \ref{tab:param_SAMiISB}, respectively.
Even though differences of two parameter sets, published and precise, are less than $ 0.1 \, \% $ level at most,
the calculated total energies differ around several hundred $ \mathrm{keV} $ in SAMi
($ -1636.1648 \, \mathrm{MeV} $ by the published parameter set
and
$ -1636.6149 \, \mathrm{MeV} $ by the precise parameter set for $ \nuc{Pb}{208}{} $) 
and
even several $ \mathrm{MeV} $ in SAMi-ISB
($ -1629.2878 \, \mathrm{MeV} $ by the published parameter set
and
$ -1635.6319 \, \mathrm{MeV} $ by the precise parameter set for $ \nuc{Pb}{208}{} $).
Although both parameter sets give similar root-mean-square radii,
the results may differ in the order of $ 0.001 \, \mathrm{fm} $ level
($ 5.5187 \, \mathrm{fm} $ by the published SAMi and 
$ 5.5185 \, \mathrm{fm} $ by the precise SAMi,
while $ 5.5092 \, \mathrm{fm} $ by the published SAMi-ISB
and
$ 5.5071 \, \mathrm{fm} $ by the precise SAMi-ISB for $ \nuc{Pb}{208}{} $).
Thus, one needs to use the ``precise'' parameter sets to achieve enough accuracy for the present aim,
so that we use the ``precise'' parameter sets in this paper.
\begin{table}[tb]
  \centering
  \caption{Parameters of SAMi EDF~\cite{
      Roca-Maza2012Phys.Rev.C86_031306}.
    Columns named ``Published'' and ``Precise''
    show 
    the parameters given in the published paper
    and
    those with precise values, respectively.}
  \label{tab:param_SAMi}
  \begin{ruledtabular}
    \begin{tabular}{ldd}
      & \multicolumn{1}{c}{Published} & \multicolumn{1}{c}{Precise} \\
      \hline
      $ t_0 $ ($ \mathrm{MeV} \, \mathrm{fm}^3 $)              & -1877.75    & -1877.746     \\
      $ t_1 $ ($ \mathrm{MeV} \, \mathrm{fm}^5 $)              &   475.6     &   475.5856    \\
      $ t_2 $ ($ \mathrm{MeV} \, \mathrm{fm}^5 $)              &   -85.2     &   -85.20021   \\
      $ t_3 $ ($ \mathrm{MeV} \, \mathrm{fm}^{3 + 3 \alpha} $) & 10219.6     & 10219.58      \\
      $ x_0 $                                                  &     0.320   &     0.3197176 \\
      $ x_1 $                                                  &    -0.532   &    -0.5319419 \\
      $ x_2 $                                                  &    -0.014   &    -0.0137857 \\
      $ x_3 $                                                  &     0.688   &     0.6883226 \\
      $ W_0 $ ($ \mathrm{MeV} \, \mathrm{fm}^5 $)              &   137       &  137.0603     \\
      $ W'_0 $ ($ \mathrm{MeV} \, \mathrm{fm}^5 $)             &    42       &   42.32571    \\
      $ \alpha $                                               &     0.25614 &    0.2561388  \\
    \end{tabular}
  \end{ruledtabular}
\end{table}
\begin{table}[tb]
  \centering
  \caption{The same as Table~\ref{tab:param_SAMi} but for SAMi-ISB EDF~\cite{
      Roca-Maza2018Phys.Rev.Lett.120_202501}.}
  \label{tab:param_SAMiISB}
  \begin{ruledtabular}
    \begin{tabular}{ldd}
      & \multicolumn{1}{c}{Published} & \multicolumn{1}{c}{Precise} \\ \hline
      $ t_0 $ ($ \mathrm{MeV} \, \mathrm{fm}^3 $)              & -2098.3   & -2098.259  \\
      $ t_1 $ ($ \mathrm{MeV} \, \mathrm{fm}^5 $)              &   394.7   &   394.7479 \\
      $ t_2 $ ($ \mathrm{MeV} \, \mathrm{fm}^5 $)              &  -136.4   &  -136.4254 \\
      $ t_3 $ ($ \mathrm{MeV} \, \mathrm{fm}^{3 + 3 \alpha} $) & 11995     & 11995.53   \\
      $ x_0 $                                                  &     0.242 &     0.2419145 \\
      $ x_1 $                                                  &    -0.17  &    -0.1711566 \\
      $ x_2 $                                                  &    -0.470 &    -0.4702394 \\
      $ x_3 $                                                  &     0.32  &     0.3208390 \\
      $ W_0 $ ($ \mathrm{MeV} \, \mathrm{fm}^5 $)              &   294     &   294.7846 \\
      $ W'_0 $ ($ \mathrm{MeV} \, \mathrm{fm}^5 $)             &  -367     &  -367.3859 \\
      $ \alpha $                                               &     0.223 &     0.2233004 \\
      $ s_0 $ ($ \mathrm{MeV} \, \mathrm{fm}^3 $)              &   -26.3   &   -26.3       \\
      $ u_0 $ ($ \mathrm{MeV} \, \mathrm{fm}^3 $)              &    25.8   &    25.8       \\
      $ y_0 $                                                  &    -1     &    -1         \\
      $ z_0 $                                                  &    -1     &    -1         \\
    \end{tabular}
  \end{ruledtabular}
\end{table}
%
%
%

\begin{thebibliography}{91}%
\makeatletter
\providecommand \@ifxundefined [1]{%
 \@ifx{#1\undefined}
}%
\providecommand \@ifnum [1]{%
 \ifnum #1\expandafter \@firstoftwo
 \else \expandafter \@secondoftwo
 \fi
}%
\providecommand \@ifx [1]{%
 \ifx #1\expandafter \@firstoftwo
 \else \expandafter \@secondoftwo
 \fi
}%
\providecommand \natexlab [1]{#1}%
\providecommand \enquote  [1]{``#1''}%
\providecommand \bibnamefont  [1]{#1}%
\providecommand \bibfnamefont [1]{#1}%
\providecommand \citenamefont [1]{#1}%
\providecommand \href@noop [0]{\@secondoftwo}%
\providecommand \href [0]{\begingroup \@sanitize@url \@href}%
\providecommand \@href[1]{\@@startlink{#1}\@@href}%
\providecommand \@@href[1]{\endgroup#1\@@endlink}%
\providecommand \@sanitize@url [0]{\catcode `\\12\catcode `\$12\catcode
  `\&12\catcode `\#12\catcode `\^12\catcode `\_12\catcode `\%12\relax}%
\providecommand \@@startlink[1]{}%
\providecommand \@@endlink[0]{}%
\providecommand \url  [0]{\begingroup\@sanitize@url \@url }%
\providecommand \@url [1]{\endgroup\@href {#1}{\urlprefix }}%
\providecommand \urlprefix  [0]{URL }%
\providecommand \Eprint [0]{\href }%
\providecommand \doibase [0]{https://doi.org/}%
\providecommand \selectlanguage [0]{\@gobble}%
\providecommand \bibinfo  [0]{\@secondoftwo}%
\providecommand \bibfield  [0]{\@secondoftwo}%
\providecommand \translation [1]{[#1]}%
\providecommand \BibitemOpen [0]{}%
\providecommand \bibitemStop [0]{}%
\providecommand \bibitemNoStop [0]{.\EOS\space}%
\providecommand \EOS [0]{\spacefactor3000\relax}%
\providecommand \BibitemShut  [1]{\csname bibitem#1\endcsname}%
\let\auto@bib@innerbib\@empty
\bibitem [{\citenamefont {Heisenberg}(1932)}]{Heisenberg1932Z.Phys.77_1}%
  \BibitemOpen
  \bibfield  {author} {\bibinfo {author} {\bibfnamefont {W.}~\bibnamefont
  {Heisenberg}},\ }\bibfield  {title} {\bibinfo {title} {{\"{U}ber den Bau der
  Atomkerne. I}},\ }\href {https://doi.org/10.1007/BF01342433} {\bibfield
  {journal} {\bibinfo  {journal} {Z. Phys.}\ }\textbf {\bibinfo {volume}
  {77}},\ \bibinfo {pages} {1} (\bibinfo {year} {1932})}\BibitemShut {NoStop}%
\bibitem [{\citenamefont {Coon}\ and\ \citenamefont
  {Scadron}(1982)}]{Coon1982Phys.Rev.C26_2402}%
  \BibitemOpen
  \bibfield  {author} {\bibinfo {author} {\bibfnamefont {S.~A.}\ \bibnamefont
  {Coon}}\ and\ \bibinfo {author} {\bibfnamefont {M.~D.}\ \bibnamefont
  {Scadron}},\ }\bibfield  {title} {\bibinfo {title} {{Two-pion exchange
  contributions to charge asymmetric and charge dependent nuclear forces}},\
  }\href {https://doi.org/10.1103/PhysRevC.26.2402} {\bibfield  {journal}
  {\bibinfo  {journal} {Phys. Rev. C}\ }\textbf {\bibinfo {volume} {26}},\
  \bibinfo {pages} {2402} (\bibinfo {year} {1982})}\BibitemShut {NoStop}%
\bibitem [{\citenamefont {Okamoto}(1964)}]{Okamoto1964Phys.Lett.11_150}%
  \BibitemOpen
  \bibfield  {author} {\bibinfo {author} {\bibfnamefont {K.}~\bibnamefont
  {Okamoto}},\ }\bibfield  {title} {\bibinfo {title} {{Coulomb energy of $
  \mathrm{He}^3 $ and possible charge asymmetry of nuclear forces}},\ }\href
  {https://doi.org/10.1016/0031-9163(64)90650-X} {\bibfield  {journal}
  {\bibinfo  {journal} {Phys. Lett.}\ }\textbf {\bibinfo {volume} {11}},\
  \bibinfo {pages} {150} (\bibinfo {year} {1964})}\BibitemShut {NoStop}%
\bibitem [{\citenamefont {Nolen}\ and\ \citenamefont
  {Schiffer}(1969)}]{Nolen1969Annu.Rev.Nucl.Sci.19_471}%
  \BibitemOpen
  \bibfield  {author} {\bibinfo {author} {\bibfnamefont {J.~A.}\ \bibnamefont
  {Nolen}, \bibfnamefont {Jr.}}\ and\ \bibinfo {author} {\bibfnamefont {J.~P.}\
  \bibnamefont {Schiffer}},\ }\bibfield  {title} {\bibinfo {title} {{Coulomb
  Energies}},\ }\href {https://doi.org/10.1146/annurev.ns.19.120169.002351}
  {\bibfield  {journal} {\bibinfo  {journal} {Annu. Rev. Nucl. Sci.}\ }\textbf
  {\bibinfo {volume} {19}},\ \bibinfo {pages} {471} (\bibinfo {year}
  {1969})}\BibitemShut {NoStop}%
\bibitem [{\citenamefont {Shlomo}(1978)}]{Shlomo1978Rep.Prog.Phys.41_957}%
  \BibitemOpen
  \bibfield  {author} {\bibinfo {author} {\bibfnamefont {S.}~\bibnamefont
  {Shlomo}},\ }\bibfield  {title} {\bibinfo {title} {{Nuclear Coulomb
  energies}},\ }\href {https://doi.org/10.1088/0034-4885/41/7/001} {\bibfield
  {journal} {\bibinfo  {journal} {Rep. Prog. Phys.}\ }\textbf {\bibinfo
  {volume} {41}},\ \bibinfo {pages} {957} (\bibinfo {year} {1978})}\BibitemShut
  {NoStop}%
\bibitem [{\citenamefont {Hatsuda}\ \emph {et~al.}(1991)\citenamefont
  {Hatsuda}, \citenamefont {H\o{}gaasen},\ and\ \citenamefont
  {Prakash}}]{Hatsuda1991Phys.Rev.Lett.66_2851}%
  \BibitemOpen
  \bibfield  {author} {\bibinfo {author} {\bibfnamefont {T.}~\bibnamefont
  {Hatsuda}}, \bibinfo {author} {\bibfnamefont {H.}~\bibnamefont
  {H\o{}gaasen}},\ and\ \bibinfo {author} {\bibfnamefont {M.}~\bibnamefont
  {Prakash}},\ }\bibfield  {title} {\bibinfo {title} {{QCD sum rules in medium
  and the Okamoto-Nolen-Schiffer anomaly}},\ }\href
  {https://doi.org/10.1103/PhysRevLett.66.2851} {\bibfield  {journal} {\bibinfo
   {journal} {Phys. Rev. Lett.}\ }\textbf {\bibinfo {volume} {66}},\ \bibinfo
  {pages} {2851} (\bibinfo {year} {1991})}\BibitemShut {NoStop}%
\bibitem [{\citenamefont {Sagawa}\ \emph {et~al.}(1995)\citenamefont {Sagawa},
  \citenamefont {Van~Giai},\ and\ \citenamefont
  {Suzuki}}]{Sagawa1995Phys.Lett.B353_7}%
  \BibitemOpen
  \bibfield  {author} {\bibinfo {author} {\bibfnamefont {H.}~\bibnamefont
  {Sagawa}}, \bibinfo {author} {\bibfnamefont {N.}~\bibnamefont {Van~Giai}},\
  and\ \bibinfo {author} {\bibfnamefont {T.}~\bibnamefont {Suzuki}},\
  }\bibfield  {title} {\bibinfo {title} {{Isospin mixing and the sum rule of
  super-allowed fermi $ \beta $ decay}},\ }\href
  {https://doi.org/10.1016/0370-2693(95)00498-A} {\bibfield  {journal}
  {\bibinfo  {journal} {Phys. Lett. B}\ }\textbf {\bibinfo {volume} {353}},\
  \bibinfo {pages} {7} (\bibinfo {year} {1995})}\BibitemShut {NoStop}%
\bibitem [{\citenamefont {Brown}(1998)}]{Brown1998Phys.Rev.C58_220}%
  \BibitemOpen
  \bibfield  {author} {\bibinfo {author} {\bibfnamefont {B.~A.}\ \bibnamefont
  {Brown}},\ }\bibfield  {title} {\bibinfo {title} {{New Skyrme interaction for
  normal and exotic nuclei}},\ }\href {https://doi.org/10.1103/PhysRevC.58.220}
  {\bibfield  {journal} {\bibinfo  {journal} {Phys. Rev. C}\ }\textbf {\bibinfo
  {volume} {58}},\ \bibinfo {pages} {220} (\bibinfo {year} {1998})}\BibitemShut
  {NoStop}%
\bibitem [{\citenamefont {Brown}\ \emph {et~al.}(2000)\citenamefont {Brown},
  \citenamefont {Richter},\ and\ \citenamefont
  {Lindsay}}]{Brown2000Phys.Lett.B483_49}%
  \BibitemOpen
  \bibfield  {author} {\bibinfo {author} {\bibfnamefont {B.~A.}\ \bibnamefont
  {Brown}}, \bibinfo {author} {\bibfnamefont {W.~A.}\ \bibnamefont {Richter}},\
  and\ \bibinfo {author} {\bibfnamefont {R.}~\bibnamefont {Lindsay}},\
  }\bibfield  {title} {\bibinfo {title} {{Displacement energies with the Skyrme
  Hartree-Fock method}},\ }\href
  {https://doi.org/10.1016/S0370-2693(00)00589-X} {\bibfield  {journal}
  {\bibinfo  {journal} {Phys. Lett. B}\ }\textbf {\bibinfo {volume} {483}},\
  \bibinfo {pages} {49} (\bibinfo {year} {2000})}\BibitemShut {NoStop}%
\bibitem [{\citenamefont {Liang}\ \emph {et~al.}(2009)\citenamefont {Liang},
  \citenamefont {Van~Giai},\ and\ \citenamefont
  {Meng}}]{Liang2009Phys.Rev.C79_064316}%
  \BibitemOpen
  \bibfield  {author} {\bibinfo {author} {\bibfnamefont {H.}~\bibnamefont
  {Liang}}, \bibinfo {author} {\bibfnamefont {N.}~\bibnamefont {Van~Giai}},\
  and\ \bibinfo {author} {\bibfnamefont {J.}~\bibnamefont {Meng}},\ }\bibfield
  {title} {\bibinfo {title} {{Isospin corrections for superallowed Fermi $
  \beta $ decay in self-consistent relativistic random-phase approximation
  approaches}},\ }\href {https://doi.org/10.1103/PhysRevC.79.064316} {\bibfield
   {journal} {\bibinfo  {journal} {Phys. Rev. C}\ }\textbf {\bibinfo {volume}
  {79}},\ \bibinfo {pages} {064316} (\bibinfo {year} {2009})}\BibitemShut
  {NoStop}%
\bibitem [{\citenamefont {Kaneko}\ \emph {et~al.}(2010)\citenamefont {Kaneko},
  \citenamefont {Tazaki}, \citenamefont {Mizusaki}, \citenamefont {Sun},
  \citenamefont {Hasegawa},\ and\ \citenamefont
  {de~Angelis}}]{Kaneko2010Phys.Rev.C82_061301}%
  \BibitemOpen
  \bibfield  {author} {\bibinfo {author} {\bibfnamefont {K.}~\bibnamefont
  {Kaneko}}, \bibinfo {author} {\bibfnamefont {S.}~\bibnamefont {Tazaki}},
  \bibinfo {author} {\bibfnamefont {T.}~\bibnamefont {Mizusaki}}, \bibinfo
  {author} {\bibfnamefont {Y.}~\bibnamefont {Sun}}, \bibinfo {author}
  {\bibfnamefont {M.}~\bibnamefont {Hasegawa}},\ and\ \bibinfo {author}
  {\bibfnamefont {G.}~\bibnamefont {de~Angelis}},\ }\bibfield  {title}
  {\bibinfo {title} {{Isospin symmetry breaking at high spins in the mirror
  pair $^{67}\mathrm{Se}$ and $^{67}\mathrm{As}$}},\ }\href
  {https://doi.org/10.1103/PhysRevC.82.061301} {\bibfield  {journal} {\bibinfo
  {journal} {Phys. Rev. C}\ }\textbf {\bibinfo {volume} {82}},\ \bibinfo
  {pages} {061301} (\bibinfo {year} {2010})}\BibitemShut {NoStop}%
\bibitem [{\citenamefont {Kaneko}\ \emph {et~al.}(2012)\citenamefont {Kaneko},
  \citenamefont {Mizusaki}, \citenamefont {Sun}, \citenamefont {Tazaki},\ and\
  \citenamefont {de~Angelis}}]{Kaneko2012Phys.Rev.Lett.109_092504}%
  \BibitemOpen
  \bibfield  {author} {\bibinfo {author} {\bibfnamefont {K.}~\bibnamefont
  {Kaneko}}, \bibinfo {author} {\bibfnamefont {T.}~\bibnamefont {Mizusaki}},
  \bibinfo {author} {\bibfnamefont {Y.}~\bibnamefont {Sun}}, \bibinfo {author}
  {\bibfnamefont {S.}~\bibnamefont {Tazaki}},\ and\ \bibinfo {author}
  {\bibfnamefont {G.}~\bibnamefont {de~Angelis}},\ }\bibfield  {title}
  {\bibinfo {title} {{Coulomb Energy Difference as a Probe of Isospin-Symmetry
  Breaking in the Upper $fp$-Shell Nuclei}},\ }\href
  {https://doi.org/10.1103/PhysRevLett.109.092504} {\bibfield  {journal}
  {\bibinfo  {journal} {Phys. Rev. Lett.}\ }\textbf {\bibinfo {volume} {109}},\
  \bibinfo {pages} {092504} (\bibinfo {year} {2012})}\BibitemShut {NoStop}%
\bibitem [{\citenamefont {Satu\l{}a}\ \emph {et~al.}(2012)\citenamefont
  {Satu\l{}a}, \citenamefont {Dobaczewski}, \citenamefont {Nazarewicz},\ and\
  \citenamefont {Werner}}]{Satula2012Phys.Rev.C86_054316}%
  \BibitemOpen
  \bibfield  {author} {\bibinfo {author} {\bibfnamefont {W.}~\bibnamefont
  {Satu\l{}a}}, \bibinfo {author} {\bibfnamefont {J.}~\bibnamefont
  {Dobaczewski}}, \bibinfo {author} {\bibfnamefont {W.}~\bibnamefont
  {Nazarewicz}},\ and\ \bibinfo {author} {\bibfnamefont {T.~R.}\ \bibnamefont
  {Werner}},\ }\bibfield  {title} {\bibinfo {title} {{Isospin-breaking
  corrections to superallowed Fermi $ \beta $ decay in isospin- and
  angular-momentum-projected nuclear density functional theory}},\ }\href
  {https://doi.org/10.1103/PhysRevC.86.054316} {\bibfield  {journal} {\bibinfo
  {journal} {Phys. Rev. C}\ }\textbf {\bibinfo {volume} {86}},\ \bibinfo
  {pages} {054316} (\bibinfo {year} {2012})}\BibitemShut {NoStop}%
\bibitem [{\citenamefont {Wiringa}\ \emph {et~al.}(2013)\citenamefont
  {Wiringa}, \citenamefont {Pastore}, \citenamefont {Pieper},\ and\
  \citenamefont {Miller}}]{Wiringa2013Phys.Rev.C88_044333}%
  \BibitemOpen
  \bibfield  {author} {\bibinfo {author} {\bibfnamefont {R.~B.}\ \bibnamefont
  {Wiringa}}, \bibinfo {author} {\bibfnamefont {S.}~\bibnamefont {Pastore}},
  \bibinfo {author} {\bibfnamefont {S.~C.}\ \bibnamefont {Pieper}},\ and\
  \bibinfo {author} {\bibfnamefont {G.~A.}\ \bibnamefont {Miller}},\ }\bibfield
   {title} {\bibinfo {title} {{Charge-symmetry breaking forces and isospin
  mixing in ${}^{8} \mathrm{Be} $}},\ }\href
  {https://doi.org/10.1103/PhysRevC.88.044333} {\bibfield  {journal} {\bibinfo
  {journal} {Phys. Rev. C}\ }\textbf {\bibinfo {volume} {88}},\ \bibinfo
  {pages} {044333} (\bibinfo {year} {2013})}\BibitemShut {NoStop}%
\bibitem [{\citenamefont {Kaneko}\ \emph {et~al.}(2013)\citenamefont {Kaneko},
  \citenamefont {Sun}, \citenamefont {Mizusaki},\ and\ \citenamefont
  {Tazaki}}]{Kaneko2013Phys.Rev.Lett.110_172505}%
  \BibitemOpen
  \bibfield  {author} {\bibinfo {author} {\bibfnamefont {K.}~\bibnamefont
  {Kaneko}}, \bibinfo {author} {\bibfnamefont {Y.}~\bibnamefont {Sun}},
  \bibinfo {author} {\bibfnamefont {T.}~\bibnamefont {Mizusaki}},\ and\
  \bibinfo {author} {\bibfnamefont {S.}~\bibnamefont {Tazaki}},\ }\bibfield
  {title} {\bibinfo {title} {{Variation in Displacement Energies Due to
  Isospin-Nonconserving Forces}},\ }\href
  {https://doi.org/10.1103/PhysRevLett.110.172505} {\bibfield  {journal}
  {\bibinfo  {journal} {Phys. Rev. Lett.}\ }\textbf {\bibinfo {volume} {110}},\
  \bibinfo {pages} {172505} (\bibinfo {year} {2013})}\BibitemShut {NoStop}%
\bibitem [{\citenamefont {Kaneko}\ \emph {et~al.}(2014)\citenamefont {Kaneko},
  \citenamefont {Sun}, \citenamefont {Mizusaki},\ and\ \citenamefont
  {Tazaki}}]{Kaneko2014Phys.Rev.C89_031302}%
  \BibitemOpen
  \bibfield  {author} {\bibinfo {author} {\bibfnamefont {K.}~\bibnamefont
  {Kaneko}}, \bibinfo {author} {\bibfnamefont {Y.}~\bibnamefont {Sun}},
  \bibinfo {author} {\bibfnamefont {T.}~\bibnamefont {Mizusaki}},\ and\
  \bibinfo {author} {\bibfnamefont {S.}~\bibnamefont {Tazaki}},\ }\bibfield
  {title} {\bibinfo {title} {{Isospin nonconserving interaction in the $T=1$
  analogue states of the mass-70 region}},\ }\href
  {https://doi.org/10.1103/PhysRevC.89.031302} {\bibfield  {journal} {\bibinfo
  {journal} {Phys. Rev. C}\ }\textbf {\bibinfo {volume} {89}},\ \bibinfo
  {pages} {031302} (\bibinfo {year} {2014})}\BibitemShut {NoStop}%
\bibitem [{\citenamefont {Kaneko}\ \emph {et~al.}(2015)\citenamefont {Kaneko},
  \citenamefont {Sun}, \citenamefont {Mizusaki},\ and\ \citenamefont
  {Tazaki}}]{Kaneko2015Phys.Scr.T166_014011}%
  \BibitemOpen
  \bibfield  {author} {\bibinfo {author} {\bibfnamefont {K.}~\bibnamefont
  {Kaneko}}, \bibinfo {author} {\bibfnamefont {Y.}~\bibnamefont {Sun}},
  \bibinfo {author} {\bibfnamefont {T.}~\bibnamefont {Mizusaki}},\ and\
  \bibinfo {author} {\bibfnamefont {S.}~\bibnamefont {Tazaki}},\ }\bibfield
  {title} {\bibinfo {title} {Probing isospin-symmetry breaking with the modern
  storage rings and other advanced facilities},\ }\href
  {https://doi.org/10.1088/0031-8949/2015/t166/014011} {\bibfield  {journal}
  {\bibinfo  {journal} {Phys. Scr.}\ }\textbf {\bibinfo {volume} {T166}},\
  \bibinfo {pages} {014011} (\bibinfo {year} {2015})}\BibitemShut {NoStop}%
\bibitem [{\citenamefont {Hardy}\ and\ \citenamefont
  {Towner}(2015)}]{Hardy2015Phys.Rev.C91_025501}%
  \BibitemOpen
  \bibfield  {author} {\bibinfo {author} {\bibfnamefont {J.~C.}\ \bibnamefont
  {Hardy}}\ and\ \bibinfo {author} {\bibfnamefont {I.~S.}\ \bibnamefont
  {Towner}},\ }\bibfield  {title} {\bibinfo {title} {{Superallowed $ 0^+
  \rightarrow 0^+ $ nuclear $ \beta $ decays: 2014 critical survey, with
  precise results for $ V_{ud} $ and CKM unitarity}},\ }\href
  {https://doi.org/10.1103/PhysRevC.91.025501} {\bibfield  {journal} {\bibinfo
  {journal} {Phys. Rev. C}\ }\textbf {\bibinfo {volume} {91}},\ \bibinfo
  {pages} {025501} (\bibinfo {year} {2015})}\BibitemShut {NoStop}%
\bibitem [{\citenamefont {Kaneko}\ \emph {et~al.}(2017)\citenamefont {Kaneko},
  \citenamefont {Sun}, \citenamefont {Mizusaki}, \citenamefont {Tazaki},\ and\
  \citenamefont {Ghorui}}]{Kaneko2017Phys.Lett.B773_521}%
  \BibitemOpen
  \bibfield  {author} {\bibinfo {author} {\bibfnamefont {K.}~\bibnamefont
  {Kaneko}}, \bibinfo {author} {\bibfnamefont {Y.}~\bibnamefont {Sun}},
  \bibinfo {author} {\bibfnamefont {T.}~\bibnamefont {Mizusaki}}, \bibinfo
  {author} {\bibfnamefont {S.}~\bibnamefont {Tazaki}},\ and\ \bibinfo {author}
  {\bibfnamefont {S.~K.}\ \bibnamefont {Ghorui}},\ }\bibfield  {title}
  {\bibinfo {title} {{Isospin-symmetry breaking in superallowed Fermi $ \beta
  $-decay due to isospin-nonconserving forces}},\ }\href
  {https://doi.org/10.1016/j.physletb.2017.08.056} {\bibfield  {journal}
  {\bibinfo  {journal} {Phys. Lett. B}\ }\textbf {\bibinfo {volume} {773}},\
  \bibinfo {pages} {521} (\bibinfo {year} {2017})}\BibitemShut {NoStop}%
\bibitem [{\citenamefont {Dong}\ \emph {et~al.}(2018)\citenamefont {Dong},
  \citenamefont {Zhang}, \citenamefont {Zuo}, \citenamefont {Gu}, \citenamefont
  {Wang},\ and\ \citenamefont {Sun}}]{Dong2018Phys.Rev.C97_021301}%
  \BibitemOpen
  \bibfield  {author} {\bibinfo {author} {\bibfnamefont {J.~M.}\ \bibnamefont
  {Dong}}, \bibinfo {author} {\bibfnamefont {Y.~H.}\ \bibnamefont {Zhang}},
  \bibinfo {author} {\bibfnamefont {W.}~\bibnamefont {Zuo}}, \bibinfo {author}
  {\bibfnamefont {J.~Z.}\ \bibnamefont {Gu}}, \bibinfo {author} {\bibfnamefont
  {L.~J.}\ \bibnamefont {Wang}},\ and\ \bibinfo {author} {\bibfnamefont
  {Y.}~\bibnamefont {Sun}},\ }\bibfield  {title} {\bibinfo {title}
  {{Generalized isobaric multiplet mass equation and its application to the
  Nolen-Schiffer anomaly}},\ }\href
  {https://doi.org/10.1103/PhysRevC.97.021301} {\bibfield  {journal} {\bibinfo
  {journal} {Phys. Rev. C}\ }\textbf {\bibinfo {volume} {97}},\ \bibinfo
  {pages} {021301} (\bibinfo {year} {2018})}\BibitemShut {NoStop}%
\bibitem [{\citenamefont {Kaneko}\ \emph {et~al.}(2018)\citenamefont {Kaneko},
  \citenamefont {Sun},\ and\ \citenamefont
  {Mizusaki}}]{Kaneko2018Phys.Rev.C97_054326}%
  \BibitemOpen
  \bibfield  {author} {\bibinfo {author} {\bibfnamefont {K.}~\bibnamefont
  {Kaneko}}, \bibinfo {author} {\bibfnamefont {Y.}~\bibnamefont {Sun}},\ and\
  \bibinfo {author} {\bibfnamefont {T.}~\bibnamefont {Mizusaki}},\ }\bibfield
  {title} {\bibinfo {title} {{Isoscalar neutron-proton pairing and
  SU(4)-symmetry breaking in Gamow-Teller transitions}},\ }\href
  {https://doi.org/10.1103/PhysRevC.97.054326} {\bibfield  {journal} {\bibinfo
  {journal} {Phys. Rev. C}\ }\textbf {\bibinfo {volume} {97}},\ \bibinfo
  {pages} {054326} (\bibinfo {year} {2018})}\BibitemShut {NoStop}%
\bibitem [{\citenamefont {Roca-Maza}\ \emph {et~al.}(2018)\citenamefont
  {Roca-Maza}, \citenamefont {Col\`{o}},\ and\ \citenamefont
  {Sagawa}}]{Roca-Maza2018Phys.Rev.Lett.120_202501}%
  \BibitemOpen
  \bibfield  {author} {\bibinfo {author} {\bibfnamefont {X.}~\bibnamefont
  {Roca-Maza}}, \bibinfo {author} {\bibfnamefont {G.}~\bibnamefont
  {Col\`{o}}},\ and\ \bibinfo {author} {\bibfnamefont {H.}~\bibnamefont
  {Sagawa}},\ }\bibfield  {title} {\bibinfo {title} {{Nuclear Symmetry Energy
  and the Breaking of the Isospin Symmetry: How Do They Reconcile with Each
  Other?}},\ }\href {https://doi.org/10.1103/PhysRevLett.120.202501} {\bibfield
   {journal} {\bibinfo  {journal} {Phys. Rev. Lett.}\ }\textbf {\bibinfo
  {volume} {120}},\ \bibinfo {pages} {202501} (\bibinfo {year}
  {2018})}\BibitemShut {NoStop}%
\bibitem [{\citenamefont {Loc}\ \emph {et~al.}(2019)\citenamefont {Loc},
  \citenamefont {Auerbach},\ and\ \citenamefont
  {Col\`o}}]{Loc2019Phys.Rev.C99_014311}%
  \BibitemOpen
  \bibfield  {author} {\bibinfo {author} {\bibfnamefont {B.~M.}\ \bibnamefont
  {Loc}}, \bibinfo {author} {\bibfnamefont {N.}~\bibnamefont {Auerbach}},\ and\
  \bibinfo {author} {\bibfnamefont {G.}~\bibnamefont {Col\`o}},\ }\bibfield
  {title} {\bibinfo {title} {{Isospin mixing and Coulomb mixing in ground
  states of even-even nuclei}},\ }\href
  {https://doi.org/10.1103/PhysRevC.99.014311} {\bibfield  {journal} {\bibinfo
  {journal} {Phys. Rev. C}\ }\textbf {\bibinfo {volume} {99}},\ \bibinfo
  {pages} {014311} (\bibinfo {year} {2019})}\BibitemShut {NoStop}%
\bibitem [{\citenamefont {Dong}\ \emph {et~al.}(2019)\citenamefont {Dong},
  \citenamefont {Gu}, \citenamefont {Zhang}, \citenamefont {Zuo}, \citenamefont
  {Wang}, \citenamefont {Litvinov},\ and\ \citenamefont
  {Sun}}]{Dong2019Phys.Rev.C99_014319}%
  \BibitemOpen
  \bibfield  {author} {\bibinfo {author} {\bibfnamefont {J.~M.}\ \bibnamefont
  {Dong}}, \bibinfo {author} {\bibfnamefont {J.~Z.}\ \bibnamefont {Gu}},
  \bibinfo {author} {\bibfnamefont {Y.~H.}\ \bibnamefont {Zhang}}, \bibinfo
  {author} {\bibfnamefont {W.}~\bibnamefont {Zuo}}, \bibinfo {author}
  {\bibfnamefont {L.~J.}\ \bibnamefont {Wang}}, \bibinfo {author}
  {\bibfnamefont {Y.~A.}\ \bibnamefont {Litvinov}},\ and\ \bibinfo {author}
  {\bibfnamefont {Y.}~\bibnamefont {Sun}},\ }\bibfield  {title} {\bibinfo
  {title} {{Beyond Wigner's isobaric multiplet mass equation: Effect of
  charge-symmetry-breaking interaction and Coulomb polarization}},\ }\href
  {https://doi.org/10.1103/PhysRevC.99.014319} {\bibfield  {journal} {\bibinfo
  {journal} {Phys. Rev. C}\ }\textbf {\bibinfo {volume} {99}},\ \bibinfo
  {pages} {014319} (\bibinfo {year} {2019})}\BibitemShut {NoStop}%
\bibitem [{\citenamefont {Roca-Maza}\ \emph {et~al.}(2020)\citenamefont
  {Roca-Maza}, \citenamefont {Sagawa},\ and\ \citenamefont
  {Col\`o}}]{Roca-Maza2020Phys.Rev.C101_014320}%
  \BibitemOpen
  \bibfield  {author} {\bibinfo {author} {\bibfnamefont {X.}~\bibnamefont
  {Roca-Maza}}, \bibinfo {author} {\bibfnamefont {H.}~\bibnamefont {Sagawa}},\
  and\ \bibinfo {author} {\bibfnamefont {G.}~\bibnamefont {Col\`o}},\
  }\bibfield  {title} {\bibinfo {title} {{Double charge-exchange phonon
  states}},\ }\href {https://doi.org/10.1103/PhysRevC.101.014320} {\bibfield
  {journal} {\bibinfo  {journal} {Phys. Rev. C}\ }\textbf {\bibinfo {volume}
  {101}},\ \bibinfo {pages} {014320} (\bibinfo {year} {2020})}\BibitemShut
  {NoStop}%
\bibitem [{\citenamefont {Novario}\ \emph {et~al.}(2023)\citenamefont
  {Novario}, \citenamefont {Lonardoni}, \citenamefont {Gandolfi},\ and\
  \citenamefont {Hagen}}]{Novario2023Phys.Rev.Lett.130_032501}%
  \BibitemOpen
  \bibfield  {author} {\bibinfo {author} {\bibfnamefont {S.~J.}\ \bibnamefont
  {Novario}}, \bibinfo {author} {\bibfnamefont {D.}~\bibnamefont {Lonardoni}},
  \bibinfo {author} {\bibfnamefont {S.}~\bibnamefont {Gandolfi}},\ and\
  \bibinfo {author} {\bibfnamefont {G.}~\bibnamefont {Hagen}},\ }\bibfield
  {title} {\bibinfo {title} {{Trends of Neutron Skins and Radii of Mirror
  Nuclei from First Principles}},\ }\href
  {https://doi.org/10.1103/PhysRevLett.130.032501} {\bibfield  {journal}
  {\bibinfo  {journal} {Phys. Rev. Lett.}\ }\textbf {\bibinfo {volume} {130}},\
  \bibinfo {pages} {032501} (\bibinfo {year} {2023})}\BibitemShut {NoStop}%
\bibitem [{\citenamefont {Selva}\ \emph {et~al.}(2021)\citenamefont {Selva},
  \citenamefont {Roca-Maza},\ and\ \citenamefont
  {Col\`{o}}}]{Selva2021Symmetry13_144}%
  \BibitemOpen
  \bibfield  {author} {\bibinfo {author} {\bibfnamefont {G.}~\bibnamefont
  {Selva}}, \bibinfo {author} {\bibfnamefont {X.}~\bibnamefont {Roca-Maza}},\
  and\ \bibinfo {author} {\bibfnamefont {G.}~\bibnamefont {Col\`{o}}},\
  }\bibfield  {title} {\bibinfo {title} {{Isospin Symmetry Breaking Effects on
  the Mass-Radius Relation of a Neutron Star}},\ }\href
  {https://doi.org/10.3390/sym13010144} {\bibfield  {journal} {\bibinfo
  {journal} {Symmetry}\ }\textbf {\bibinfo {volume} {13}},\ \bibinfo {pages}
  {144} (\bibinfo {year} {2021})}\BibitemShut {NoStop}%
\bibitem [{\citenamefont {Auerbach}(1983)}]{Auerbach1983Phys.Rep.98_273}%
  \BibitemOpen
  \bibfield  {author} {\bibinfo {author} {\bibfnamefont {N.}~\bibnamefont
  {Auerbach}},\ }\bibfield  {title} {\bibinfo {title} {{Coulomb effects in
  nuclear structure}},\ }\href {https://doi.org/10.1016/0370-1573(83)90008-X}
  {\bibfield  {journal} {\bibinfo  {journal} {Phys. Rep.}\ }\textbf {\bibinfo
  {volume} {98}},\ \bibinfo {pages} {273} (\bibinfo {year} {1983})}\BibitemShut
  {NoStop}%
\bibitem [{\citenamefont {Naito}\ \emph {et~al.}(2018)\citenamefont {Naito},
  \citenamefont {Akashi},\ and\ \citenamefont
  {Liang}}]{Naito2018Phys.Rev.C97_044319}%
  \BibitemOpen
  \bibfield  {author} {\bibinfo {author} {\bibfnamefont {T.}~\bibnamefont
  {Naito}}, \bibinfo {author} {\bibfnamefont {R.}~\bibnamefont {Akashi}},\ and\
  \bibinfo {author} {\bibfnamefont {H.}~\bibnamefont {Liang}},\ }\bibfield
  {title} {\bibinfo {title} {{Application of a Coulomb energy density
  functional for atomic nuclei: Case studies of local density approximation and
  generalized gradient approximation}},\ }\href
  {https://doi.org/10.1103/PhysRevC.97.044319} {\bibfield  {journal} {\bibinfo
  {journal} {Phys. Rev. C}\ }\textbf {\bibinfo {volume} {97}},\ \bibinfo
  {pages} {044319} (\bibinfo {year} {2018})}\BibitemShut {NoStop}%
\bibitem [{\citenamefont {Naito}\ \emph {et~al.}(2019)\citenamefont {Naito},
  \citenamefont {Roca-Maza}, \citenamefont {Col\`o},\ and\ \citenamefont
  {Liang}}]{Naito2019Phys.Rev.C99_024309}%
  \BibitemOpen
  \bibfield  {author} {\bibinfo {author} {\bibfnamefont {T.}~\bibnamefont
  {Naito}}, \bibinfo {author} {\bibfnamefont {X.}~\bibnamefont {Roca-Maza}},
  \bibinfo {author} {\bibfnamefont {G.}~\bibnamefont {Col\`o}},\ and\ \bibinfo
  {author} {\bibfnamefont {H.}~\bibnamefont {Liang}},\ }\bibfield  {title}
  {\bibinfo {title} {{Coulomb exchange functional with generalized gradient
  approximation for self-consistent Skyrme Hartree-Fock calculations}},\ }\href
  {https://doi.org/10.1103/PhysRevC.99.024309} {\bibfield  {journal} {\bibinfo
  {journal} {Phys. Rev. C}\ }\textbf {\bibinfo {volume} {99}},\ \bibinfo
  {pages} {024309} (\bibinfo {year} {2019})}\BibitemShut {NoStop}%
\bibitem [{\citenamefont {Naito}\ \emph {et~al.}(2020)\citenamefont {Naito},
  \citenamefont {Roca-Maza}, \citenamefont {Col\`o},\ and\ \citenamefont
  {Liang}}]{Naito2020Phys.Rev.C101_064311}%
  \BibitemOpen
  \bibfield  {author} {\bibinfo {author} {\bibfnamefont {T.}~\bibnamefont
  {Naito}}, \bibinfo {author} {\bibfnamefont {X.}~\bibnamefont {Roca-Maza}},
  \bibinfo {author} {\bibfnamefont {G.}~\bibnamefont {Col\`o}},\ and\ \bibinfo
  {author} {\bibfnamefont {H.}~\bibnamefont {Liang}},\ }\bibfield  {title}
  {\bibinfo {title} {{Effects of finite nucleon size, vacuum polarization, and
  electromagnetic spin-orbit interaction on nuclear binding energies and radii
  in spherical nuclei}},\ }\href {https://doi.org/10.1103/PhysRevC.101.064311}
  {\bibfield  {journal} {\bibinfo  {journal} {Phys. Rev. C}\ }\textbf {\bibinfo
  {volume} {101}},\ \bibinfo {pages} {064311} (\bibinfo {year}
  {2020})}\BibitemShut {NoStop}%
\bibitem [{\citenamefont {Vautherin}\ and\ \citenamefont
  {Brink}(1972)}]{Vautherin1972Phys.Rev.C5_626}%
  \BibitemOpen
  \bibfield  {author} {\bibinfo {author} {\bibfnamefont {D.}~\bibnamefont
  {Vautherin}}\ and\ \bibinfo {author} {\bibfnamefont {D.~M.}\ \bibnamefont
  {Brink}},\ }\bibfield  {title} {\bibinfo {title} {{Hartree-Fock Calculations
  with Skyrme's Interaction. I. Spherical Nuclei}},\ }\href
  {https://doi.org/10.1103/PhysRevC.5.626} {\bibfield  {journal} {\bibinfo
  {journal} {Phys. Rev. C}\ }\textbf {\bibinfo {volume} {5}},\ \bibinfo {pages}
  {626} (\bibinfo {year} {1972})}\BibitemShut {NoStop}%
\bibitem [{\citenamefont {B{\k{a}}czyk}\ \emph {et~al.}(2018)\citenamefont
  {B{\k{a}}czyk}, \citenamefont {Dobaczewski}, \citenamefont {Konieczka},
  \citenamefont {Satu\l{}a}, \citenamefont {Nakatsukasa},\ and\ \citenamefont
  {Sato}}]{Baczyk2018Phys.Lett.B778_178}%
  \BibitemOpen
  \bibfield  {author} {\bibinfo {author} {\bibfnamefont {P.}~\bibnamefont
  {B{\k{a}}czyk}}, \bibinfo {author} {\bibfnamefont {J.}~\bibnamefont
  {Dobaczewski}}, \bibinfo {author} {\bibfnamefont {M.}~\bibnamefont
  {Konieczka}}, \bibinfo {author} {\bibfnamefont {W.}~\bibnamefont
  {Satu\l{}a}}, \bibinfo {author} {\bibfnamefont {T.}~\bibnamefont
  {Nakatsukasa}},\ and\ \bibinfo {author} {\bibfnamefont {K.}~\bibnamefont
  {Sato}},\ }\bibfield  {title} {\bibinfo {title} {{Isospin-symmetry breaking
  in masses of $ N \simeq Z $ nuclei}},\ }\href
  {https://doi.org/10.1016/j.physletb.2017.12.068} {\bibfield  {journal}
  {\bibinfo  {journal} {Phys. Lett. B}\ }\textbf {\bibinfo {volume} {778}},\
  \bibinfo {pages} {178} (\bibinfo {year} {2018})}\BibitemShut {NoStop}%
\bibitem [{\citenamefont {B{\k{a}}czyk}\ \emph {et~al.}(2019)\citenamefont
  {B{\k{a}}czyk}, \citenamefont {Satu{\l}a}, \citenamefont {Dobaczewski},\ and\
  \citenamefont {Konieczka}}]{Baczyk2019J.Phys.G46_03LT01}%
  \BibitemOpen
  \bibfield  {author} {\bibinfo {author} {\bibfnamefont {P.}~\bibnamefont
  {B{\k{a}}czyk}}, \bibinfo {author} {\bibfnamefont {W.}~\bibnamefont
  {Satu{\l}a}}, \bibinfo {author} {\bibfnamefont {J.}~\bibnamefont
  {Dobaczewski}},\ and\ \bibinfo {author} {\bibfnamefont {M.}~\bibnamefont
  {Konieczka}},\ }\bibfield  {title} {\bibinfo {title} {{Isobaric multiplet
  mass equation within nuclear density functional theory}},\ }\href
  {https://doi.org/10.1088/1361-6471/aaffe4} {\bibfield  {journal} {\bibinfo
  {journal} {J. Phys. G}\ }\textbf {\bibinfo {volume} {46}},\ \bibinfo {pages}
  {03LT01} (\bibinfo {year} {2019})}\BibitemShut {NoStop}%
\bibitem [{\citenamefont {B\k{a}czyk}\ and\ \citenamefont
  {Satu\l{}a}(2021)}]{Baczyk2021Phys.Rev.C103_054320}%
  \BibitemOpen
  \bibfield  {author} {\bibinfo {author} {\bibfnamefont {P.}~\bibnamefont
  {B\k{a}czyk}}\ and\ \bibinfo {author} {\bibfnamefont {W.}~\bibnamefont
  {Satu\l{}a}},\ }\bibfield  {title} {\bibinfo {title} {{Mirror energy
  differences in $T=1/2\phantom{\rule{4pt}{0ex}}{f}_{7/2}$-shell nuclei within
  isospin-dependent density functional theory}},\ }\href
  {https://doi.org/10.1103/PhysRevC.103.054320} {\bibfield  {journal} {\bibinfo
   {journal} {Phys. Rev. C}\ }\textbf {\bibinfo {volume} {103}},\ \bibinfo
  {pages} {054320} (\bibinfo {year} {2021})}\BibitemShut {NoStop}%
\bibitem [{\citenamefont {Naito}\ \emph
  {et~al.}(2022{\natexlab{a}})\citenamefont {Naito}, \citenamefont {Col\`o},
  \citenamefont {Liang}, \citenamefont {Roca-Maza},\ and\ \citenamefont
  {Sagawa}}]{Naito2022Phys.Rev.C105_L021304}%
  \BibitemOpen
  \bibfield  {author} {\bibinfo {author} {\bibfnamefont {T.}~\bibnamefont
  {Naito}}, \bibinfo {author} {\bibfnamefont {G.}~\bibnamefont {Col\`o}},
  \bibinfo {author} {\bibfnamefont {H.}~\bibnamefont {Liang}}, \bibinfo
  {author} {\bibfnamefont {X.}~\bibnamefont {Roca-Maza}},\ and\ \bibinfo
  {author} {\bibfnamefont {H.}~\bibnamefont {Sagawa}},\ }\bibfield  {title}
  {\bibinfo {title} {{Toward \textit{ab initio} charge symmetry breaking in
  nuclear energy density functionals}},\ }\href
  {https://doi.org/10.1103/PhysRevC.105.L021304} {\bibfield  {journal}
  {\bibinfo  {journal} {Phys. Rev. C}\ }\textbf {\bibinfo {volume} {105}},\
  \bibinfo {pages} {L021304} (\bibinfo {year}
  {2022}{\natexlab{a}})}\BibitemShut {NoStop}%
\bibitem [{\citenamefont {Naito}\ \emph
  {et~al.}(2022{\natexlab{b}})\citenamefont {Naito}, \citenamefont {Roca-Maza},
  \citenamefont {Col\`o}, \citenamefont {Liang},\ and\ \citenamefont
  {Sagawa}}]{Naito:2022hyb}%
  \BibitemOpen
  \bibfield  {author} {\bibinfo {author} {\bibfnamefont {T.}~\bibnamefont
  {Naito}}, \bibinfo {author} {\bibfnamefont {X.}~\bibnamefont {Roca-Maza}},
  \bibinfo {author} {\bibfnamefont {G.}~\bibnamefont {Col\`o}}, \bibinfo
  {author} {\bibfnamefont {H.}~\bibnamefont {Liang}},\ and\ \bibinfo {author}
  {\bibfnamefont {H.}~\bibnamefont {Sagawa}},\ }\bibfield  {title} {\bibinfo
  {title} {{Isospin symmetry breaking in the charge radius difference of mirror
  nuclei}},\ }\href {https://doi.org/10.1103/PhysRevC.106.L061306} {\bibfield
  {journal} {\bibinfo  {journal} {Phys. Rev. C}\ }\textbf {\bibinfo {volume}
  {106}},\ \bibinfo {pages} {L061306} (\bibinfo {year}
  {2022}{\natexlab{b}})}\BibitemShut {NoStop}%
\bibitem [{\citenamefont {Perera}\ \emph {et~al.}(2021)\citenamefont {Perera},
  \citenamefont {Afanasjev},\ and\ \citenamefont
  {Ring}}]{Perera2021Phys.Rev.C104_064313}%
  \BibitemOpen
  \bibfield  {author} {\bibinfo {author} {\bibfnamefont {U.~C.}\ \bibnamefont
  {Perera}}, \bibinfo {author} {\bibfnamefont {A.~V.}\ \bibnamefont
  {Afanasjev}},\ and\ \bibinfo {author} {\bibfnamefont {P.}~\bibnamefont
  {Ring}},\ }\bibfield  {title} {\bibinfo {title} {{Charge radii in covariant
  density functional theory: A global view}},\ }\href
  {https://doi.org/10.1103/PhysRevC.104.064313} {\bibfield  {journal} {\bibinfo
   {journal} {Phys. Rev. C}\ }\textbf {\bibinfo {volume} {104}},\ \bibinfo
  {pages} {064313} (\bibinfo {year} {2021})}\BibitemShut {NoStop}%
\bibitem [{\citenamefont {Myers}\ and\ \citenamefont
  {Swiatecki}(1969)}]{Myers1969Ann.Phys.55_395}%
  \BibitemOpen
  \bibfield  {author} {\bibinfo {author} {\bibfnamefont {W.~D.}\ \bibnamefont
  {Myers}}\ and\ \bibinfo {author} {\bibfnamefont {W.~J.}\ \bibnamefont
  {Swiatecki}},\ }\bibfield  {title} {\bibinfo {title} {{Average nuclear
  properties}},\ }\href {https://doi.org/10.1016/0003-4916(69)90202-4}
  {\bibfield  {journal} {\bibinfo  {journal} {Ann. Phys.}\ }\textbf {\bibinfo
  {volume} {55}},\ \bibinfo {pages} {395} (\bibinfo {year} {1969})}\BibitemShut
  {NoStop}%
\bibitem [{\citenamefont {Roca-Maza}\ \emph {et~al.}(2011)\citenamefont
  {Roca-Maza}, \citenamefont {Centelles}, \citenamefont {Vi\~nas},\ and\
  \citenamefont {Warda}}]{Roca-Maza2011Phys.Rev.Lett.106_252501}%
  \BibitemOpen
  \bibfield  {author} {\bibinfo {author} {\bibfnamefont {X.}~\bibnamefont
  {Roca-Maza}}, \bibinfo {author} {\bibfnamefont {M.}~\bibnamefont
  {Centelles}}, \bibinfo {author} {\bibfnamefont {X.}~\bibnamefont {Vi\~nas}},\
  and\ \bibinfo {author} {\bibfnamefont {M.}~\bibnamefont {Warda}},\ }\bibfield
   {title} {\bibinfo {title} {{Neutron Skin of $^{208}\mathrm{Pb}$, Nuclear
  Symmetry Energy, and the Parity Radius Experiment}},\ }\href
  {https://doi.org/10.1103/PhysRevLett.106.252501} {\bibfield  {journal}
  {\bibinfo  {journal} {Phys. Rev. Lett.}\ }\textbf {\bibinfo {volume} {106}},\
  \bibinfo {pages} {252501} (\bibinfo {year} {2011})}\BibitemShut {NoStop}%
\bibitem [{\citenamefont {Reinhard}\ and\ \citenamefont
  {Nazarewicz}(2022)}]{Reinhard2022Phys.Rev.C105_L021301}%
  \BibitemOpen
  \bibfield  {author} {\bibinfo {author} {\bibfnamefont {P.-G.}\ \bibnamefont
  {Reinhard}}\ and\ \bibinfo {author} {\bibfnamefont {W.}~\bibnamefont
  {Nazarewicz}},\ }\bibfield  {title} {\bibinfo {title} {{Information content
  of the differences in the charge radii of mirror nuclei}},\ }\href
  {https://doi.org/10.1103/PhysRevC.105.L021301} {\bibfield  {journal}
  {\bibinfo  {journal} {Phys. Rev. C}\ }\textbf {\bibinfo {volume} {105}},\
  \bibinfo {pages} {L021301} (\bibinfo {year} {2022})}\BibitemShut {NoStop}%
\bibitem [{\citenamefont {Skyrme}(1958)}]{Skyrme1958Nucl.Phys.9_615}%
  \BibitemOpen
  \bibfield  {author} {\bibinfo {author} {\bibfnamefont {T.~H.~R.}\
  \bibnamefont {Skyrme}},\ }\bibfield  {title} {\bibinfo {title} {{The
  effective nuclear potential}},\ }\href
  {https://doi.org/10.1016/0029-5582(58)90345-6} {\bibfield  {journal}
  {\bibinfo  {journal} {Nucl. Phys.}\ }\textbf {\bibinfo {volume} {9}},\
  \bibinfo {pages} {615} (\bibinfo {year} {1958})}\BibitemShut {NoStop}%
\bibitem [{\citenamefont {Roca-Maza}\ and\ \citenamefont
  {Paar}(2018)}]{Roca-Maza2018Prog.Part.Nucl.Phys.101_96}%
  \BibitemOpen
  \bibfield  {author} {\bibinfo {author} {\bibfnamefont {X.}~\bibnamefont
  {Roca-Maza}}\ and\ \bibinfo {author} {\bibfnamefont {N.}~\bibnamefont
  {Paar}},\ }\bibfield  {title} {\bibinfo {title} {{Nuclear equation of state
  from ground and collective excited state properties of nuclei}},\ }\href
  {https://doi.org/10.1016/j.ppnp.2018.04.001} {\bibfield  {journal} {\bibinfo
  {journal} {Prog. Part. Nucl. Phys.}\ }\textbf {\bibinfo {volume} {101}},\
  \bibinfo {pages} {96} (\bibinfo {year} {2018})}\BibitemShut {NoStop}%
\bibitem [{\citenamefont {Miller}\ \emph {et~al.}(2006)\citenamefont {Miller},
  \citenamefont {Opper},\ and\ \citenamefont
  {Stephenson}}]{Miller2006Annu.Rev.Nucl.Part.Sci.56_253}%
  \BibitemOpen
  \bibfield  {author} {\bibinfo {author} {\bibfnamefont {G.~A.}\ \bibnamefont
  {Miller}}, \bibinfo {author} {\bibfnamefont {A.~K.}\ \bibnamefont {Opper}},\
  and\ \bibinfo {author} {\bibfnamefont {E.~J.}\ \bibnamefont {Stephenson}},\
  }\bibfield  {title} {\bibinfo {title} {{Charge Symmetry Breaking and QCD}},\
  }\href {https://doi.org/10.1146/annurev.nucl.56.080805.140446} {\bibfield
  {journal} {\bibinfo  {journal} {Annu. Rev. Nucl. Part. Sci.}\ }\textbf
  {\bibinfo {volume} {56}},\ \bibinfo {pages} {253} (\bibinfo {year}
  {2006})}\BibitemShut {NoStop}%
\bibitem [{\citenamefont {Stoitsov}\ \emph {et~al.}(2005)\citenamefont
  {Stoitsov}, \citenamefont {Dobaczewski}, \citenamefont {Nazarewicz},\ and\
  \citenamefont {Ring}}]{Stoitsov2005Comput.Phys.Commun.167_43}%
  \BibitemOpen
  \bibfield  {author} {\bibinfo {author} {\bibfnamefont {M.~V.}\ \bibnamefont
  {Stoitsov}}, \bibinfo {author} {\bibfnamefont {J.}~\bibnamefont
  {Dobaczewski}}, \bibinfo {author} {\bibfnamefont {W.}~\bibnamefont
  {Nazarewicz}},\ and\ \bibinfo {author} {\bibfnamefont {P.}~\bibnamefont
  {Ring}},\ }\bibfield  {title} {\bibinfo {title} {{Axially deformed solution
  of the Skyrme-Hartree-Fock-Bogolyubov equations using the transformed
  harmonic oscillator basis. The program HFBTHO (v1.66p)}},\ }\href
  {https://doi.org/10.1016/j.cpc.2005.01.001} {\bibfield  {journal} {\bibinfo
  {journal} {Comput. Phys. Commun.}\ }\textbf {\bibinfo {volume} {167}},\
  \bibinfo {pages} {43} (\bibinfo {year} {2005})}\BibitemShut {NoStop}%
\bibitem [{\citenamefont {Sato}\ \emph {et~al.}(2013)\citenamefont {Sato},
  \citenamefont {Dobaczewski}, \citenamefont {Nakatsukasa},\ and\ \citenamefont
  {Satu\l{}a}}]{Sato2013Phys.Rev.C88_061301}%
  \BibitemOpen
  \bibfield  {author} {\bibinfo {author} {\bibfnamefont {K.}~\bibnamefont
  {Sato}}, \bibinfo {author} {\bibfnamefont {J.}~\bibnamefont {Dobaczewski}},
  \bibinfo {author} {\bibfnamefont {T.}~\bibnamefont {Nakatsukasa}},\ and\
  \bibinfo {author} {\bibfnamefont {W.}~\bibnamefont {Satu\l{}a}},\ }\bibfield
  {title} {\bibinfo {title} {{Energy-density-functional calculations including
  proton-neutron mixing}},\ }\href {https://doi.org/10.1103/PhysRevC.88.061301}
  {\bibfield  {journal} {\bibinfo  {journal} {Phys. Rev. C}\ }\textbf {\bibinfo
  {volume} {88}},\ \bibinfo {pages} {061301} (\bibinfo {year}
  {2013})}\BibitemShut {NoStop}%
\bibitem [{\citenamefont {Sheikh}\ \emph {et~al.}(2014)\citenamefont {Sheikh},
  \citenamefont {Hinohara}, \citenamefont {Dobaczewski}, \citenamefont
  {Nakatsukasa}, \citenamefont {Nazarewicz},\ and\ \citenamefont
  {Sato}}]{Sheikh2014Phys.Rev.C89_054317}%
  \BibitemOpen
  \bibfield  {author} {\bibinfo {author} {\bibfnamefont {J.~A.}\ \bibnamefont
  {Sheikh}}, \bibinfo {author} {\bibfnamefont {N.}~\bibnamefont {Hinohara}},
  \bibinfo {author} {\bibfnamefont {J.}~\bibnamefont {Dobaczewski}}, \bibinfo
  {author} {\bibfnamefont {T.}~\bibnamefont {Nakatsukasa}}, \bibinfo {author}
  {\bibfnamefont {W.}~\bibnamefont {Nazarewicz}},\ and\ \bibinfo {author}
  {\bibfnamefont {K.}~\bibnamefont {Sato}},\ }\bibfield  {title} {\bibinfo
  {title} {{Isospin-invariant Skyrme energy-density-functional approach with
  axial symmetry}},\ }\href {https://doi.org/10.1103/PhysRevC.89.054317}
  {\bibfield  {journal} {\bibinfo  {journal} {Phys. Rev. C}\ }\textbf {\bibinfo
  {volume} {89}},\ \bibinfo {pages} {054317} (\bibinfo {year}
  {2014})}\BibitemShut {NoStop}%
\bibitem [{\citenamefont {Dobaczewski}\ \emph {et~al.}(2021)\citenamefont
  {Dobaczewski}, \citenamefont {B\k{a}czyk}, \citenamefont {Becker},
  \citenamefont {Bender}, \citenamefont {Bennaceur}, \citenamefont {Bonnard},
  \citenamefont {Gao}, \citenamefont {Idini}, \citenamefont {Konieczka},
  \citenamefont {Kortelainen}, \citenamefont {Pr\'{o}chniak}, \citenamefont
  {Romero}, \citenamefont {Satu{\l}a}, \citenamefont {Shi}, \citenamefont
  {Werner},\ and\ \citenamefont {Yu}}]{Dobaczewski2021J.Phys.G48_102001}%
  \BibitemOpen
  \bibfield  {author} {\bibinfo {author} {\bibfnamefont {J.}~\bibnamefont
  {Dobaczewski}}, \bibinfo {author} {\bibfnamefont {P.}~\bibnamefont
  {B\k{a}czyk}}, \bibinfo {author} {\bibfnamefont {P.}~\bibnamefont {Becker}},
  \bibinfo {author} {\bibfnamefont {M.}~\bibnamefont {Bender}}, \bibinfo
  {author} {\bibfnamefont {K.}~\bibnamefont {Bennaceur}}, \bibinfo {author}
  {\bibfnamefont {J.}~\bibnamefont {Bonnard}}, \bibinfo {author} {\bibfnamefont
  {Y.}~\bibnamefont {Gao}}, \bibinfo {author} {\bibfnamefont {A.}~\bibnamefont
  {Idini}}, \bibinfo {author} {\bibfnamefont {M.}~\bibnamefont {Konieczka}},
  \bibinfo {author} {\bibfnamefont {M.}~\bibnamefont {Kortelainen}}, \bibinfo
  {author} {\bibfnamefont {L.}~\bibnamefont {Pr\'{o}chniak}}, \bibinfo {author}
  {\bibfnamefont {A.~M.}\ \bibnamefont {Romero}}, \bibinfo {author}
  {\bibfnamefont {W.}~\bibnamefont {Satu{\l}a}}, \bibinfo {author}
  {\bibfnamefont {Y.}~\bibnamefont {Shi}}, \bibinfo {author} {\bibfnamefont
  {T.~R.}\ \bibnamefont {Werner}},\ and\ \bibinfo {author} {\bibfnamefont
  {L.~F.}\ \bibnamefont {Yu}},\ }\bibfield  {title} {\bibinfo {title}
  {{Solution of universal nonrelativistic nuclear DFT equations in the
  Cartesian deformed harmonic-oscillator basis. (IX) HFODD (v3.06h): a new
  version of the program}},\ }\href {https://doi.org/10.1088/1361-6471/ac0a82}
  {\bibfield  {journal} {\bibinfo  {journal} {J. Phys. G}\ }\textbf {\bibinfo
  {volume} {48}},\ \bibinfo {pages} {102001} (\bibinfo {year}
  {2021})}\BibitemShut {NoStop}%
\bibitem [{\citenamefont {Sagawa}\ \emph {et~al.}(2019)\citenamefont {Sagawa},
  \citenamefont {Col{\`o}}, \citenamefont {Roca-Maza},\ and\ \citenamefont
  {Niu}}]{Sagawa2019Eur.Phys.J.A55_227}%
  \BibitemOpen
  \bibfield  {author} {\bibinfo {author} {\bibfnamefont {H.}~\bibnamefont
  {Sagawa}}, \bibinfo {author} {\bibfnamefont {G.}~\bibnamefont {Col{\`o}}},
  \bibinfo {author} {\bibfnamefont {X.}~\bibnamefont {Roca-Maza}},\ and\
  \bibinfo {author} {\bibfnamefont {Y.}~\bibnamefont {Niu}},\ }\bibfield
  {title} {\bibinfo {title} {{Collective excitations involving spin and isospin
  degrees of freedom}},\ }\href {https://doi.org/10.1140/epja/i2019-12923-y}
  {\bibfield  {journal} {\bibinfo  {journal} {Eur. Phys. J. A}\ }\textbf
  {\bibinfo {volume} {55}},\ \bibinfo {pages} {227} (\bibinfo {year}
  {2019})}\BibitemShut {NoStop}%
\bibitem [{\citenamefont {Reinhard}\ \emph {et~al.}(1999)\citenamefont
  {Reinhard}, \citenamefont {Dean}, \citenamefont {Nazarewicz}, \citenamefont
  {Dobaczewski}, \citenamefont {Maruhn},\ and\ \citenamefont
  {Strayer}}]{Reinhard1999Phys.Rev.C60_014316}%
  \BibitemOpen
  \bibfield  {author} {\bibinfo {author} {\bibfnamefont {P.-G.}\ \bibnamefont
  {Reinhard}}, \bibinfo {author} {\bibfnamefont {D.~J.}\ \bibnamefont {Dean}},
  \bibinfo {author} {\bibfnamefont {W.}~\bibnamefont {Nazarewicz}}, \bibinfo
  {author} {\bibfnamefont {J.}~\bibnamefont {Dobaczewski}}, \bibinfo {author}
  {\bibfnamefont {J.~A.}\ \bibnamefont {Maruhn}},\ and\ \bibinfo {author}
  {\bibfnamefont {M.~R.}\ \bibnamefont {Strayer}},\ }\bibfield  {title}
  {\bibinfo {title} {{Shape coexistence and the effective nucleon-nucleon
  interaction}},\ }\href {https://doi.org/10.1103/PhysRevC.60.014316}
  {\bibfield  {journal} {\bibinfo  {journal} {Phys. Rev. C}\ }\textbf {\bibinfo
  {volume} {60}},\ \bibinfo {pages} {014316} (\bibinfo {year}
  {1999})}\BibitemShut {NoStop}%
\bibitem [{\citenamefont {Danielewicz}\ and\ \citenamefont
  {Lee}(2009)}]{Danielewicz2009Nucl.Phys.A818_36}%
  \BibitemOpen
  \bibfield  {author} {\bibinfo {author} {\bibfnamefont {P.}~\bibnamefont
  {Danielewicz}}\ and\ \bibinfo {author} {\bibfnamefont {J.}~\bibnamefont
  {Lee}},\ }\bibfield  {title} {\bibinfo {title} {{Symmetry energy I:
  Semi-infinite matter}},\ }\href
  {https://doi.org/10.1016/j.nuclphysa.2008.11.007} {\bibfield  {journal}
  {\bibinfo  {journal} {Nucl. Phys. A}\ }\textbf {\bibinfo {volume} {818}},\
  \bibinfo {pages} {36} (\bibinfo {year} {2009})}\BibitemShut {NoStop}%
\bibitem [{\citenamefont {Mondal}\ \emph {et~al.}(2018)\citenamefont {Mondal},
  \citenamefont {Agrawal}, \citenamefont {De},\ and\ \citenamefont
  {Samaddar}}]{Mondal2018Int.J.Mod.Phys.E27_1850078}%
  \BibitemOpen
  \bibfield  {author} {\bibinfo {author} {\bibfnamefont {C.}~\bibnamefont
  {Mondal}}, \bibinfo {author} {\bibfnamefont {B.~K.}\ \bibnamefont {Agrawal}},
  \bibinfo {author} {\bibfnamefont {J.~N.}\ \bibnamefont {De}},\ and\ \bibinfo
  {author} {\bibfnamefont {S.~K.}\ \bibnamefont {Samaddar}},\ }\bibfield
  {title} {\bibinfo {title} {{Correlations among symmetry energy elements in
  Skyrme models}},\ }\href {https://doi.org/10.1142/S0218301318500787}
  {\bibfield  {journal} {\bibinfo  {journal} {Int. J. Mod. Phys. E}\ }\textbf
  {\bibinfo {volume} {27}},\ \bibinfo {pages} {1850078} (\bibinfo {year}
  {2018})}\BibitemShut {NoStop}%
\bibitem [{\citenamefont {Hohenberg}\ and\ \citenamefont
  {Kohn}(1964)}]{Hohenberg1964Phys.Rev.136_B864}%
  \BibitemOpen
  \bibfield  {author} {\bibinfo {author} {\bibfnamefont {P.}~\bibnamefont
  {Hohenberg}}\ and\ \bibinfo {author} {\bibfnamefont {W.}~\bibnamefont
  {Kohn}},\ }\bibfield  {title} {\bibinfo {title} {{Inhomogeneous Electron
  Gas}},\ }\href {https://doi.org/10.1103/PhysRev.136.B864} {\bibfield
  {journal} {\bibinfo  {journal} {Phys. Rev.}\ }\textbf {\bibinfo {volume}
  {136}},\ \bibinfo {pages} {B864} (\bibinfo {year} {1964})}\BibitemShut
  {NoStop}%
\bibitem [{\citenamefont {Kohn}\ and\ \citenamefont
  {Sham}(1965)}]{Kohn1965Phys.Rev.140_A1133}%
  \BibitemOpen
  \bibfield  {author} {\bibinfo {author} {\bibfnamefont {W.}~\bibnamefont
  {Kohn}}\ and\ \bibinfo {author} {\bibfnamefont {L.~J.}\ \bibnamefont
  {Sham}},\ }\bibfield  {title} {\bibinfo {title} {{Self-Consistent Equations
  Including Exchange and Correlation Effects}},\ }\href
  {https://doi.org/10.1103/PhysRev.140.A1133} {\bibfield  {journal} {\bibinfo
  {journal} {Phys. Rev.}\ }\textbf {\bibinfo {volume} {140}},\ \bibinfo {pages}
  {A1133} (\bibinfo {year} {1965})}\BibitemShut {NoStop}%
\bibitem [{\citenamefont {Kohn}(1999)}]{Kohn1999Rev.Mod.Phys.71_1253}%
  \BibitemOpen
  \bibfield  {author} {\bibinfo {author} {\bibfnamefont {W.}~\bibnamefont
  {Kohn}},\ }\bibfield  {title} {\bibinfo {title} {{Nobel Lecture: Electronic
  structure of matter---wave functions and density functionals}},\ }\href
  {https://doi.org/10.1103/RevModPhys.71.1253} {\bibfield  {journal} {\bibinfo
  {journal} {Rev. Mod. Phys.}\ }\textbf {\bibinfo {volume} {71}},\ \bibinfo
  {pages} {1253} (\bibinfo {year} {1999})}\BibitemShut {NoStop}%
\bibitem [{\citenamefont {Bender}\ \emph {et~al.}(2003)\citenamefont {Bender},
  \citenamefont {Heenen},\ and\ \citenamefont
  {Reinhard}}]{Bender2003Rev.Mod.Phys.75_121}%
  \BibitemOpen
  \bibfield  {author} {\bibinfo {author} {\bibfnamefont {M.}~\bibnamefont
  {Bender}}, \bibinfo {author} {\bibfnamefont {P.-H.}\ \bibnamefont {Heenen}},\
  and\ \bibinfo {author} {\bibfnamefont {P.-G.}\ \bibnamefont {Reinhard}},\
  }\bibfield  {title} {\bibinfo {title} {{Self-consistent mean-field models for
  nuclear structure}},\ }\href {https://doi.org/10.1103/RevModPhys.75.121}
  {\bibfield  {journal} {\bibinfo  {journal} {Rev. Mod. Phys.}\ }\textbf
  {\bibinfo {volume} {75}},\ \bibinfo {pages} {121} (\bibinfo {year}
  {2003})}\BibitemShut {NoStop}%
\bibitem [{\citenamefont {Meng}\ \emph {et~al.}(2006)\citenamefont {Meng},
  \citenamefont {Toki}, \citenamefont {Zhou}, \citenamefont {Zhang},
  \citenamefont {Long},\ and\ \citenamefont
  {Geng}}]{Meng2006Prog.Part.Nucl.Phys.57_470}%
  \BibitemOpen
  \bibfield  {author} {\bibinfo {author} {\bibfnamefont {J.}~\bibnamefont
  {Meng}}, \bibinfo {author} {\bibfnamefont {H.}~\bibnamefont {Toki}}, \bibinfo
  {author} {\bibfnamefont {S.~G.}\ \bibnamefont {Zhou}}, \bibinfo {author}
  {\bibfnamefont {S.~Q.}\ \bibnamefont {Zhang}}, \bibinfo {author}
  {\bibfnamefont {W.~H.}\ \bibnamefont {Long}},\ and\ \bibinfo {author}
  {\bibfnamefont {L.~S.}\ \bibnamefont {Geng}},\ }\bibfield  {title} {\bibinfo
  {title} {{Relativistic continuum Hartree Bogoliubov theory for ground-state
  properties of exotic nuclei}},\ }\href
  {https://doi.org/10.1016/j.ppnp.2005.06.001} {\bibfield  {journal} {\bibinfo
  {journal} {Prog. Part. Nucl. Phys.}\ }\textbf {\bibinfo {volume} {57}},\
  \bibinfo {pages} {470} (\bibinfo {year} {2006})}\BibitemShut {NoStop}%
\bibitem [{\citenamefont {Liang}\ \emph {et~al.}(2015)\citenamefont {Liang},
  \citenamefont {Meng},\ and\ \citenamefont {Zhou}}]{Liang2015Phys.Rep.570_1}%
  \BibitemOpen
  \bibfield  {author} {\bibinfo {author} {\bibfnamefont {H.}~\bibnamefont
  {Liang}}, \bibinfo {author} {\bibfnamefont {J.}~\bibnamefont {Meng}},\ and\
  \bibinfo {author} {\bibfnamefont {S.-G.}\ \bibnamefont {Zhou}},\ }\bibfield
  {title} {\bibinfo {title} {{Hidden pseudospin and spin symmetries and their
  origins in atomic nuclei}},\ }\href
  {https://doi.org/10.1016/j.physrep.2014.12.005} {\bibfield  {journal}
  {\bibinfo  {journal} {Phys. Rep.}\ }\textbf {\bibinfo {volume} {570}},\
  \bibinfo {pages} {1} (\bibinfo {year} {2015})}\BibitemShut {NoStop}%
\bibitem [{\citenamefont {Roca-Maza}\ \emph {et~al.}(2012)\citenamefont
  {Roca-Maza}, \citenamefont {Col\`o},\ and\ \citenamefont
  {Sagawa}}]{Roca-Maza2012Phys.Rev.C86_031306}%
  \BibitemOpen
  \bibfield  {author} {\bibinfo {author} {\bibfnamefont {X.}~\bibnamefont
  {Roca-Maza}}, \bibinfo {author} {\bibfnamefont {G.}~\bibnamefont {Col\`o}},\
  and\ \bibinfo {author} {\bibfnamefont {H.}~\bibnamefont {Sagawa}},\
  }\bibfield  {title} {\bibinfo {title} {{New Skyrme interaction with improved
  spin-isospin properties}},\ }\href
  {https://doi.org/10.1103/PhysRevC.86.031306} {\bibfield  {journal} {\bibinfo
  {journal} {Phys. Rev. C}\ }\textbf {\bibinfo {volume} {86}},\ \bibinfo
  {pages} {031306} (\bibinfo {year} {2012})}\BibitemShut {NoStop}%
\bibitem [{\citenamefont {Roca-Maza}\ \emph {et~al.}(2013)\citenamefont
  {Roca-Maza}, \citenamefont {Brenna}, \citenamefont {Agrawal}, \citenamefont
  {Bortignon}, \citenamefont {Col\`{o}}, \citenamefont {Cao}, \citenamefont
  {Paar},\ and\ \citenamefont {Vretenar}}]{Roca-Maza2013Phys.Rev.C87_034301}%
  \BibitemOpen
  \bibfield  {author} {\bibinfo {author} {\bibfnamefont {X.}~\bibnamefont
  {Roca-Maza}}, \bibinfo {author} {\bibfnamefont {M.}~\bibnamefont {Brenna}},
  \bibinfo {author} {\bibfnamefont {B.~K.}\ \bibnamefont {Agrawal}}, \bibinfo
  {author} {\bibfnamefont {P.~F.}\ \bibnamefont {Bortignon}}, \bibinfo {author}
  {\bibfnamefont {G.}~\bibnamefont {Col\`{o}}}, \bibinfo {author}
  {\bibfnamefont {L.-G.}\ \bibnamefont {Cao}}, \bibinfo {author} {\bibfnamefont
  {N.}~\bibnamefont {Paar}},\ and\ \bibinfo {author} {\bibfnamefont
  {D.}~\bibnamefont {Vretenar}},\ }\bibfield  {title} {\bibinfo {title} {{Giant
  quadrupole resonances in ${}^{208} \mathrm{Pb} $, the nuclear symmetry
  energy, and the neutron skin thickness}},\ }\href
  {https://doi.org/10.1103/PhysRevC.87.034301} {\bibfield  {journal} {\bibinfo
  {journal} {Phys. Rev. C}\ }\textbf {\bibinfo {volume} {87}},\ \bibinfo
  {pages} {034301} (\bibinfo {year} {2013})}\BibitemShut {NoStop}%
\bibitem [{\citenamefont {Bulgac}\ and\ \citenamefont
  {Shaginyan}(1996)}]{Bulgac1996Nucl.Phys.A601_103}%
  \BibitemOpen
  \bibfield  {author} {\bibinfo {author} {\bibfnamefont {A.}~\bibnamefont
  {Bulgac}}\ and\ \bibinfo {author} {\bibfnamefont {V.~R.}\ \bibnamefont
  {Shaginyan}},\ }\bibfield  {title} {\bibinfo {title} {A systematic surface
  contribution to the ground-state binding energies},\ }\href
  {https://doi.org/10.1016/0375-9474(96)00094-2} {\bibfield  {journal}
  {\bibinfo  {journal} {Nucl. Phys. A}\ }\textbf {\bibinfo {volume} {601}},\
  \bibinfo {pages} {103} (\bibinfo {year} {1996})}\BibitemShut {NoStop}%
\bibitem [{\citenamefont {Bulgac}\ and\ \citenamefont
  {Shaginyan}(1999{\natexlab{a}})}]{Bulgac1999Eur.Phys.J.A5_247}%
  \BibitemOpen
  \bibfield  {author} {\bibinfo {author} {\bibfnamefont {A.}~\bibnamefont
  {Bulgac}}\ and\ \bibinfo {author} {\bibfnamefont {V.~R.}\ \bibnamefont
  {Shaginyan}},\ }\bibfield  {title} {\bibinfo {title} {{Influence of Coulomb
  correlations on the location of drip line, single particle spectra and
  effective mass}},\ }\href {https://doi.org/10.1007/s100500050282} {\bibfield
  {journal} {\bibinfo  {journal} {Eur. Phys. J. A}\ }\textbf {\bibinfo {volume}
  {5}},\ \bibinfo {pages} {247} (\bibinfo {year}
  {1999}{\natexlab{a}})}\BibitemShut {NoStop}%
\bibitem [{\citenamefont {Bulgac}\ and\ \citenamefont
  {Shaginyan}(1999{\natexlab{b}})}]{Bulgac1999Phys.Lett.B469_1}%
  \BibitemOpen
  \bibfield  {author} {\bibinfo {author} {\bibfnamefont {A.}~\bibnamefont
  {Bulgac}}\ and\ \bibinfo {author} {\bibfnamefont {V.~R.}\ \bibnamefont
  {Shaginyan}},\ }\bibfield  {title} {\bibinfo {title} {{Proton single-particle
  energy shifts due to Coulomb correlations}},\ }\href
  {https://doi.org/10.1016/S0370-2693(99)01262-9} {\bibfield  {journal}
  {\bibinfo  {journal} {Phys. Lett. B}\ }\textbf {\bibinfo {volume} {469}},\
  \bibinfo {pages} {1} (\bibinfo {year} {1999}{\natexlab{b}})}\BibitemShut
  {NoStop}%
\bibitem [{\citenamefont {Dirac}(1930)}]{Dirac1930Proc.Camb.Phil.Soc.26_376}%
  \BibitemOpen
  \bibfield  {author} {\bibinfo {author} {\bibfnamefont {P.~A.~M.}\
  \bibnamefont {Dirac}},\ }\bibfield  {title} {\bibinfo {title} {{Note on
  Exchange Phenomena in the Thomas Atom}},\ }\href
  {https://doi.org/10.1017/S0305004100016108} {\bibfield  {journal} {\bibinfo
  {journal} {Proc. Camb. Phil. Soc.}\ }\textbf {\bibinfo {volume} {26}},\
  \bibinfo {pages} {376} (\bibinfo {year} {1930})}\BibitemShut {NoStop}%
\bibitem [{\citenamefont {Slater}(1951)}]{Slater1951Phys.Rev.81_385}%
  \BibitemOpen
  \bibfield  {author} {\bibinfo {author} {\bibfnamefont {J.~C.}\ \bibnamefont
  {Slater}},\ }\bibfield  {title} {\bibinfo {title} {{A Simplification of the
  Hartree-Fock Method}},\ }\href {https://doi.org/10.1103/PhysRev.81.385}
  {\bibfield  {journal} {\bibinfo  {journal} {Phys. Rev.}\ }\textbf {\bibinfo
  {volume} {81}},\ \bibinfo {pages} {385} (\bibinfo {year} {1951})}\BibitemShut
  {NoStop}%
\bibitem [{\citenamefont {Perdew}\ \emph {et~al.}(1996)\citenamefont {Perdew},
  \citenamefont {Burke},\ and\ \citenamefont
  {Ernzerhof}}]{Perdew1996Phys.Rev.Lett.77_3865}%
  \BibitemOpen
  \bibfield  {author} {\bibinfo {author} {\bibfnamefont {J.~P.}\ \bibnamefont
  {Perdew}}, \bibinfo {author} {\bibfnamefont {K.}~\bibnamefont {Burke}},\ and\
  \bibinfo {author} {\bibfnamefont {M.}~\bibnamefont {Ernzerhof}},\ }\bibfield
  {title} {\bibinfo {title} {{Generalized Gradient Approximation Made
  Simple}},\ }\href {https://doi.org/10.1103/PhysRevLett.77.3865} {\bibfield
  {journal} {\bibinfo  {journal} {Phys. Rev. Lett.}\ }\textbf {\bibinfo
  {volume} {77}},\ \bibinfo {pages} {3865} (\bibinfo {year}
  {1996})}\BibitemShut {NoStop}%
\bibitem [{\citenamefont {Friedrich}\ and\ \citenamefont
  {Walcher}(2003)}]{Friedrich2003Eur.Phys.J.A17_607}%
  \BibitemOpen
  \bibfield  {author} {\bibinfo {author} {\bibfnamefont {J.}~\bibnamefont
  {Friedrich}}\ and\ \bibinfo {author} {\bibfnamefont {T.}~\bibnamefont
  {Walcher}},\ }\bibfield  {title} {\bibinfo {title} {{A coherent
  interpretation of the form factors of the nucleon in terms of a pion cloud
  and constituent quarks}},\ }\href
  {https://doi.org/10.1140/epja/i2003-10025-3} {\bibfield  {journal} {\bibinfo
  {journal} {Eur. Phys. J. A}\ }\textbf {\bibinfo {volume} {17}},\ \bibinfo
  {pages} {607} (\bibinfo {year} {2003})}\BibitemShut {NoStop}%
\bibitem [{\citenamefont {Col\`{o}}\ \emph {et~al.}(2013)\citenamefont
  {Col\`{o}}, \citenamefont {Cao}, \citenamefont {Van~Giai},\ and\
  \citenamefont {Capelli}}]{Colo2013Comput.Phys.Commun.184_142}%
  \BibitemOpen
  \bibfield  {author} {\bibinfo {author} {\bibfnamefont {G.}~\bibnamefont
  {Col\`{o}}}, \bibinfo {author} {\bibfnamefont {L.}~\bibnamefont {Cao}},
  \bibinfo {author} {\bibfnamefont {N.}~\bibnamefont {Van~Giai}},\ and\
  \bibinfo {author} {\bibfnamefont {L.}~\bibnamefont {Capelli}},\ }\bibfield
  {title} {\bibinfo {title} {{Self-consistent RPA calculations with Skyrme-type
  interactions: The skyrme\_rpa program}},\ }\href
  {https://doi.org/10.1016/j.cpc.2012.07.016} {\bibfield  {journal} {\bibinfo
  {journal} {Comput. Phys. Commun.}\ }\textbf {\bibinfo {volume} {184}},\
  \bibinfo {pages} {142} (\bibinfo {year} {2013})}\BibitemShut {NoStop}%
\bibitem [{\citenamefont {Brown}(2000)}]{Brown2000Phys.Rev.Lett.85_5296}%
  \BibitemOpen
  \bibfield  {author} {\bibinfo {author} {\bibfnamefont {B.~A.}\ \bibnamefont
  {Brown}},\ }\bibfield  {title} {\bibinfo {title} {{Neutron Radii in Nuclei
  and the Neutron Equation of State}},\ }\href
  {https://doi.org/10.1103/PhysRevLett.85.5296} {\bibfield  {journal} {\bibinfo
   {journal} {Phys. Rev. Lett.}\ }\textbf {\bibinfo {volume} {85}},\ \bibinfo
  {pages} {5296} (\bibinfo {year} {2000})}\BibitemShut {NoStop}%
\bibitem [{\citenamefont {Chen}\ \emph {et~al.}(2005)\citenamefont {Chen},
  \citenamefont {Ko},\ and\ \citenamefont {Li}}]{Chen2005Phys.Rev.C72_064309}%
  \BibitemOpen
  \bibfield  {author} {\bibinfo {author} {\bibfnamefont {L.-W.}\ \bibnamefont
  {Chen}}, \bibinfo {author} {\bibfnamefont {C.~M.}\ \bibnamefont {Ko}},\ and\
  \bibinfo {author} {\bibfnamefont {B.-A.}\ \bibnamefont {Li}},\ }\bibfield
  {title} {\bibinfo {title} {{Nuclear matter symmetry energy and the neutron
  skin thickness of heavy nuclei}},\ }\href
  {https://doi.org/10.1103/PhysRevC.72.064309} {\bibfield  {journal} {\bibinfo
  {journal} {Phys. Rev. C}\ }\textbf {\bibinfo {volume} {72}},\ \bibinfo
  {pages} {064309} (\bibinfo {year} {2005})}\BibitemShut {NoStop}%
\bibitem [{\citenamefont {Zenihiro}\ \emph {et~al.}(2018)\citenamefont
  {Zenihiro}, \citenamefont {Sakaguchi}, \citenamefont {Terashima},
  \citenamefont {Uesaka}, \citenamefont {Hagen}, \citenamefont {Itoh},
  \citenamefont {Murakami}, \citenamefont {Nakatsugawa}, \citenamefont
  {Ohnishi}, \citenamefont {Sagawa}, \citenamefont {Takeda}, \citenamefont
  {Uchida}, \citenamefont {Yoshida}, \citenamefont {Yoshida},\ and\
  \citenamefont {Yosoi}}]{Zenihiro_2018}%
  \BibitemOpen
  \bibfield  {author} {\bibinfo {author} {\bibfnamefont {J.}~\bibnamefont
  {Zenihiro}}, \bibinfo {author} {\bibfnamefont {H.}~\bibnamefont {Sakaguchi}},
  \bibinfo {author} {\bibfnamefont {S.}~\bibnamefont {Terashima}}, \bibinfo
  {author} {\bibfnamefont {T.}~\bibnamefont {Uesaka}}, \bibinfo {author}
  {\bibfnamefont {G.}~\bibnamefont {Hagen}}, \bibinfo {author} {\bibfnamefont
  {M.}~\bibnamefont {Itoh}}, \bibinfo {author} {\bibfnamefont {T.}~\bibnamefont
  {Murakami}}, \bibinfo {author} {\bibfnamefont {Y.}~\bibnamefont
  {Nakatsugawa}}, \bibinfo {author} {\bibfnamefont {T.}~\bibnamefont
  {Ohnishi}}, \bibinfo {author} {\bibfnamefont {H.}~\bibnamefont {Sagawa}},
  \bibinfo {author} {\bibfnamefont {H.}~\bibnamefont {Takeda}}, \bibinfo
  {author} {\bibfnamefont {M.}~\bibnamefont {Uchida}}, \bibinfo {author}
  {\bibfnamefont {H.~P.}\ \bibnamefont {Yoshida}}, \bibinfo {author}
  {\bibfnamefont {S.}~\bibnamefont {Yoshida}},\ and\ \bibinfo {author}
  {\bibfnamefont {M.}~\bibnamefont {Yosoi}},\ }\bibfield  {title} {\bibinfo
  {title} {{Direct determination of the neutron skin thicknesses in
  $^{40,48}$Ca from proton elastic scattering at $E_p = 295$ MeV}},\ }\Eprint
  {https://arxiv.org/abs/1810.11796} {arXiv:1810.11796 [nucl-ex]}  (\bibinfo
  {year} {2018})\BibitemShut {NoStop}%
\bibitem [{\citenamefont {Adhikari}\ \emph {et~al.}(2022)\citenamefont
  {Adhikari}, \citenamefont {Albataineh}, \citenamefont {Androic},
  \citenamefont {Aniol}, \citenamefont {Armstrong}, \citenamefont {Averett},
  \citenamefont {Ayerbe~Gayoso}, \citenamefont {Barcus}, \citenamefont
  {Bellini}, \citenamefont {Beminiwattha}, \citenamefont {Benesch},
  \citenamefont {Bhatt}, \citenamefont {Bhatta~Pathak}, \citenamefont
  {Bhetuwal}, \citenamefont {Blaikie}, \citenamefont {Boyd}, \citenamefont
  {Campagna}, \citenamefont {Camsonne}, \citenamefont {Cates}, \citenamefont
  {Chen}, \citenamefont {Clarke}, \citenamefont {Cornejo}, \citenamefont
  {Covrig~Dusa}, \citenamefont {Dalton}, \citenamefont {Datta}, \citenamefont
  {Deshpande}, \citenamefont {Dutta}, \citenamefont {Feldman}, \citenamefont
  {Fuchey}, \citenamefont {Gal}, \citenamefont {Gaskell}, \citenamefont
  {Gautam}, \citenamefont {Gericke}, \citenamefont {Ghosh}, \citenamefont
  {Halilovic}, \citenamefont {Hansen}, \citenamefont {Hassan}, \citenamefont
  {Hauenstein}, \citenamefont {Henry}, \citenamefont {Horowitz}, \citenamefont
  {Jantzi}, \citenamefont {Jian}, \citenamefont {Johnston}, \citenamefont
  {Jones}, \citenamefont {Kakkar}, \citenamefont {Katugampola}, \citenamefont
  {Keppel}, \citenamefont {King}, \citenamefont {King}, \citenamefont {Kumar},
  \citenamefont {Kutz}, \citenamefont {Lashley-Colthirst}, \citenamefont
  {Leverick}, \citenamefont {Liu}, \citenamefont {Liyanage}, \citenamefont
  {Mammei}, \citenamefont {Mammei}, \citenamefont {McCaughan}, \citenamefont
  {McNulty}, \citenamefont {Meekins}, \citenamefont {Metts}, \citenamefont
  {Michaels}, \citenamefont {Mihovilovic}, \citenamefont {Mondal},
  \citenamefont {Napolitano}, \citenamefont {Narayan}, \citenamefont
  {Nikolaev}, \citenamefont {Owen}, \citenamefont {Palatchi}, \citenamefont
  {Pan}, \citenamefont {Pandey}, \citenamefont {Park}, \citenamefont {Paschke},
  \citenamefont {Petrusky}, \citenamefont {Pitt}, \citenamefont {Premathilake},
  \citenamefont {Quinn}, \citenamefont {Radloff}, \citenamefont {Rahman},
  \citenamefont {Rashad}, \citenamefont {Rathnayake}, \citenamefont {Reed},
  \citenamefont {Reimer}, \citenamefont {Richards}, \citenamefont {Riordan},
  \citenamefont {Roblin}, \citenamefont {Seeds}, \citenamefont {Shahinyan},
  \citenamefont {Souder}, \citenamefont {Thiel}, \citenamefont {Tian},
  \citenamefont {Urciuoli}, \citenamefont {Wertz}, \citenamefont
  {Wojtsekhowski}, \citenamefont {Yale}, \citenamefont {Ye}, \citenamefont
  {Yoon}, \citenamefont {Xiong}, \citenamefont {Zec}, \citenamefont {Zhang},
  \citenamefont {Zhang},\ and\ \citenamefont {Zheng}}]{CREX:2022kgg}%
  \BibitemOpen
  \bibfield  {author} {\bibinfo {author} {\bibfnamefont {D.}~\bibnamefont
  {Adhikari}}, \bibinfo {author} {\bibfnamefont {H.}~\bibnamefont
  {Albataineh}}, \bibinfo {author} {\bibfnamefont {D.}~\bibnamefont {Androic}},
  \bibinfo {author} {\bibfnamefont {K.~A.}\ \bibnamefont {Aniol}}, \bibinfo
  {author} {\bibfnamefont {D.~S.}\ \bibnamefont {Armstrong}}, \bibinfo {author}
  {\bibfnamefont {T.}~\bibnamefont {Averett}}, \bibinfo {author} {\bibfnamefont
  {C.}~\bibnamefont {Ayerbe~Gayoso}}, \bibinfo {author} {\bibfnamefont {S.~K.}\
  \bibnamefont {Barcus}}, \bibinfo {author} {\bibfnamefont {V.}~\bibnamefont
  {Bellini}}, \bibinfo {author} {\bibfnamefont {R.~S.}\ \bibnamefont
  {Beminiwattha}}, \bibinfo {author} {\bibfnamefont {J.~F.}\ \bibnamefont
  {Benesch}}, \bibinfo {author} {\bibfnamefont {H.}~\bibnamefont {Bhatt}},
  \bibinfo {author} {\bibfnamefont {D.}~\bibnamefont {Bhatta~Pathak}}, \bibinfo
  {author} {\bibfnamefont {D.}~\bibnamefont {Bhetuwal}}, \bibinfo {author}
  {\bibfnamefont {B.}~\bibnamefont {Blaikie}}, \bibinfo {author} {\bibfnamefont
  {J.}~\bibnamefont {Boyd}}, \bibinfo {author} {\bibfnamefont {Q.}~\bibnamefont
  {Campagna}}, \bibinfo {author} {\bibfnamefont {A.}~\bibnamefont {Camsonne}},
  \bibinfo {author} {\bibfnamefont {G.~D.}\ \bibnamefont {Cates}}, \bibinfo
  {author} {\bibfnamefont {Y.}~\bibnamefont {Chen}}, \bibinfo {author}
  {\bibfnamefont {C.}~\bibnamefont {Clarke}}, \bibinfo {author} {\bibfnamefont
  {J.~C.}\ \bibnamefont {Cornejo}}, \bibinfo {author} {\bibfnamefont
  {S.}~\bibnamefont {Covrig~Dusa}}, \bibinfo {author} {\bibfnamefont {M.~M.}\
  \bibnamefont {Dalton}}, \bibinfo {author} {\bibfnamefont {P.}~\bibnamefont
  {Datta}}, \bibinfo {author} {\bibfnamefont {A.}~\bibnamefont {Deshpande}},
  \bibinfo {author} {\bibfnamefont {D.}~\bibnamefont {Dutta}}, \bibinfo
  {author} {\bibfnamefont {C.}~\bibnamefont {Feldman}}, \bibinfo {author}
  {\bibfnamefont {E.}~\bibnamefont {Fuchey}}, \bibinfo {author} {\bibfnamefont
  {C.}~\bibnamefont {Gal}}, \bibinfo {author} {\bibfnamefont {D.}~\bibnamefont
  {Gaskell}}, \bibinfo {author} {\bibfnamefont {T.}~\bibnamefont {Gautam}},
  \bibinfo {author} {\bibfnamefont {M.}~\bibnamefont {Gericke}}, \bibinfo
  {author} {\bibfnamefont {C.}~\bibnamefont {Ghosh}}, \bibinfo {author}
  {\bibfnamefont {I.}~\bibnamefont {Halilovic}}, \bibinfo {author}
  {\bibfnamefont {J.-O.}\ \bibnamefont {Hansen}}, \bibinfo {author}
  {\bibfnamefont {O.}~\bibnamefont {Hassan}}, \bibinfo {author} {\bibfnamefont
  {F.}~\bibnamefont {Hauenstein}}, \bibinfo {author} {\bibfnamefont
  {W.}~\bibnamefont {Henry}}, \bibinfo {author} {\bibfnamefont {C.~J.}\
  \bibnamefont {Horowitz}}, \bibinfo {author} {\bibfnamefont {C.}~\bibnamefont
  {Jantzi}}, \bibinfo {author} {\bibfnamefont {S.}~\bibnamefont {Jian}},
  \bibinfo {author} {\bibfnamefont {S.}~\bibnamefont {Johnston}}, \bibinfo
  {author} {\bibfnamefont {D.~C.}\ \bibnamefont {Jones}}, \bibinfo {author}
  {\bibfnamefont {S.}~\bibnamefont {Kakkar}}, \bibinfo {author} {\bibfnamefont
  {S.}~\bibnamefont {Katugampola}}, \bibinfo {author} {\bibfnamefont
  {C.}~\bibnamefont {Keppel}}, \bibinfo {author} {\bibfnamefont {P.~M.}\
  \bibnamefont {King}}, \bibinfo {author} {\bibfnamefont {D.~E.}\ \bibnamefont
  {King}}, \bibinfo {author} {\bibfnamefont {K.~S.}\ \bibnamefont {Kumar}},
  \bibinfo {author} {\bibfnamefont {T.}~\bibnamefont {Kutz}}, \bibinfo {author}
  {\bibfnamefont {N.}~\bibnamefont {Lashley-Colthirst}}, \bibinfo {author}
  {\bibfnamefont {G.}~\bibnamefont {Leverick}}, \bibinfo {author}
  {\bibfnamefont {H.}~\bibnamefont {Liu}}, \bibinfo {author} {\bibfnamefont
  {N.}~\bibnamefont {Liyanage}}, \bibinfo {author} {\bibfnamefont
  {J.}~\bibnamefont {Mammei}}, \bibinfo {author} {\bibfnamefont
  {R.}~\bibnamefont {Mammei}}, \bibinfo {author} {\bibfnamefont
  {M.}~\bibnamefont {McCaughan}}, \bibinfo {author} {\bibfnamefont
  {D.}~\bibnamefont {McNulty}}, \bibinfo {author} {\bibfnamefont
  {D.}~\bibnamefont {Meekins}}, \bibinfo {author} {\bibfnamefont
  {C.}~\bibnamefont {Metts}}, \bibinfo {author} {\bibfnamefont
  {R.}~\bibnamefont {Michaels}}, \bibinfo {author} {\bibfnamefont
  {M.}~\bibnamefont {Mihovilovic}}, \bibinfo {author} {\bibfnamefont {M.~M.}\
  \bibnamefont {Mondal}}, \bibinfo {author} {\bibfnamefont {J.}~\bibnamefont
  {Napolitano}}, \bibinfo {author} {\bibfnamefont {A.}~\bibnamefont {Narayan}},
  \bibinfo {author} {\bibfnamefont {D.}~\bibnamefont {Nikolaev}}, \bibinfo
  {author} {\bibfnamefont {V.}~\bibnamefont {Owen}}, \bibinfo {author}
  {\bibfnamefont {C.}~\bibnamefont {Palatchi}}, \bibinfo {author}
  {\bibfnamefont {J.}~\bibnamefont {Pan}}, \bibinfo {author} {\bibfnamefont
  {B.}~\bibnamefont {Pandey}}, \bibinfo {author} {\bibfnamefont
  {S.}~\bibnamefont {Park}}, \bibinfo {author} {\bibfnamefont {K.~D.}\
  \bibnamefont {Paschke}}, \bibinfo {author} {\bibfnamefont {M.}~\bibnamefont
  {Petrusky}}, \bibinfo {author} {\bibfnamefont {M.~L.}\ \bibnamefont {Pitt}},
  \bibinfo {author} {\bibfnamefont {S.}~\bibnamefont {Premathilake}}, \bibinfo
  {author} {\bibfnamefont {B.}~\bibnamefont {Quinn}}, \bibinfo {author}
  {\bibfnamefont {R.}~\bibnamefont {Radloff}}, \bibinfo {author} {\bibfnamefont
  {S.}~\bibnamefont {Rahman}}, \bibinfo {author} {\bibfnamefont {M.~N.~H.}\
  \bibnamefont {Rashad}}, \bibinfo {author} {\bibfnamefont {A.}~\bibnamefont
  {Rathnayake}}, \bibinfo {author} {\bibfnamefont {B.~T.}\ \bibnamefont
  {Reed}}, \bibinfo {author} {\bibfnamefont {P.~E.}\ \bibnamefont {Reimer}},
  \bibinfo {author} {\bibfnamefont {R.}~\bibnamefont {Richards}}, \bibinfo
  {author} {\bibfnamefont {S.}~\bibnamefont {Riordan}}, \bibinfo {author}
  {\bibfnamefont {Y.~R.}\ \bibnamefont {Roblin}}, \bibinfo {author}
  {\bibfnamefont {S.}~\bibnamefont {Seeds}}, \bibinfo {author} {\bibfnamefont
  {A.}~\bibnamefont {Shahinyan}}, \bibinfo {author} {\bibfnamefont
  {P.}~\bibnamefont {Souder}}, \bibinfo {author} {\bibfnamefont
  {M.}~\bibnamefont {Thiel}}, \bibinfo {author} {\bibfnamefont
  {Y.}~\bibnamefont {Tian}}, \bibinfo {author} {\bibfnamefont {G.~M.}\
  \bibnamefont {Urciuoli}}, \bibinfo {author} {\bibfnamefont {E.~W.}\
  \bibnamefont {Wertz}}, \bibinfo {author} {\bibfnamefont {B.}~\bibnamefont
  {Wojtsekhowski}}, \bibinfo {author} {\bibfnamefont {B.}~\bibnamefont {Yale}},
  \bibinfo {author} {\bibfnamefont {T.}~\bibnamefont {Ye}}, \bibinfo {author}
  {\bibfnamefont {A.}~\bibnamefont {Yoon}}, \bibinfo {author} {\bibfnamefont
  {W.}~\bibnamefont {Xiong}}, \bibinfo {author} {\bibfnamefont
  {A.}~\bibnamefont {Zec}}, \bibinfo {author} {\bibfnamefont {W.}~\bibnamefont
  {Zhang}}, \bibinfo {author} {\bibfnamefont {J.}~\bibnamefont {Zhang}},\ and\
  \bibinfo {author} {\bibfnamefont {X.}~\bibnamefont {Zheng}} (\bibinfo
  {collaboration} {CREX Collaboration}),\ }\bibfield  {title} {\bibinfo {title}
  {{Precision Determination of the Neutral Weak Form Factor of
  $^{48}\mathrm{Ca}$}},\ }\href
  {https://doi.org/10.1103/PhysRevLett.129.042501} {\bibfield  {journal}
  {\bibinfo  {journal} {Phys. Rev. Lett.}\ }\textbf {\bibinfo {volume} {129}},\
  \bibinfo {pages} {042501} (\bibinfo {year} {2022})}\BibitemShut {NoStop}%
\bibitem [{\citenamefont {Zenihiro}\ \emph {et~al.}(2010)\citenamefont
  {Zenihiro}, \citenamefont {Sakaguchi}, \citenamefont {Murakami},
  \citenamefont {Yosoi}, \citenamefont {Yasuda}, \citenamefont {Terashima},
  \citenamefont {Iwao}, \citenamefont {Takeda}, \citenamefont {Itoh},
  \citenamefont {Yoshida},\ and\ \citenamefont
  {Uchida}}]{Zenihiro2010Phys.Rev.C82_044611}%
  \BibitemOpen
  \bibfield  {author} {\bibinfo {author} {\bibfnamefont {J.}~\bibnamefont
  {Zenihiro}}, \bibinfo {author} {\bibfnamefont {H.}~\bibnamefont {Sakaguchi}},
  \bibinfo {author} {\bibfnamefont {T.}~\bibnamefont {Murakami}}, \bibinfo
  {author} {\bibfnamefont {M.}~\bibnamefont {Yosoi}}, \bibinfo {author}
  {\bibfnamefont {Y.}~\bibnamefont {Yasuda}}, \bibinfo {author} {\bibfnamefont
  {S.}~\bibnamefont {Terashima}}, \bibinfo {author} {\bibfnamefont
  {Y.}~\bibnamefont {Iwao}}, \bibinfo {author} {\bibfnamefont {H.}~\bibnamefont
  {Takeda}}, \bibinfo {author} {\bibfnamefont {M.}~\bibnamefont {Itoh}},
  \bibinfo {author} {\bibfnamefont {H.~P.}\ \bibnamefont {Yoshida}},\ and\
  \bibinfo {author} {\bibfnamefont {M.}~\bibnamefont {Uchida}},\ }\bibfield
  {title} {\bibinfo {title} {{Neutron density distributions of
  $^{204,206,208}\mathrm{Pb}$ deduced via proton elastic scattering at
  ${E}_{p}=295$ MeV}},\ }\href {https://doi.org/10.1103/PhysRevC.82.044611}
  {\bibfield  {journal} {\bibinfo  {journal} {Phys. Rev. C}\ }\textbf {\bibinfo
  {volume} {82}},\ \bibinfo {pages} {044611} (\bibinfo {year}
  {2010})}\BibitemShut {NoStop}%
\bibitem [{\citenamefont {Adhikari}\ \emph {et~al.}(2021)\citenamefont
  {Adhikari}, \citenamefont {Albataineh}, \citenamefont {Androic},
  \citenamefont {Aniol}, \citenamefont {Armstrong}, \citenamefont {Averett},
  \citenamefont {Ayerbe~Gayoso}, \citenamefont {Barcus}, \citenamefont
  {Bellini}, \citenamefont {Beminiwattha}, \citenamefont {Benesch},
  \citenamefont {Bhatt}, \citenamefont {Bhatta~Pathak}, \citenamefont
  {Bhetuwal}, \citenamefont {Blaikie}, \citenamefont {Campagna}, \citenamefont
  {Camsonne}, \citenamefont {Cates}, \citenamefont {Chen}, \citenamefont
  {Clarke}, \citenamefont {Cornejo}, \citenamefont {Covrig~Dusa}, \citenamefont
  {Datta}, \citenamefont {Deshpande}, \citenamefont {Dutta}, \citenamefont
  {Feldman}, \citenamefont {Fuchey}, \citenamefont {Gal}, \citenamefont
  {Gaskell}, \citenamefont {Gautam}, \citenamefont {Gericke}, \citenamefont
  {Ghosh}, \citenamefont {Halilovic}, \citenamefont {Hansen}, \citenamefont
  {Hauenstein}, \citenamefont {Henry}, \citenamefont {Horowitz}, \citenamefont
  {Jantzi}, \citenamefont {Jian}, \citenamefont {Johnston}, \citenamefont
  {Jones}, \citenamefont {Karki}, \citenamefont {Katugampola}, \citenamefont
  {Keppel}, \citenamefont {King}, \citenamefont {King}, \citenamefont {Knauss},
  \citenamefont {Kumar}, \citenamefont {Kutz}, \citenamefont
  {Lashley-Colthirst}, \citenamefont {Leverick}, \citenamefont {Liu},
  \citenamefont {Liyange}, \citenamefont {Malace}, \citenamefont {Mammei},
  \citenamefont {Mammei}, \citenamefont {McCaughan}, \citenamefont {McNulty},
  \citenamefont {Meekins}, \citenamefont {Metts}, \citenamefont {Michaels},
  \citenamefont {Mondal}, \citenamefont {Napolitano}, \citenamefont {Narayan},
  \citenamefont {Nikolaev}, \citenamefont {Rashad}, \citenamefont {Owen},
  \citenamefont {Palatchi}, \citenamefont {Pan}, \citenamefont {Pandey},
  \citenamefont {Park}, \citenamefont {Paschke}, \citenamefont {Petrusky},
  \citenamefont {Pitt}, \citenamefont {Premathilake}, \citenamefont {Puckett},
  \citenamefont {Quinn}, \citenamefont {Radloff}, \citenamefont {Rahman},
  \citenamefont {Rathnayake}, \citenamefont {Reed}, \citenamefont {Reimer},
  \citenamefont {Richards}, \citenamefont {Riordan}, \citenamefont {Roblin},
  \citenamefont {Seeds}, \citenamefont {Shahinyan}, \citenamefont {Souder},
  \citenamefont {Tang}, \citenamefont {Thiel}, \citenamefont {Tian},
  \citenamefont {Urciuoli}, \citenamefont {Wertz}, \citenamefont
  {Wojtsekhowski}, \citenamefont {Yale}, \citenamefont {Ye}, \citenamefont
  {Yoon}, \citenamefont {Zec}, \citenamefont {Zhang}, \citenamefont {Zhang},\
  and\ \citenamefont {Zheng}}]{Adhikari2021Phys.Rev.Lett.126_172502}%
  \BibitemOpen
  \bibfield  {author} {\bibinfo {author} {\bibfnamefont {D.}~\bibnamefont
  {Adhikari}}, \bibinfo {author} {\bibfnamefont {H.}~\bibnamefont
  {Albataineh}}, \bibinfo {author} {\bibfnamefont {D.}~\bibnamefont {Androic}},
  \bibinfo {author} {\bibfnamefont {K.}~\bibnamefont {Aniol}}, \bibinfo
  {author} {\bibfnamefont {D.~S.}\ \bibnamefont {Armstrong}}, \bibinfo {author}
  {\bibfnamefont {T.}~\bibnamefont {Averett}}, \bibinfo {author} {\bibfnamefont
  {C.}~\bibnamefont {Ayerbe~Gayoso}}, \bibinfo {author} {\bibfnamefont
  {S.}~\bibnamefont {Barcus}}, \bibinfo {author} {\bibfnamefont
  {V.}~\bibnamefont {Bellini}}, \bibinfo {author} {\bibfnamefont {R.~S.}\
  \bibnamefont {Beminiwattha}}, \bibinfo {author} {\bibfnamefont {J.~F.}\
  \bibnamefont {Benesch}}, \bibinfo {author} {\bibfnamefont {H.}~\bibnamefont
  {Bhatt}}, \bibinfo {author} {\bibfnamefont {D.}~\bibnamefont
  {Bhatta~Pathak}}, \bibinfo {author} {\bibfnamefont {D.}~\bibnamefont
  {Bhetuwal}}, \bibinfo {author} {\bibfnamefont {B.}~\bibnamefont {Blaikie}},
  \bibinfo {author} {\bibfnamefont {Q.}~\bibnamefont {Campagna}}, \bibinfo
  {author} {\bibfnamefont {A.}~\bibnamefont {Camsonne}}, \bibinfo {author}
  {\bibfnamefont {G.~D.}\ \bibnamefont {Cates}}, \bibinfo {author}
  {\bibfnamefont {Y.}~\bibnamefont {Chen}}, \bibinfo {author} {\bibfnamefont
  {C.}~\bibnamefont {Clarke}}, \bibinfo {author} {\bibfnamefont {J.~C.}\
  \bibnamefont {Cornejo}}, \bibinfo {author} {\bibfnamefont {S.}~\bibnamefont
  {Covrig~Dusa}}, \bibinfo {author} {\bibfnamefont {P.}~\bibnamefont {Datta}},
  \bibinfo {author} {\bibfnamefont {A.}~\bibnamefont {Deshpande}}, \bibinfo
  {author} {\bibfnamefont {D.}~\bibnamefont {Dutta}}, \bibinfo {author}
  {\bibfnamefont {C.}~\bibnamefont {Feldman}}, \bibinfo {author} {\bibfnamefont
  {E.}~\bibnamefont {Fuchey}}, \bibinfo {author} {\bibfnamefont
  {C.}~\bibnamefont {Gal}}, \bibinfo {author} {\bibfnamefont {D.}~\bibnamefont
  {Gaskell}}, \bibinfo {author} {\bibfnamefont {T.}~\bibnamefont {Gautam}},
  \bibinfo {author} {\bibfnamefont {M.}~\bibnamefont {Gericke}}, \bibinfo
  {author} {\bibfnamefont {C.}~\bibnamefont {Ghosh}}, \bibinfo {author}
  {\bibfnamefont {I.}~\bibnamefont {Halilovic}}, \bibinfo {author}
  {\bibfnamefont {J.-O.}\ \bibnamefont {Hansen}}, \bibinfo {author}
  {\bibfnamefont {F.}~\bibnamefont {Hauenstein}}, \bibinfo {author}
  {\bibfnamefont {W.}~\bibnamefont {Henry}}, \bibinfo {author} {\bibfnamefont
  {C.~J.}\ \bibnamefont {Horowitz}}, \bibinfo {author} {\bibfnamefont
  {C.}~\bibnamefont {Jantzi}}, \bibinfo {author} {\bibfnamefont
  {S.}~\bibnamefont {Jian}}, \bibinfo {author} {\bibfnamefont {S.}~\bibnamefont
  {Johnston}}, \bibinfo {author} {\bibfnamefont {D.~C.}\ \bibnamefont {Jones}},
  \bibinfo {author} {\bibfnamefont {B.}~\bibnamefont {Karki}}, \bibinfo
  {author} {\bibfnamefont {S.}~\bibnamefont {Katugampola}}, \bibinfo {author}
  {\bibfnamefont {C.}~\bibnamefont {Keppel}}, \bibinfo {author} {\bibfnamefont
  {P.~M.}\ \bibnamefont {King}}, \bibinfo {author} {\bibfnamefont {D.~E.}\
  \bibnamefont {King}}, \bibinfo {author} {\bibfnamefont {M.}~\bibnamefont
  {Knauss}}, \bibinfo {author} {\bibfnamefont {K.~S.}\ \bibnamefont {Kumar}},
  \bibinfo {author} {\bibfnamefont {T.}~\bibnamefont {Kutz}}, \bibinfo {author}
  {\bibfnamefont {N.}~\bibnamefont {Lashley-Colthirst}}, \bibinfo {author}
  {\bibfnamefont {G.}~\bibnamefont {Leverick}}, \bibinfo {author}
  {\bibfnamefont {H.}~\bibnamefont {Liu}}, \bibinfo {author} {\bibfnamefont
  {N.}~\bibnamefont {Liyange}}, \bibinfo {author} {\bibfnamefont
  {S.}~\bibnamefont {Malace}}, \bibinfo {author} {\bibfnamefont
  {R.}~\bibnamefont {Mammei}}, \bibinfo {author} {\bibfnamefont
  {J.}~\bibnamefont {Mammei}}, \bibinfo {author} {\bibfnamefont
  {M.}~\bibnamefont {McCaughan}}, \bibinfo {author} {\bibfnamefont
  {D.}~\bibnamefont {McNulty}}, \bibinfo {author} {\bibfnamefont
  {D.}~\bibnamefont {Meekins}}, \bibinfo {author} {\bibfnamefont
  {C.}~\bibnamefont {Metts}}, \bibinfo {author} {\bibfnamefont
  {R.}~\bibnamefont {Michaels}}, \bibinfo {author} {\bibfnamefont {M.~M.}\
  \bibnamefont {Mondal}}, \bibinfo {author} {\bibfnamefont {J.}~\bibnamefont
  {Napolitano}}, \bibinfo {author} {\bibfnamefont {A.}~\bibnamefont {Narayan}},
  \bibinfo {author} {\bibfnamefont {D.}~\bibnamefont {Nikolaev}}, \bibinfo
  {author} {\bibfnamefont {M.~N.~H.}\ \bibnamefont {Rashad}}, \bibinfo {author}
  {\bibfnamefont {V.}~\bibnamefont {Owen}}, \bibinfo {author} {\bibfnamefont
  {C.}~\bibnamefont {Palatchi}}, \bibinfo {author} {\bibfnamefont
  {J.}~\bibnamefont {Pan}}, \bibinfo {author} {\bibfnamefont {B.}~\bibnamefont
  {Pandey}}, \bibinfo {author} {\bibfnamefont {S.}~\bibnamefont {Park}},
  \bibinfo {author} {\bibfnamefont {K.~D.}\ \bibnamefont {Paschke}}, \bibinfo
  {author} {\bibfnamefont {M.}~\bibnamefont {Petrusky}}, \bibinfo {author}
  {\bibfnamefont {M.~L.}\ \bibnamefont {Pitt}}, \bibinfo {author}
  {\bibfnamefont {S.}~\bibnamefont {Premathilake}}, \bibinfo {author}
  {\bibfnamefont {A.~J.~R.}\ \bibnamefont {Puckett}}, \bibinfo {author}
  {\bibfnamefont {B.}~\bibnamefont {Quinn}}, \bibinfo {author} {\bibfnamefont
  {R.}~\bibnamefont {Radloff}}, \bibinfo {author} {\bibfnamefont
  {S.}~\bibnamefont {Rahman}}, \bibinfo {author} {\bibfnamefont
  {A.}~\bibnamefont {Rathnayake}}, \bibinfo {author} {\bibfnamefont {B.~T.}\
  \bibnamefont {Reed}}, \bibinfo {author} {\bibfnamefont {P.~E.}\ \bibnamefont
  {Reimer}}, \bibinfo {author} {\bibfnamefont {R.}~\bibnamefont {Richards}},
  \bibinfo {author} {\bibfnamefont {S.}~\bibnamefont {Riordan}}, \bibinfo
  {author} {\bibfnamefont {Y.}~\bibnamefont {Roblin}}, \bibinfo {author}
  {\bibfnamefont {S.}~\bibnamefont {Seeds}}, \bibinfo {author} {\bibfnamefont
  {A.}~\bibnamefont {Shahinyan}}, \bibinfo {author} {\bibfnamefont
  {P.}~\bibnamefont {Souder}}, \bibinfo {author} {\bibfnamefont
  {L.}~\bibnamefont {Tang}}, \bibinfo {author} {\bibfnamefont {M.}~\bibnamefont
  {Thiel}}, \bibinfo {author} {\bibfnamefont {Y.}~\bibnamefont {Tian}},
  \bibinfo {author} {\bibfnamefont {G.~M.}\ \bibnamefont {Urciuoli}}, \bibinfo
  {author} {\bibfnamefont {E.~W.}\ \bibnamefont {Wertz}}, \bibinfo {author}
  {\bibfnamefont {B.}~\bibnamefont {Wojtsekhowski}}, \bibinfo {author}
  {\bibfnamefont {B.}~\bibnamefont {Yale}}, \bibinfo {author} {\bibfnamefont
  {T.}~\bibnamefont {Ye}}, \bibinfo {author} {\bibfnamefont {A.}~\bibnamefont
  {Yoon}}, \bibinfo {author} {\bibfnamefont {A.}~\bibnamefont {Zec}}, \bibinfo
  {author} {\bibfnamefont {W.}~\bibnamefont {Zhang}}, \bibinfo {author}
  {\bibfnamefont {J.}~\bibnamefont {Zhang}},\ and\ \bibinfo {author}
  {\bibfnamefont {X.}~\bibnamefont {Zheng}} (\bibinfo {collaboration} {PREX
  Collaboration}),\ }\bibfield  {title} {\bibinfo {title} {{Accurate
  Determination of the Neutron Skin Thickness of $^{208}\mathrm{Pb}$ through
  Parity-Violation in Electron Scattering}},\ }\href
  {https://doi.org/10.1103/PhysRevLett.126.172502} {\bibfield  {journal}
  {\bibinfo  {journal} {Phys. Rev. Lett.}\ }\textbf {\bibinfo {volume} {126}},\
  \bibinfo {pages} {172502} (\bibinfo {year} {2021})}\BibitemShut {NoStop}%
\bibitem [{\citenamefont {Reinhard}\ \emph {et~al.}(2021)\citenamefont
  {Reinhard}, \citenamefont {Roca-Maza},\ and\ \citenamefont
  {Nazarewicz}}]{Reinhard2021Phys.Rev.Lett.127_232501}%
  \BibitemOpen
  \bibfield  {author} {\bibinfo {author} {\bibfnamefont {P.-G.}\ \bibnamefont
  {Reinhard}}, \bibinfo {author} {\bibfnamefont {X.}~\bibnamefont
  {Roca-Maza}},\ and\ \bibinfo {author} {\bibfnamefont {W.}~\bibnamefont
  {Nazarewicz}},\ }\bibfield  {title} {\bibinfo {title} {{Information Content
  of the Parity-Violating Asymmetry in $^{208}\mathrm{Pb}$}},\ }\href
  {https://doi.org/10.1103/PhysRevLett.127.232501} {\bibfield  {journal}
  {\bibinfo  {journal} {Phys. Rev. Lett.}\ }\textbf {\bibinfo {volume} {127}},\
  \bibinfo {pages} {232501} (\bibinfo {year} {2021})}\BibitemShut {NoStop}%
\bibitem [{\citenamefont {Baldo}\ and\ \citenamefont
  {Burgio}(2016)}]{Baldo2016Prog.Part.Nucl.Phys.91_203}%
  \BibitemOpen
  \bibfield  {author} {\bibinfo {author} {\bibfnamefont {M.}~\bibnamefont
  {Baldo}}\ and\ \bibinfo {author} {\bibfnamefont {G.~F.}\ \bibnamefont
  {Burgio}},\ }\bibfield  {title} {\bibinfo {title} {{The nuclear symmetry
  energy}},\ }\href {https://doi.org/10.1016/j.ppnp.2016.06.006} {\bibfield
  {journal} {\bibinfo  {journal} {Prog. Part. Nucl. Phys.}\ }\textbf {\bibinfo
  {volume} {91}},\ \bibinfo {pages} {203} (\bibinfo {year} {2016})}\BibitemShut
  {NoStop}%
\bibitem [{\citenamefont {Hamamoto}\ and\ \citenamefont
  {Sagawa}(1993)}]{Hamamoto1993Phys.Rev.C48_R960}%
  \BibitemOpen
  \bibfield  {author} {\bibinfo {author} {\bibfnamefont {I.}~\bibnamefont
  {Hamamoto}}\ and\ \bibinfo {author} {\bibfnamefont {H.}~\bibnamefont
  {Sagawa}},\ }\bibfield  {title} {\bibinfo {title} {{Gamow-Teller beta decay
  and isospin impurity in nuclei near the proton drip line}},\ }\href
  {https://doi.org/10.1103/PhysRevC.48.R960} {\bibfield  {journal} {\bibinfo
  {journal} {Phys. Rev. C}\ }\textbf {\bibinfo {volume} {48}},\ \bibinfo
  {pages} {R960} (\bibinfo {year} {1993})}\BibitemShut {NoStop}%
\bibitem [{\citenamefont {Sagawa}\ \emph {et~al.}(1996)\citenamefont {Sagawa},
  \citenamefont {Giai},\ and\ \citenamefont
  {Suzuki}}]{Sagawa1996Phys.Rev.C53_2163}%
  \BibitemOpen
  \bibfield  {author} {\bibinfo {author} {\bibfnamefont {H.}~\bibnamefont
  {Sagawa}}, \bibinfo {author} {\bibfnamefont {N.~V.}\ \bibnamefont {Giai}},\
  and\ \bibinfo {author} {\bibfnamefont {T.}~\bibnamefont {Suzuki}},\
  }\bibfield  {title} {\bibinfo {title} {{Effect of isospin mixing on
  superallowed Fermi $ \beta $ decay}},\ }\href
  {https://doi.org/10.1103/PhysRevC.53.2163} {\bibfield  {journal} {\bibinfo
  {journal} {Phys. Rev. C}\ }\textbf {\bibinfo {volume} {53}},\ \bibinfo
  {pages} {2163} (\bibinfo {year} {1996})}\BibitemShut {NoStop}%
\bibitem [{\citenamefont {Sagawa}\ \emph {et~al.}(2022)\citenamefont {Sagawa},
  \citenamefont {Yoshida}, \citenamefont {Naito}, \citenamefont {Uesaka},
  \citenamefont {Zenihiro}, \citenamefont {Tanaka},\ and\ \citenamefont
  {Suzuki}}]{Sagawa2022Phys.Lett.B829_137072}%
  \BibitemOpen
  \bibfield  {author} {\bibinfo {author} {\bibfnamefont {H.}~\bibnamefont
  {Sagawa}}, \bibinfo {author} {\bibfnamefont {S.}~\bibnamefont {Yoshida}},
  \bibinfo {author} {\bibfnamefont {T.}~\bibnamefont {Naito}}, \bibinfo
  {author} {\bibfnamefont {T.}~\bibnamefont {Uesaka}}, \bibinfo {author}
  {\bibfnamefont {J.}~\bibnamefont {Zenihiro}}, \bibinfo {author}
  {\bibfnamefont {J.}~\bibnamefont {Tanaka}},\ and\ \bibinfo {author}
  {\bibfnamefont {T.}~\bibnamefont {Suzuki}},\ }\bibfield  {title} {\bibinfo
  {title} {{Isovector density and isospin impurity in $ {}^{40} \mathrm{Ca}
  $}},\ }\href {https://doi.org/10.1016/j.physletb.2022.137072} {\bibfield
  {journal} {\bibinfo  {journal} {Phys. Lett. B}\ }\textbf {\bibinfo {volume}
  {829}},\ \bibinfo {pages} {137072} (\bibinfo {year} {2022})}\BibitemShut
  {NoStop}%
\bibitem [{\citenamefont {Centelles}\ \emph {et~al.}(2010)\citenamefont
  {Centelles}, \citenamefont {Roca-Maza}, \citenamefont {Vi\~nas},\ and\
  \citenamefont {Warda}}]{Centelles2010Phys.Rev.C82_054314}%
  \BibitemOpen
  \bibfield  {author} {\bibinfo {author} {\bibfnamefont {M.}~\bibnamefont
  {Centelles}}, \bibinfo {author} {\bibfnamefont {X.}~\bibnamefont
  {Roca-Maza}}, \bibinfo {author} {\bibfnamefont {X.}~\bibnamefont {Vi\~nas}},\
  and\ \bibinfo {author} {\bibfnamefont {M.}~\bibnamefont {Warda}},\ }\bibfield
   {title} {\bibinfo {title} {{Origin of the neutron skin thickness of
  $^{208}\mathrm{Pb}$ in nuclear mean-field models}},\ }\href
  {https://doi.org/10.1103/PhysRevC.82.054314} {\bibfield  {journal} {\bibinfo
  {journal} {Phys. Rev. C}\ }\textbf {\bibinfo {volume} {82}},\ \bibinfo
  {pages} {054314} (\bibinfo {year} {2010})}\BibitemShut {NoStop}%
\bibitem [{\citenamefont {Goriely}\ and\ \citenamefont
  {Pearson}(2008)}]{Goriely2008Phys.Rev.C77_031301}%
  \BibitemOpen
  \bibfield  {author} {\bibinfo {author} {\bibfnamefont {S.}~\bibnamefont
  {Goriely}}\ and\ \bibinfo {author} {\bibfnamefont {J.~M.}\ \bibnamefont
  {Pearson}},\ }\bibfield  {title} {\bibinfo {title} {{Further explorations of
  Skyrme-Hartree-Fock-Bogoliubov mass formulas. VIII. Role of Coulomb
  exchange}},\ }\href {https://doi.org/10.1103/PhysRevC.77.031301} {\bibfield
  {journal} {\bibinfo  {journal} {Phys. Rev. C}\ }\textbf {\bibinfo {volume}
  {77}},\ \bibinfo {pages} {031301} (\bibinfo {year} {2008})}\BibitemShut
  {NoStop}%
\bibitem [{\citenamefont {Horowitz}\ and\ \citenamefont
  {Piekarewicz}(2012)}]{Horowitz2012Phys.Rev.C86_045503}%
  \BibitemOpen
  \bibfield  {author} {\bibinfo {author} {\bibfnamefont {C.~J.}\ \bibnamefont
  {Horowitz}}\ and\ \bibinfo {author} {\bibfnamefont {J.}~\bibnamefont
  {Piekarewicz}},\ }\bibfield  {title} {\bibinfo {title} {{Impact of spin-orbit
  currents on the electroweak skin of neutron-rich nuclei}},\ }\href
  {https://doi.org/10.1103/PhysRevC.86.045503} {\bibfield  {journal} {\bibinfo
  {journal} {Phys. Rev. C}\ }\textbf {\bibinfo {volume} {86}},\ \bibinfo
  {pages} {045503} (\bibinfo {year} {2012})}\BibitemShut {NoStop}%
\bibitem [{\citenamefont {Reinhard}\ and\ \citenamefont
  {Nazarewicz}(2021)}]{Reinhard2021Phys.Rev.C103_054310}%
  \BibitemOpen
  \bibfield  {author} {\bibinfo {author} {\bibfnamefont {P.-G.}\ \bibnamefont
  {Reinhard}}\ and\ \bibinfo {author} {\bibfnamefont {W.}~\bibnamefont
  {Nazarewicz}},\ }\bibfield  {title} {\bibinfo {title} {{Nuclear charge
  densities in spherical and deformed nuclei: Toward precise calculations of
  charge radii}},\ }\href {https://doi.org/10.1103/PhysRevC.103.054310}
  {\bibfield  {journal} {\bibinfo  {journal} {Phys. Rev. C}\ }\textbf {\bibinfo
  {volume} {103}},\ \bibinfo {pages} {054310} (\bibinfo {year}
  {2021})}\BibitemShut {NoStop}%
\bibitem [{\citenamefont {Naito}\ \emph {et~al.}(2021)\citenamefont {Naito},
  \citenamefont {Col\`o}, \citenamefont {Liang},\ and\ \citenamefont
  {Roca-Maza}}]{Naito2021Phys.Rev.C104_024316}%
  \BibitemOpen
  \bibfield  {author} {\bibinfo {author} {\bibfnamefont {T.}~\bibnamefont
  {Naito}}, \bibinfo {author} {\bibfnamefont {G.}~\bibnamefont {Col\`o}},
  \bibinfo {author} {\bibfnamefont {H.}~\bibnamefont {Liang}},\ and\ \bibinfo
  {author} {\bibfnamefont {X.}~\bibnamefont {Roca-Maza}},\ }\bibfield  {title}
  {\bibinfo {title} {{Second and fourth moments of the charge density and
  neutron-skin thickness of atomic nuclei}},\ }\href
  {https://doi.org/10.1103/PhysRevC.104.024316} {\bibfield  {journal} {\bibinfo
   {journal} {Phys. Rev. C}\ }\textbf {\bibinfo {volume} {104}},\ \bibinfo
  {pages} {024316} (\bibinfo {year} {2021})}\BibitemShut {NoStop}%
\bibitem [{\citenamefont {Angeli}\ and\ \citenamefont
  {Marinova}(2013)}]{Angeli2013At.DataNucl.DataTables99_69}%
  \BibitemOpen
  \bibfield  {author} {\bibinfo {author} {\bibfnamefont {I.}~\bibnamefont
  {Angeli}}\ and\ \bibinfo {author} {\bibfnamefont {K.~P.}\ \bibnamefont
  {Marinova}},\ }\bibfield  {title} {\bibinfo {title} {{Table of experimental
  nuclear ground state charge radii: An update}},\ }\href
  {https://doi.org/10.1016/j.adt.2011.12.006} {\bibfield  {journal} {\bibinfo
  {journal} {At. Data Nucl. Data Tables}\ }\textbf {\bibinfo {volume} {99}},\
  \bibinfo {pages} {69} (\bibinfo {year} {2013})}\BibitemShut {NoStop}%
\bibitem [{\citenamefont {Huang}\ \emph {et~al.}(2021)\citenamefont {Huang},
  \citenamefont {Wang}, \citenamefont {Kondev}, \citenamefont {Audi},\ and\
  \citenamefont {Naimi}}]{Huang2021Chin.Phys.C45_030002}%
  \BibitemOpen
  \bibfield  {author} {\bibinfo {author} {\bibfnamefont {W.~J.}\ \bibnamefont
  {Huang}}, \bibinfo {author} {\bibfnamefont {M.}~\bibnamefont {Wang}},
  \bibinfo {author} {\bibfnamefont {F.}~\bibnamefont {Kondev}}, \bibinfo
  {author} {\bibfnamefont {G.}~\bibnamefont {Audi}},\ and\ \bibinfo {author}
  {\bibfnamefont {S.}~\bibnamefont {Naimi}},\ }\bibfield  {title} {\bibinfo
  {title} {{The AME 2020 atomic mass evaluation (I). Evaluation of input data,
  and adjustment procedures}},\ }\href
  {https://doi.org/10.1088/1674-1137/abddb0} {\bibfield  {journal} {\bibinfo
  {journal} {Chin. Phys. C}\ }\textbf {\bibinfo {volume} {45}},\ \bibinfo
  {pages} {030002} (\bibinfo {year} {2021})}\BibitemShut {NoStop}%
\bibitem [{\citenamefont {Reinhard}\ and\ \citenamefont
  {Nazarewicz}(2017)}]{Reinhard2017Phys.Rev.C95_064328}%
  \BibitemOpen
  \bibfield  {author} {\bibinfo {author} {\bibfnamefont {P.-G.}\ \bibnamefont
  {Reinhard}}\ and\ \bibinfo {author} {\bibfnamefont {W.}~\bibnamefont
  {Nazarewicz}},\ }\bibfield  {title} {\bibinfo {title} {{Toward a global
  description of nuclear charge radii: Exploring the Fayans energy density
  functional}},\ }\href {https://doi.org/10.1103/PhysRevC.95.064328} {\bibfield
   {journal} {\bibinfo  {journal} {Phys. Rev. C}\ }\textbf {\bibinfo {volume}
  {95}},\ \bibinfo {pages} {064328} (\bibinfo {year} {2017})}\BibitemShut
  {NoStop}%
\bibitem [{\citenamefont {Vi\~{n}as}\ \emph {et~al.}(2014)\citenamefont
  {Vi\~{n}as}, \citenamefont {Centelles}, \citenamefont {Roca-Maza},\ and\
  \citenamefont {Warda}}]{Vinas2014Eur.Phys.J.A50_27}%
  \BibitemOpen
  \bibfield  {author} {\bibinfo {author} {\bibfnamefont {X.}~\bibnamefont
  {Vi\~{n}as}}, \bibinfo {author} {\bibfnamefont {M.}~\bibnamefont
  {Centelles}}, \bibinfo {author} {\bibfnamefont {X.}~\bibnamefont
  {Roca-Maza}},\ and\ \bibinfo {author} {\bibfnamefont {M.}~\bibnamefont
  {Warda}},\ }\bibfield  {title} {\bibinfo {title} {{Density dependence of the
  symmetry energy from neutron skin thickness in finite nuclei}},\ }\href
  {https://doi.org/10.1140/epja/i2014-14027-8} {\bibfield  {journal} {\bibinfo
  {journal} {Eur. Phys. J. A}\ }\textbf {\bibinfo {volume} {50}},\ \bibinfo
  {pages} {27} (\bibinfo {year} {2014})}\BibitemShut {NoStop}%
\bibitem [{\citenamefont {Sotani}\ \emph {et~al.}(2022)\citenamefont {Sotani},
  \citenamefont {Nishimura},\ and\ \citenamefont
  {Naito}}]{Sotani2022Prog.Theor.Exp.Phys.2022_041D01}%
  \BibitemOpen
  \bibfield  {author} {\bibinfo {author} {\bibfnamefont {H.}~\bibnamefont
  {Sotani}}, \bibinfo {author} {\bibfnamefont {N.}~\bibnamefont {Nishimura}},\
  and\ \bibinfo {author} {\bibfnamefont {T.}~\bibnamefont {Naito}},\ }\bibfield
   {title} {\bibinfo {title} {{New constraints on the neutron-star mass and
  radius relation from terrestrial nuclear experiments}},\ }\href
  {https://doi.org/10.1093/ptep/ptac055} {\bibfield  {journal} {\bibinfo
  {journal} {Prog. Theor. Exp. Phys.}\ }\textbf {\bibinfo {volume} {2022}},\
  \bibinfo {pages} {041D01} (\bibinfo {year} {2022})}\BibitemShut {NoStop}%
\bibitem [{\citenamefont
  {Yukawa}(1935)}]{Yukawa1935Proc.Phys.Math.Soc.Jpn.Third17_48}%
  \BibitemOpen
  \bibfield  {author} {\bibinfo {author} {\bibfnamefont {H.}~\bibnamefont
  {Yukawa}},\ }\bibfield  {title} {\bibinfo {title} {{On the Interaction of
  Elementary Particles. I}},\ }\href
  {https://doi.org/10.11429/ppmsj1919.17.0_48} {\bibfield  {journal} {\bibinfo
  {journal} {Proc. Phys. Math. Soc. Jpn. Third}\ }\textbf {\bibinfo {volume}
  {17}},\ \bibinfo {pages} {48} (\bibinfo {year} {1935})}\BibitemShut {NoStop}%
\bibitem [{\citenamefont {Machleidt}\ \emph {et~al.}(1987)\citenamefont
  {Machleidt}, \citenamefont {Holinde},\ and\ \citenamefont
  {Elster}}]{Machleidt1987Phys.Rep.149_1}%
  \BibitemOpen
  \bibfield  {author} {\bibinfo {author} {\bibfnamefont {R.}~\bibnamefont
  {Machleidt}}, \bibinfo {author} {\bibfnamefont {K.}~\bibnamefont {Holinde}},\
  and\ \bibinfo {author} {\bibfnamefont {C.}~\bibnamefont {Elster}},\
  }\bibfield  {title} {\bibinfo {title} {{The Bonn meson-exchange model for the
  nucleon--nucleon interaction}},\ }\href
  {https://doi.org/10.1016/S0370-1573(87)80002-9} {\bibfield  {journal}
  {\bibinfo  {journal} {Phys. Rep.}\ }\textbf {\bibinfo {volume} {149}},\
  \bibinfo {pages} {1} (\bibinfo {year} {1987})}\BibitemShut {NoStop}%
\end{thebibliography}
\end{document}